\documentclass[twocolumn,floatfix,superscriptaddress,longbibliography,PRL]{revtex4-2}
\usepackage{amsfonts,amssymb,amsmath,hyperref}
\usepackage[T1]{fontenc}
\usepackage[utf8]{inputenc}

\renewcommand{\.}{{\cdot}}

\newcommand{\beq}{\begin{equation}}
\newcommand{\eeq}{\end{equation}}
\newcommand{\bea}{\begin{eqnarray}}
\newcommand{\eea}{\end{eqnarray}}

\renewcommand{\a}{\alpha}

\renewcommand{\l}{\lambda}

\usepackage{mathtools}
\usepackage{etoolbox}
\def\indentit{\mbox{l\hspace{-0.55em}1}}

\usepackage[paperwidth=210mm,paperheight=297mm,centering,hmargin=2cm,vmargin=2cm]{geometry}

\newif\ifhyper
\hypertrue
\ifhyper
\hypersetup{
citecolor = {green},
colorlinks = {true}, 
urlcolor = {blue} 
}
\fi

\newlength{\ldag}
\renewcommand{\l}{\lambda}

\newcommand{\dbar}{\mathchar'26\mkern-12mu d}

\def\be{\begin{equation}}
\def\ee{\end{equation}}
\def\bea{\begin{eqnarray}}
\def\eea{\end{eqnarray}}
\def\bse{\begin{subequations}}
\def\ese{\end{subequations}}
\def\bc{\begin{center}}
\def\ec{\end{center}}

\allowdisplaybreaks 

\begin{document}

\title{Renormalization group approach to the elastic properties of graphene bilayers}

\author{L. Delzescaux} 
\email{louise.delzescaux@sorbonne-universite.fr}
\affiliation{Sorbonne Universit\'e, CNRS, Laboratoire de Physique Th\'eorique de la Mati\`ere Condens\'ee, LPTMC, 75005 Paris, France}
\author{D. Mouhanna} 
\email{mouhanna@lptmc.jussieu.fr}
\affiliation{Sorbonne Universit\'e, CNRS, Laboratoire de Physique Th\'eorique de la Mati\`ere Condens\'ee, LPTMC, 75005 Paris, France}


\begin{abstract}

We investigate the effects of thermal fluctuations in graphene bilayers by means of a nonperturbative renormalization group (NPRG) approach, following the pioneering work of Mauri {\it et al.} [Phys. Rev. B {\bf 102}, 165421 (2020)] based on a self-consistent screening approximation (SCSA). We consider a model of two continuum polymerized membranes, separated by a distance $\ell$, in their flat phase, coupled by interlayer shear, compression/dilatation and elastic terms. Within a controlled truncation of the effective average  action, we retain only the contributions that generate a pronounced crossover of the effective bending rigidity along the renormalization group flow between two regimes: at high running scale $k$, the rigidity is dominated by the in-plane elastic properties, with $\kappa_{\mathrm{eff}}\sim \ell^{2}(\lambda+2\mu)/2$, whereas at low $k$ it is controlled by the bending rigidity of two independent monolayers, $\kappa_{\mathrm{eff}}\sim 2\kappa$. This crossover is reminiscent of that observed by Mauri {\it et al.} as a function of the wavevector scale $q$, but here it is obtained within a renormalization group framework. This has several advantages. First, although approximations are performed, the NPRG approach allows one, in principle, to take into account all nonlinearities present in the elastic theory, in contrast to the SCSA treatment which requires, already at the formal level, significant simplifications. Second, it demonstrates that the bilayer problem can be treated as a straightforward extension of the monolayer case, with flow equations that keep the same structure and differ only by bilayer-specific adjustments. Third, unlike the SCSA, the NPRG framework  admits a controlled, systematically improvable, hierarchy of approximations.

\end{abstract}

\maketitle

\section{Introduction}

Since its isolation in 2004, graphene \cite{novoselov04,novoselov05} has attracted immense interest owing to its outstanding physical properties, which set it apart from most other materials. Its key features include extremely high electrical conductivity, arising from the large mobility of its charge carriers, exceptional thermal conductivity, and mechanical strength far exceeding that of steel while remaining extremely lightweight  (see e.g., Refs.~\cite{katsnelson12,akinwande17}). Moreover, despite its finite electronic density, graphene is almost perfectly transparent. Its unique electronic structure, with Dirac fermions behaving as relativistic quasiparticles, also enables the exploration of unconventional quantum regimes \cite{castroneto09,katsnelson12}.

While graphene monolayers have been extensively studied, interest in graphene bilayers has grown steadily in recent years (see, e.g., Refs.~\cite{gao20,nimbalkar20,pantaleon23}). Composed of two stacked graphene sheets, bilayers exhibit tunable electronic properties that distinguish them from monolayer graphene (see, e.g., Refs.~\cite{mcCann13,rozhkov16}). These structures offer novel possibilities, including band-gap modulation under a perpendicular electric field, as well as the emergence of exotic phases such as superconductivity and Mott insulating states in specific configurations, notably in twisted bilayer graphene (see, e.g., Refs.~\cite{andrei20,chu20,nimbalkar20,calderon20}).

Here we focus on the elastic properties of these systems, which, as is well known from the study of monolayers, already provide an extremely fruitful interplay between two-dimensional fluctuating geometry and thermal fluctuations \cite{NelsonPiranWeinberg2004}. We therefore now turn to the modeling of the elasto-mechanical properties of graphene. At long distances, graphene can be described as a two-dimensional polymerized membrane embedded in three-dimensional space (see, e.g., Refs.~\cite{fasolino07,katsnelson13,los17}), where out-of-plane flexural fluctuations, ${\bf h}$, are coupled to in-plane elastic phonon fluctuations, ${\bf u}$. Within a Gaussian approximation, integrating out the phonon degrees of freedom generates an effective long-range interaction between flexural modes, which plays a crucial role in stabilizing long-range order \cite{nelson87,aronovitz89,coquand19b}, in spite of the Mermin--Wagner theorem \cite{mermin66,hohenberg67}. This mechanism underlies, in particular, the remarkable mechanical stability of graphene.

Moreover, it has been shown that in the ordered, flat phase of two-dimensional polymerized membranes, which is controlled from the renormalization group (RG) point of view by an infrared fixed point \cite{aronovitz88,david88}, one observes power-law behaviors for both the phonon--phonon and flexuron--flexuron correlation functions \cite{aronovitz88,guitter88,aronovitz89,guitter89}:
\begin{equation}
G_{uu}(q)\underset{\bar {\bf q}\to 0} \sim q^{-(4-2\eta)}  \hspace{0.5cm} G_{hh}(q)\underset{\bar {\bf q}\to 0}\sim q^{-(4-\eta)},
\label{correlation}
\end{equation}
where $\eta$ is the unique and emblematic exponent characterizing this infrared fixed point. It also governs the scale dependence of the renormalized bending rigidity:
\begin{equation}
\kappa_R(q)\sim q^{-\eta}.
\label{kappaRn}
\end{equation}
The  exponent $\eta$ has been estimated by various analytical and numerical techniques. Self-consistent screening approximation (SCSA) calculations \cite{ledoussal92,zakharchenko10,roldan11,ledoussal18} already predicted a strong stiffening of $\kappa_R(q)$, with $\eta \simeq 0.821$ at leading order and $\eta \simeq 0.789$  at next-to-leading order \cite{gazit09}. High-precision Monte Carlo and molecular dynamics simulations \cite{zhang93,bowick96,troster13} later reported values in the range $\eta \simeq 0.72$-$0.81$, e.g. $\eta = 0.795(10)$ \cite{troster13}, confirming both the existence of nontrivial anomalous elasticity and the importance of corrections to scaling. More recently, high-order field-theoretic RG \cite{coquand20a,metayer22,metayer22d,metayer23,pikelner22} and nonperturbative RG (NPRG) approaches \cite{kownacki09,braghin10,hasselmann11,essafi14} have yielded estimates clustered in the range $\eta \approx 0.8$-$0.9$. In particular, three- and four-loop results and NPRG analyses are mutually consistent, reinforcing the picture of a universal, strongly renormalized flat phase. These last values are also in  very close agreement with Monte Carlo and molecular-dynamics simulations using realistic interaction potentials for graphene \cite{los09}.

Regarding graphene bilayers, let us first discuss the experimental situation; see
Ref.~\cite{mauri21bis} for an earlier overview. The effective bending rigidity of suspended
graphene membranes and few-layer stacks has been extracted from nanoindentation and
snap-through measurements on buckled drums. In particular, Lindahl {\it et al.} \cite{lindahl12} have reported
a room-temperature bending rigidity of $\kappa \simeq 35.5^{+20}_{-15}\,\text{eV}$ for bilayer
graphene, significantly larger than typical monolayer estimates of order
$1$-$2\,\text{eV}$, and, more generally, a marked increase of apparent stiffness with layer number. More recent modal-analysis experiments on monolayer graphene and helium-atom-scattering measurements on AB-stacked bilayer graphene have shown that the effective bending rigidity depends sensitively on both temperature and system size and, in the bilayer case, can exceed significantly twice the microscopic monolayer value, in agreement with the idea of a thermally
renormalized $\kappa_{\mathrm{eff}}(q,T)$~\cite{Sajadi18,Eder21}.

On the numerical side, atomistic simulations have provided a complementary microscopic picture. Monte Carlo simulations of bilayer graphene by Zakharchenko {\it et al.} \cite{zakharchenko10} have shown that the bending rigidity of a free bilayer is approximately twice that of a monolayer, and that there is a crossover from correlated to uncorrelated out-of-plane fluctuations of the two layers at a wave vector of order $q\sim 3\,\mathrm{nm}^{-1}$, highlighting the role of interlayer coupling in the long-wavelength flexural response.More recently, Herrero and Ram{\'i}rez~\cite{herrero23} have carried out extensive molecular-dynamics simulations of bilayer graphene over a wide temperature range, showing that the in-plane compression modulus of the bilayer is larger than that of a monolayer but decreases with increasing temperature, and that the onset of mechanical instability under in-plane compression is controlled by a size-dependent spinodal pressure and pronounced finite-size
effects in the out-of-plane fluctuations and elastic constants. Their analysis of height fluctuations also indicates that out-of-plane corrugations are significantly reduced in the bilayer compared to a monolayer of the same lateral size and temperature, consistently with an enhanced effective bending rigidity. In addition, molecular-dynamics and Monte Carlo simulations of few-layer graphene and related van der Waals stacks reveal a rich thickness dependence of the flexural response, including a strong softening of the apparent bending stiffness because of interlayer shear and slip at large bending angles~\cite{Han20,Jiang23}. Taken together, these studies indicate that the mechanical response of bilayer and multilayer membranes is governed by
a nontrivial interplay between in-plane elasticity, interlayer coupling, and thermal fluctuations, and call for a theoretical framework capable of treating these ingredients on an equal footing and across scales.

On the theoretical side, only a few investigations have been carried out. A long-distance elastic action incorporating interlayer interactions was first proposed by de~Andr\'es {\it et al.} in Ref.~\cite{andres12}, and later derived more systematically by Mauri {\it et al.} in Ref.~\cite{mauri21bis}. The latter authors studied this
model within a SCSA framework and identified a crossover for the (bare) bending rigidity between two regimes: at short distances, a behavior controlled by the in-plane elastic properties, and at long distances, a behavior dominated by the bending rigidity of the two layers. This specific rigidity crossover comes in addition to the usual one already known for monolayers, namely the crossover between a quasi-harmonic regime at short distances and a strong-coupling, anharmonic
regime at long distances, induced by nonlinear interactions between flexural modes.

However, the approach developed and used by Mauri {\it et al.} \cite{mauri21bis} disregards contributions that could play a significant role, in particular the nonlinearities associated with fluctuations of the relative coordinates of the two monolayers. This simplification is the price to pay to construct an effective flexural-mode theory formally analogous to that used for graphene monolayers, which moreover involves a wavevector-dependent bare bending rigidity. As a consequence, the resulting SCSA equations become rather cumbersome and  offer limited physical transparency. Finally,  at least for membranes, it has been argued that the SCSA, in its current formulation, cannot be systematically improved beyond the leading-order approximation \cite{gazit09}.

In this work, we extend the nonperturbative renormalization group (NPRG) framework developed for monolayers to symmetric bilayer crystalline membranes \cite{kownacki09,braghin10, hasselmann11}. We formulate the problem directly in terms of the average position field and the relative displacement field, which allows us to preserve full rotational invariance and, in principle, to retain all nonlinearities of the elastic theory. Within a simple truncation of the average  effective  action, the flow equations keep the same structure as in the monolayer case, and the bilayer nature enters only through a modified flexural propagator for the mean-height mode. This enables us to compute the scale dependence of the effective bending rigidity of a bilayer, $\kappa_{\mathrm{eff}}(q)$, and to recover explicitly the full crossover from the short-scale regime $\kappa_{\mathrm{eff}}(q) \sim \ell^{2}(\lambda+2\mu)/2$ to the large-scale regime $\kappa_{\mathrm{eff}}(q) \sim 2\kappa$. Our results thus provide a controlled, systematically improvable, symmetry-preserving alternative to SCSA-type treatments of bilayers \cite{mauri21bis} and clarify in which sense the bilayer problem can be viewed as a straightforward extension of the monolayer case.

Finally, we emphasize that in the present article we focus primarily on the methodological and conceptual aspects of the approach, rather than on a precise quantitative analysis, which will be addressed in a more systematic study.

The structure of this article is as follows. In Sec.~II A, we recall the long-distance elastic action relevant for describing a single polymerized membrane in its ordered, flat phase. In Sec.~II B, we extend this description to the case of membrane bilayers. Section III A provides a brief overview of the renormalization group procedure based on the Wetterich equation, which allows us to derive nonperturbative flow equations. In Sec.~III B, we revisit the RG treatment of a single-layer membrane, before deriving in Sec.~III C the corresponding flow equations for the membrane bilayer system. Finally, in Sec.~IV, we present and discuss the physical implications of our results.

\section{Elastic actions for  polymerized membranes}

\subsection{Membrane monolayers}

\subsubsection{Action}

Here we introduce the action relevant for a study of a single membrane. We consider a two-dimensional membrane embedded in a three-dimensional Euclidean space. Each point of the membrane is labeled by internal coordinates ${\bf x}\in\mathbb{R}^2$, and its position in embedding space is given by the mapping ${\bf x}\mapsto{\bf R}({\bf x})$, where ${\bf R}$ is a vector field in $\mathbb{R}^3$. The energy of a polymerized membrane can be constructed as an expansion in terms of the tangent fields $\partial_\alpha {\bf R}$, $\alpha=1,2$, and their derivatives. The corresponding long-distance elastic action reads \cite{nelson87,paczuski88}
\begin{align}
\begin{split}
S[{\bf R}] = \int \mathrm{d}^2x \, \bigg\{ &
\frac{\kappa}{2} \big(\partial_\alpha^2 {\bf R}\big)^2
+ \frac{t}{2} \big(\partial_\alpha {\bf R}\big)^2 \\
&  \hspace{-1.2cm} + \frac{\lambda}{8} \big(\partial_\alpha {\bf R}\cdot\partial_\alpha {\bf R}\big)^2
+ \frac{\mu}{4} \big(\partial_\alpha {\bf R}\cdot\partial_\beta {\bf R}\big)^2
\bigg\} \, .
\label{action1}
\end{split}
\end{align}
In Eq.~(\ref{action1}), Greek indices run from $1$ to $2$ and repeated indices are summed over.  More generally, Einstein's summation convention applies throughout, unless stated otherwise.  The action Eq.~(\ref{action1}) contains four contributions: the first term describes the bending energy, with bending rigidity $\kappa$, while the second term represents an effective surface tension, with tension parameter $t$. The last two terms account for stretching and shear elastic energies arising from the fixed connectivity, and are characterized by the two Lam\'e coefficients $\lambda$ and $\mu$. Mechanical stability requires that $\kappa$, $\mu$, and the bulk modulus $B = \lambda + \mu$ are all positive. Note that in Eq.~(\ref{action1}) the temperature has been absorbed into the definition of the field and of the coupling constants.

\subsubsection{Phase transition and phase structure}

Let us examine action Eq.(\ref{action1}) at the mean-field level, varying the parameter $t$. When $t>0$, the minimum of Eq.~(\ref{action1}) corresponds to a vanishing average value of the order parameter, i.e. $\langle \partial_{\alpha}{\bf R}_0\rangle = {\bf 0}$, which characterizes a crumpled phase. In contrast, when $t<0$, the minimum of Eq.~(\ref{action1}) is reached for a configuration
\begin{equation}
\langle {\bf R}_0  \rangle =  \zeta\, x_{\alpha}\, {\bf e}_{\alpha}\,,
\label{flatmono}
\end{equation}
where $\zeta = \sqrt{-t/(\lambda+\mu)}$ and $\{{\bf e}_{\alpha}\}$ is an orthonormal basis defining the orientation of the flat phase in the embedding space. The parameter $\zeta$, with $0<\zeta<1$, sets the magnitude of the order parameter components, $\langle\partial_{\alpha}{\bf R}\rangle = \zeta\, {\bf e}_{\alpha}$ for $\alpha=1,2$, and thus quantifies how much the membrane is stretched. The point $t=0$ therefore corresponds to a phase transition between a high-temperature crumpled phase and a low-temperature flat phase with long-range orientational order \cite{paczuski88,paczuski89}.

\subsubsection{Perturbative and nonperturbative approaches}

In order to study the flat phase within the NPRG approach, it is  convenient to rewrite Eq.(\ref{action1}) in terms of the extension parameter $\zeta$. One gets:
\begin{align}
\begin{split}
S[{\bf R}] = \int d^2x  \ \bigg\{ &\frac{\kappa}{2} ({\partial_\alpha^2}{\bf R})^2 +\frac{\lambda}{8} (\partial_\alpha {\bf R}.\partial_\alpha{\bf R}-\zeta^2\delta_{\alpha\alpha})^2 \\ 
 &  \hspace{0cm}+ \frac{\mu}{4}\,(\partial_\alpha {\bf R}.\partial_\beta{\bf R}- \zeta^2\delta_{\alpha\beta})^2  \bigg\} 
\label{action2}    
\end{split}
\end{align}
which should be understood as the leading order of an expansion around the flat phase configuration Eq.~(\ref{flatmono}). 
One can express Eq.~(\ref{action2}) more synthetically in terms of the strain tensor $u_{\alpha\beta}=1/2(\partial_\alpha {\bf R}.\partial_\beta{\bf R}-\partial_{\alpha}{\bf R}_0.\partial_{\beta}{\bf R}_0)$ that measures the deformations with respect to the flat phase configuration Eq.~(\ref{flatmono}):
\begin{align}
\begin{split}
S[{\bf R}] = \int d^2x   \ \bigg\{ &\frac{\kappa}{2} ({\partial_\alpha^2}{\bf R})^2 +\frac{\lambda}{2} u_{\alpha\alpha}^2 + \mu u_{\alpha\beta}^2  \bigg\} \ .
\label{action3}    
\end{split}
\end{align}

A fundamental point is that the actions Eqs.~(\ref{action1}), (\ref{action2}) and (\ref{action3}) are formulated entirely in terms of the position field ${\bf R}$. Moreover, although we focus on the flat phase, we deliberately refrain from decomposing ${\bf R}$ into in-plane phonon and out-of-plane flexural (flexuron) modes around the flat-phase configuration ${\bf R}_0$,
${\bf R}({\bf x}) = \langle{\bf R}_0({\bf x})\rangle + {\bf u}({\bf x}) + {\bf h}({\bf x})$. Indeed, such a decomposition is usually accompanied by the additional approximations \cite{nelson87,aronovitz88,aronovitz89}
\begin{equation}
\left\{
\begin{array}{ll}
\displaystyle (\partial_{\alpha}^2{\bf u})^2 \simeq 0, 
\\[0.35cm]
\displaystyle u_{\alpha\beta} \simeq \frac{1}{2} \big(\partial_\alpha u_{\beta} + \partial_{\beta} u_{\alpha}
+ \partial_\alpha {\bf h}\cdot\partial_{\beta} {\bf h}\big)
+ \mathcal{O}\big((\partial u)^2\big),
\end{array}
\right.
\label{expansionuh}
\end{equation}
which are well suited either for a {\sl perturbative} analysis of fluctuations around the flat phase \cite{aronovitz88,guitter88,guitter89,coquand20a,metayer22,metayer22d,metayer23,pikelner22}, or for a SCSA treatment based on an effective field theory formulated solely in terms of the flexural fields ${\bf h}$ \cite{ledoussal92,fasolino07,zakharchenko10,roldan11,mauri20,ledoussal18}. By contrast, the actions Eqs.~(\ref{action1}), (\ref{action2}) and (\ref{action3}), which display explicit and full rotational invariance, allow one to investigate within a single framework both the crumpled-to-flat transition and the flat phase of membranes in a genuinely nonperturbative way \cite{kownacki09,braghin10,hasselmann11}, i.e. without relying on expansions in the elastic couplings $\lambda$ and $\mu$ or in any of the usual perturbative parameters (inverse dimension $1/d$, temperature $T$, etc.).

\subsection{Membrane bilayers}

\subsubsection{Flat phase}

\begin{figure}[h!]
\centering
\hspace{-0.7cm}\includegraphics[trim={1cm 12cm 2cm 6cm},clip,width=0.55\textwidth]{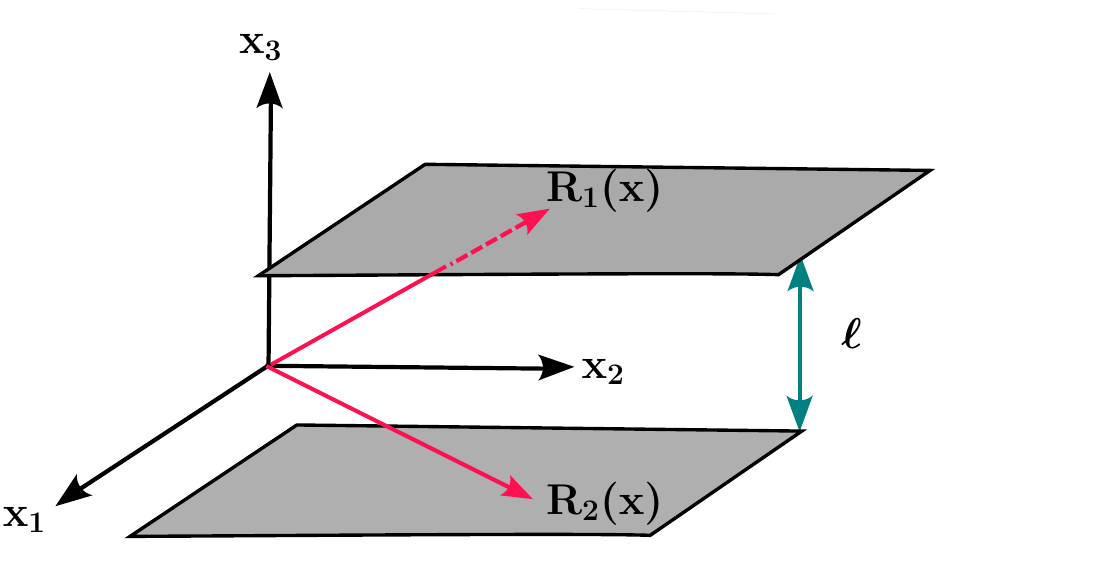}
\caption{Bilayer model: two coupled membranes, separated by a distance $\ell$. A point on membrane 1 (respectively, membrane 2) is parameterized by the position vector 
${\bf R}_1({\bf x})$ (respectively, 
${\bf R}_2({\bf x})$).}
\label{figbilayers}
\end{figure}

We now turn to the  field theoretical description of  membrane bilayers  composed of two monolayers 1 and 2 separated by a distance $\ell$. In doing so, we follow the constructions of Refs.~\cite{andres12,mauri21bis}. Each monolayer is described by a free action $S[{\bf R}_i]$, $i=1,2$, where $S[{\bf R}]$ is given by Eqs.~(\ref{action1}), (\ref{action2}) or (\ref{action3}), and ${\bf R}_i$, $i=1,2$, are the position fields of the two membranes, see Fig.~\ref{figbilayers}.

 In the flat phase, the system is characterized by the following ground state:
\begin{equation}
\left\{
\begin{array}{ll}
\displaystyle \langle {\bf R}_1\rangle = \zeta\, x_{\alpha}\, {\bf e}_{\alpha} + \frac{\ell}{2}\, {\bf n}, 
\\[0.35cm]
\displaystyle \langle {\bf R}_2\rangle = \zeta\, x_{\alpha}\, {\bf e}_{\alpha} - \frac{\ell}{2}\, {\bf n} 
\end{array}
\right.
\label{groundstate}
\end{equation}
with ${\bf n}$ the  normal to the monolayers. 

It is convenient to parametrize the interaction in terms of the mean (center-of-mass) position, ${\bf R} = \tfrac{1}{2}({\bf R}_1 + {\bf R}_2)$, and the relative position, ${\bf S} = {\bf R}_1 - {\bf R}_2$. In terms of these fields, the flat-phase configuration is described by
\begin{equation}
\left\{
\begin{array}{ll}
\displaystyle \langle {\bf R}_0 \rangle = \zeta\, x_{\alpha}\, {\bf e}_{\alpha}, 
\\[0.35cm]
\displaystyle \langle{\bf S}_0\rangle = \ell\, {\bf n}
\end{array}
\right.
\label{groundstatebi}
\end{equation}
that generalize Eq.(\ref{flatmono}).

\subsubsection{Symmetries}

Under a simultaneous translation $\mathbf R_{1,2} \to \mathbf R_{1,2} + \mathbf C$, with $\mathbf C$ an arbitrary three-dimensional vector, the center-of-mass coordinate $\mathbf R = \tfrac{1}{2}(\mathbf R_1 + \mathbf R_2)$ transforms as $\mathbf R \to \mathbf R + \mathbf C$, while the relative coordinate $\mathbf S = \mathbf R_1 - \mathbf R_2$ remains invariant. Translational invariance therefore constrains the effective theory so that $\mathbf R$ can only appear through its derivatives, $\partial_\alpha \mathbf R$. The corresponding excitations are Goldstone modes associated with the spontaneous breaking of translational symmetry by the presence of the membrane, and are consequently massless. In contrast, because $\mathbf S$ is invariant under global translations, this symmetry does not restrict the form of local potentials $V(\mathbf S)$ or couplings such as $({\bf S}^2-\ell^2)\,u_{\alpha\alpha}$ that mix $\mathbf S$ with the center-of-mass field $\mathbf R$ strain  tensor $u_{\alpha\beta}$, see below. Fluctuations in $\mathbf S$  describe internal degrees of freedom, such as local variations of thickness or relative tilt between the two monolayers, and  acquire a finite mass.

\subsubsection{Action}

The explicit form of the interaction $S_I[{\bf R},{\bf S}]$ between the two membranes can be obtained from the above symmetry considerations, together with the requirement that, in order to preserve rotational invariance in the embedding space, it must be built only from scalar combinations of $\{\partial_{\alpha} {\bf R}\}$ and ${\bf S}$. In addition, it should describe small fluctuations around the configuration Eq.~(\ref{groundstatebi}). Accordingly, it can depend on the  center-of-mass field $\mathbf R$ strain tensor $u_{\alpha\beta}$, the interlayer shear ${\bf S}\cdot \partial_{\alpha}{\bf R}$, and fluctuations of the interlayer distance, encoded in the term ${\bf S}^2-\ell^2$. The action compatible with all these symmetries is given by \cite{mauri21bis}
\begin{align}
\begin{split}
S_I[{\bf R},{\bf S}] = \int \mathrm{d}^2x \, \bigg\{ &
\frac{g_1}{8\ell^4} \big({\bf S}^2-\ell^2\big)^2
+ \frac{g_2}{2\ell^2} \big({\bf S}\cdot \partial_{\alpha}{\bf R}\big)^2 \\
& + \frac{g_3}{4\ell^2} \big({\bf S}^2-\ell^2\big)\, u_{\alpha\alpha}
\bigg\} 
\label{actioncouplage}
\end{split}
\end{align}
with $u_{\alpha\beta}=1/2(\partial_\alpha {\bf R}.\partial_\beta{\bf R}-\partial_{\alpha}{\bf R}_0.\partial_{\beta}{\bf R}_0)$.
Note that here we have neglected anisotropic terms which are relevant for realistic graphene-based materials but are expected to be subdominant compared to the contributions retained in Eq.~(\ref{actioncouplage}) (see Ref.~\cite{mauri21bis}). The interaction action $S_I[{\bf R},{\bf S}]$ contains three terms. The first one is a restoring potential that drives ${\bf S}^2$ towards $\ell^2$. The second term is a tilt (director-tangent) coupling: it penalizes the projection of $\mathbf S$ onto the local tangent directions $\partial_\alpha\mathbf R$, and thus suppresses relative tilt of the interlayer director $\mathbf S$ with respect to the membrane's tangent frame, stabilizing the alignment of $\mathbf S$ along the local normal. The last term is a volume-area coupling: it couples thickness fluctuations to areal strain, so that stretching ($u_{\alpha\alpha}>0$) favors thinning (${\bf S}^2<\ell^2$), while compression ($u_{\alpha\alpha}<0$) favors thickening (${\bf S}^2>\ell^2$), thereby encoding approximate local volume conservation.

The whole action, given by $S[{\bf R},{\bf S}]=S[{\bf R}_1]+S[{\bf R}_2]+S_I[{\bf R},{\bf S}]$ reads, in terms  of the fields ${\bf R}$ and ${\bf S}$  \cite{mauri21bis}: 
  \begin{equation}
\begin{aligned}
    S[&{\bf R},{\bf S}] = \int \mathrm{d}^2x \bigg\{ \frac{\kappa}{2}\Big(2\, (\partial^2{\bf R})^2 + \frac{1}{2}\, (\partial^2{\bf S})^2\Big) +\lambda \, u_{\alpha\alpha}^2\\
    & \hspace{-0.4cm} +  2\, \mu \, u_{\alpha\beta}^2 + \frac{\lambda}{4}\, (\partial_\alpha{\bf R}.\partial_\alpha{\bf S})^2 + \frac{\mu}{4}\, (\partial_\alpha{\bf S}.\partial_\beta{\bf R})^2\\
    & \hspace{-0.4cm}  +\frac{\lambda}{64}(\partial_\alpha{\bf S}.\partial_\alpha{\bf S})^2 + \frac{\mu}{32} (\partial_\alpha{\bf S}.\partial_\beta{\bf S})^2 + \frac{\lambda}{4}u_{\alpha \alpha}\,\partial_\beta{\bf S}.\partial_\beta{\bf S}\\
    & \hspace{-0.4cm}  + \frac{\mu}{2}u_{\alpha\beta}\,\partial_\alpha{\bf S}.\partial_\beta{\bf S} +  \frac{\mu}{4} (\partial_\alpha {\bf R}.\partial_\beta {\bf S})(\partial_\alpha {\bf S}.\partial_\beta {\bf R}) \\
    &\hspace{-0.4cm}+\frac{g_{1}}{8\ell^4} ({\bf S}^2-\ell^2)^2 + \frac{g_{2}}{2\ell^2}({\bf S}.\partial_\alpha{\bf R})^2+\frac{g_3}{4\ell^2} \big({\bf S}^2-\ell^2) u_{\alpha\alpha} \bigg\}. 
\label{bilayeraction2}
\end{aligned}
\end{equation} An important point to notice is that the coupling constants in front of  the elastic contributions $u_{\alpha\alpha}^2$ and $u_{\alpha\beta}^2$ can be deduced from those entering in action Eq.~(\ref{action3}) by the rescaling: $\lambda\to 2 \lambda$ and  $\mu \to 2 \mu$. 

Following and drawing on the works of de~Andr\'es {\it et al.} \cite{andres12} and  Mauri {\it et al.} \cite{mauri21bis}, we will focus, among the three interaction terms, on the one responsible for the rigidity crossover, namely the coupling proportional to $g_2$. Moreover we will take  the limit  $g_1\to \infty$, which fixes the interlayer separation to $|{\bf S}| = \ell$. In this limit, the term proportional to $g_3$ does not contribute, since it is multiplied by ${\bf S}^2 - \ell^2$.

\section{Nonperturbative renormalization group approach to polymerized membranes}

 For readers already familiar with the NPRG formalism and with its
application to monolayer membranes, Secs.~III~A and III~B may be skipped on a
first reading, and one can proceed directly to Sec.~III~C where the bilayer
flow equations are derived.

\subsection{Methodology}

We now present the NPRG method used to investigate graphene bilayers. It relies on the concept of a running Gibbs free energy -- or effective average  action        \cite{wetterich93,tetradis94,berges02,pawlowski07,delamotte12,dupuis21}. Fundamentally, it is based on the Wilson-Kadanoff block-spin idea \cite{wilson74}, in which one integrates out short-distance degrees of freedom in the partition function in order to obtain an effective Hamiltonian or action for the long-distance modes. It was realized in the 90's \cite{wetterich93c,morris94a}, however, that it is more convenient, both conceptually and computationally, to work with a more physical quantity than the Hamiltonian or action, namely the Gibbs free energy $\Gamma$. One therefore introduces a running Gibbs free energy -- or effective average  action -- $\Gamma_k$, where $k$ is a running wavevector  scale in the block-spin procedure. By construction, $\Gamma_k$ is the Gibbs free energy in which only fluctuations with wavevectors $q \ge k$ have been integrated out. As a consequence, at the microscopic (lattice) scale $k=\Lambda$ (often formally taken to infinity), one has $\Gamma_{k=\Lambda}=S$, where $S$ is the bare action or Hamiltonian, since no fluctuations have been integrated out. Conversely, at the macroscopic scale $k=0$, $\Gamma_k$ coincides with the usual Gibbs free energy $\Gamma$:
\begin{equation}
\left\{
\begin{array}{ll}
\Gamma_{k=\Lambda} = S, \\[0.2cm]
\Gamma_{k=0} = \Gamma,
\end{array}
\right.
\label{limites}
\end{equation}
and $\Gamma_k$ with $0 \le k \le \Lambda$ interpolates smoothly between these two limits.

We now construct the running Gibbs free energy for the field theory with a single order parameter ${\bf R}$, which is relevant for a monolayer, and postpone the discussion of the bilayer case to later.

First, one has to separate the low-wavevector  modes from the high-wavevector ones in order to integrate out the latter. To achieve this separation, we introduce a $k$-dependent quadratic ``mass'' term
\begin{equation}
     S_k[{\bf R}] = \frac{1}{2} \int_{\bf q} {\cal R}_{k,ij}({\bf q})\, R_i({\bf q})\, R_j(-{\bf q}) \,,
    \label{deltaS}
\end{equation}
with $\int_{\bf q} \equiv \int \mathrm{d}^2q/(2\pi)^2$. In Eq.({\ref{deltaS})},  ${\cal R}_{k,ij}({\bf q})$ ($i,j=1,\dots,3$) is a ${\bf q}$-dependent cut-off function, which is very often taken to be diagonal, ${\cal R}_{k,ij}({\bf q}) = {\cal R}_k({\bf q}) \delta_{ij}$. The partition function in the presence of a source ${\bf J}({\bf q})$ then reads
\begin{equation}
\begin{aligned}
     {\cal Z}_k[{\bf J}] = \int \mathcal{D}{\bf R} \,
    \exp \bigg[ & - S[{\bf R}] - \Delta S_k[{\bf R}]\\ 
    & + \int_{\bf q} {\bf J}({\bf q}) \cdot {\bf R}(-{\bf q}) \bigg]\, . 
    \label{partition}
\end{aligned}
\end{equation}
 The role of the term Eq.~(\ref{deltaS}) is to suppress low-wavevector modes so that only high-wavevector modes are effectively integrated out in Eq.~(\ref{partition}). In order to implement this program and ensure that $\Gamma_k$ satisfies the limits Eq.~(\ref{limites}), the cut-off function ${\cal R}_k({\bf q})$ must satisfy the following properties:

\vspace{0.5cm}

\underline{1) At fixed $k$:}
\begin{itemize}
    \item ${\cal R}_k({\bf q}) \sim k^{\rho}$ with $\rho>0$ when $|{\bf q}| \ll k$: low-wavevector modes acquire an effective ``mass'' of order $k^{\rho}$ and thus effectively decouple.
    \item ${\cal R}_k({\bf q}) \to 0$ when $|{\bf q}| \gg k$: high-wavevector modes are essentially unaffected by the regulator term.
\end{itemize}

\underline{2) At fixed $q$:}
\begin{itemize}
    \item ${\cal R}_k({\bf q}) \to 0$ when $k \to 0$: the regulator term vanishes when all fluctuations have been integrated out, so that the mass term no longer plays any role.
    
    \item ${\cal R}_k({\bf q}) \to \Lambda^{\rho}$ when $k \to \Lambda$ (with $\Lambda$ generally taken  infinite): at the lattice scale, ${\cal R}_k({\bf q})$ acts as a very large mass for all modes. They are effectively ``frozen'' and do not contribute to the functional integral Eq.~(\ref{partition}).
\end{itemize}

A typical cut-off function satisfying these properties is the exponential regulator \cite{berges02}:
\begin{equation}
{\cal R}_k({\bf q}) = \frac{Z_k |{\bf q}|^{\rho}}{\exp\!\big(|{\bf q}|^{\rho}/k^{\rho}\big)-1},
\label{cutoffexp}
\end{equation}
where $Z_k$ is a field renormalization factor, see below. This function is illustrated in Fig.~\ref{cutoff}. A very practical and commonly used choice is the  Litim (or $\Theta$)  cut-off \cite{litim01},
\begin{equation}
{\cal R}_k({\bf q}) = Z_k\big(k^{\rho} - |{\bf q}|^{\rho}\big)\,
\Theta\big(k^{\rho} - |{\bf q}|^{\rho}\big),
\label{Litimcutoff}
\end{equation}
where $\Theta$ denotes the Heaviside step function.


\begin{figure}[h!]
\hspace{-0.7cm}\includegraphics[trim={1cm 9cm 2cm 3cm},clip,width=0.4\textwidth]{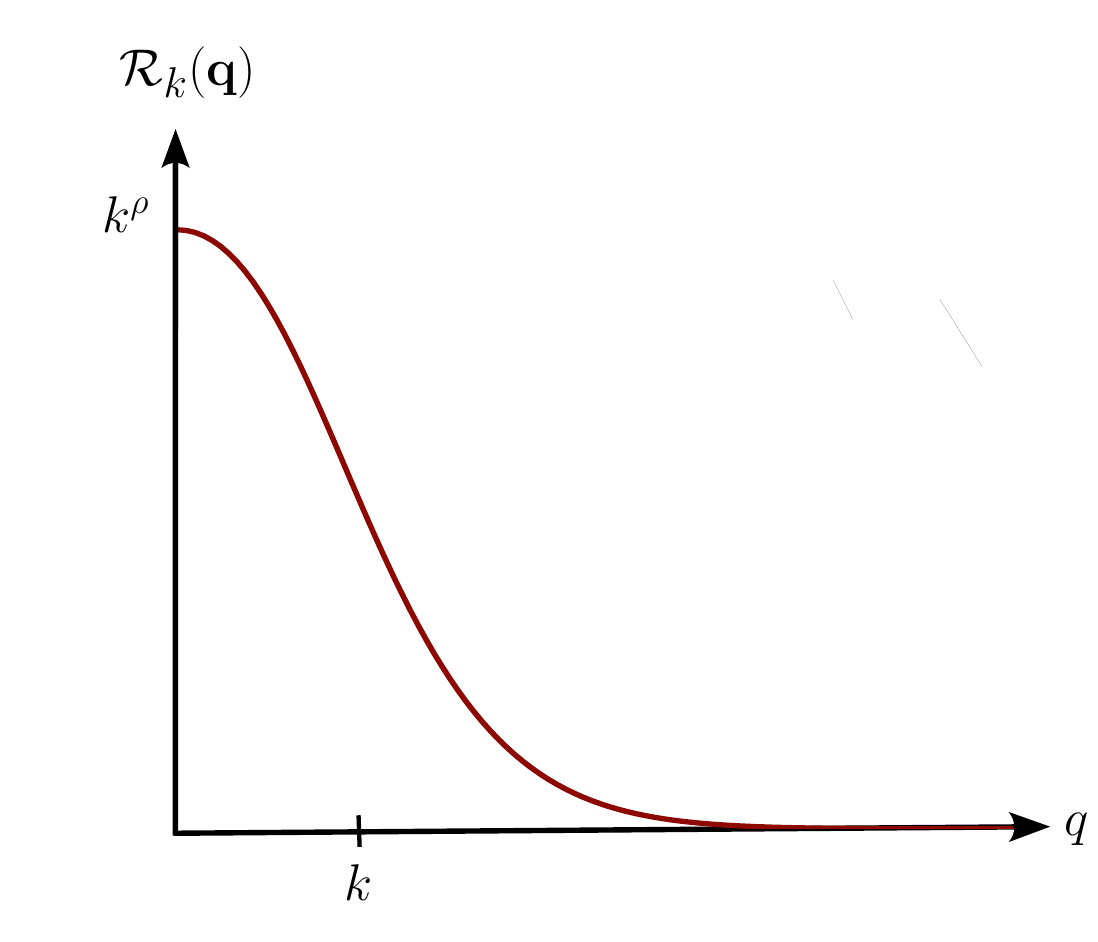}
\caption{A typical shape of the cut-off function ${\cal R}_k({\bf q})$.}
\label{cutoff}
\end{figure} 

We now define the running Gibbs free energy $\Gamma_k$ as a slightly modified Legendre transform of the Helmholtz free energy:
\be
\Gamma_k[{\bf r}] = - W_k[{\bf J}] + \int_{\bf q} {\bf J}({\bf q}) \cdot {\bf r}(-{\bf q})
- \Delta S_k[{\bf r}] \,,
\label{Legendre}
\ee
with $W_k[{\bf J}] = \log {\cal Z}_k[{\bf J}]$, where we have set $k_B T = 1$, and where ${\bf r}$ is the expectation value of the microscopic field ${\bf R}$ in the presence of the source ${\bf J}$:
\be
r_i = \langle R_i \rangle = {\delta W_k[{\bf J}] \over \delta J_i}\,.
\ee
With this definition, together with the properties of the cut-off function ${\cal R}_k({\bf q})$ specified above, it is straightforward to check that $\Gamma_k$ satisfies the limits Eq.~(\ref{limites}).

The running Gibbs free energy -- or effective average action -- $\Gamma_k$  follows an exact equation, namely the Wetterich equation \cite{wetterich93c}:
\begin{equation}
\begin{aligned}
    \partial_t \Gamma_k[{\bf r}]&=\frac{1}{2} \int_{\bf q} \partial_t {\cal R}_{k,ij}({\bf q}) \left(\Gamma_{k,ji}^{(2)}[{\bf r};{\bf q},{-{\bf q}}] + {\cal R}_{k,ji}({\bf q}) \right)^{-1} 
    \label{wetterich1}
    \\
    \\
    &\equiv \frac{1}{2} {\hbox{Tr}} \left[\partial_t {\cal R}_k\left(\Gamma_{k}^{(2)}+ {\cal R}_{k} \right)^{-1}\right]
      \end{aligned}
    \end{equation}
where the trace is performed on both discrete and continuum indices. In Eq.~(\ref{wetterich1}) one has defined a RG ``time'' $t=\ln k/ \Lambda$. A practical expression is given by: 
\begin{equation}
\partial_t \Gamma_k[{\bf r}]= \frac{1}{2} \widehat\partial_t{\hbox{Tr}} \ln\left[\Gamma_{k}^{(2)}+ {\cal R}_k\right]
    \label{wetterich2}
\end{equation}
with $\widehat\partial_t\equiv\partial_t {\cal R}_k({\bf q}){\partial_{{\cal R}_k({\bf q})}}$.

In Eqs.~(\ref{wetterich1}) and (\ref{wetterich2}),  $\Gamma_{k,ij}^{(2)}[{\bf r};{\bf q},{-{\bf q}}]$ is  the  inverse  propagator, {\it i.e.}  the second derivative of $\Gamma_k$ with respect to the order parameter field ${\bf r}$:
\be
\Gamma_{k,ij}^{(2)}[{\bf r};{\bf q},{-{\bf q}}]={\delta^2  \Gamma_k[{\bf r}]\over \delta r_i({\bf q})\delta r_j(-{\bf q})}\ 
\ee
considered  in a {\sl generic}  field configuration.

Eqs.~(\ref{wetterich1}) and (\ref{wetterich2}) govern the evolution of $\Gamma_k$ as $k$ is lowered, i.e. as fluctuations with lower and lower wavevectors are successively included. In these equations, the cut-off function ${\cal R}_k$ appears both in the numerator of the integrand, through $\partial_t {\cal R}_k$, and in the denominator as a ``mass term'' in the propagator. In the former case, combined with the generic shape of the regulated propagator (see Fig.~\ref{cutoff}), this ensures that only modes with wavevectors in a shell around $k$ contribute effectively to the RG flow, in accordance with the Wilsonian picture. In particular, ${\cal R}_k$ controls the ultraviolet (UV) sector. In the latter case, ${\cal R}_k$ acts as an infrared (IR) regulator, which guarantees that even at criticality any singular behavior is shifted to vanishing wavevector in the limit $k \to 0$. This mechanism works provided the exponent $\rho$ in $k^\rho$ is chosen appropriately: for scalar theories with $O(N)$ symmetry and a standard kinetic term of order $\partial^2$, one takes $\rho=2$, whereas for crystalline membranes, whose kinetic term is of order $\partial^4$, one chooses $\rho=4$ \cite{kownacki09}. With this choice, one can safely investigate any phase or phase transition by taking the limit $k \to 0$.

Let us comment on Eq.~(\ref{wetterich1}) in more detail.  
1) First, it is exact. As such, it can in principle be used to investigate both perturbative excitations (such as magnons, phonons, or flexurons) and genuinely nonperturbative excitations (such as topological or bound states).  
2) It involves $\Gamma_k$ which, besides being a directly physical quantity, is the generating functional of one-particle-irreducible (1PI) vertices, and therefore provides a direct link to standard perturbative computations.  
3) It has a one-loop structure, in sharp contrast with systematic loop expansions. This is possible because the propagator entering Eq.~(\ref{wetterich1}) carries the {\it full} field dependence, whereas in perturbation theory one usually deals with a propagator evaluated at vanishing order parameter.

As a consequence, for any generic theory, e.g. a scalar $\phi^4$ theory,
\be
\Gamma_k[\phi] = \int \mathrm{d}^D{\bf x} \left\{
\frac{1}{2} (\partial \phi)^2
+ \frac{1}{2} m^2 \phi^2
+ \frac{g}{4!} \phi^4 + \dots
\right\},
\ee
the propagator evaluated at a nonvanishing uniform field $\phi=\phi_0$, $G_k[\phi_0,{\bf q}]$,  exhibits a nontrivial dependence on the coupling constant $g$:
\be
\begin{aligned}
&G_k[\phi_0,{\bf q}]=\left(\Gamma_{k}^{(2)}[\phi; {\bf q},-{\bf q}]\Big\vert_{\phi=\phi_0}\right)^{-1}
= \\
& \hspace{3cm}\frac{1}{\,{\bf q}^2 + m^2 + \displaystyle \frac{g}{2} \phi_0^2 + \dots}\ .
\label{expansion}
\end{aligned}
\ee

Although exact, the running Gibbs free energy $\Gamma_k$ entering Eq.~(\ref{wetterich1}) must be approximated (truncated) in practice. A very successful approximation consists in expanding $\Gamma_k$ in powers of derivatives of the order parameter \cite{tetradis94,berges02,dupuis21}. This type of expansion relies on two main ingredients.  
1) From a technical point of view, it uses the fact that functional derivatives of $\Gamma_k[{\bf r}]$ with respect to the field $r_i$ are smooth functions of the wavevectors and can be expanded in powers of $p_i/k^2$ or $p_i/m$, where $m$ is a typical wavevector also  (improperly) called ``mass'' scale \cite{dupuis21}. This property is guaranteed by the infrared finiteness of the Wetterich equation.  
2) From a physical point of view, it assumes that the essential physics is governed by long-distance (thus low-derivative) modes. This is typically the case when the running Gibbs free energy is parametrized in terms of the relevant low-wavevector degrees of freedom, which do not form bound states.

This derivative expansion is often accompanied by an expansion in powers of the field, see  Refs.~\cite{berges02,dupuis21} for reviews, as long as the physics is not controlled by a nonanalytic dependence on the order parameter. A notable exception arises in the presence of strong disorder, e.g. in the random-field Ising model, where an infinite number of operators become marginal at the upper or lower critical dimension, see Ref.~\cite{tarjus20} for a review.

The major  advantage of these combined approximations, when they are physically justified, is that the functional equation Eq.~(\ref{wetterich1}) is converted into a closed set of differential equations for the coupling constants parametrizing the running Gibbs free energy $\Gamma_k$. This greatly simplifies the calculations while preserving the nonperturbative content of the flow, as is apparent from Eq.~(\ref{expansion}). In the case of membrane monolayers, such approximations have made it possible to treat, within a single framework, both the crumpled-to-flat transition and the flat phase of the membrane -- which are nonperturbatively related --  with good quantitative accuracy \cite{kownacki09,braghin10,hasselmann11}. However, achieving high precision in general requires extending the ansatz for $\Gamma_k$ \cite{essafi14}.

Membranes in their flat phase represent a particularly singular  case in this context. Indeed, it can be shown that all powers of the order parameter higher than $(\partial_\alpha{\bf r})^4$ do not contribute to the RG flow in the flat phase. This is because these contributions are all (pseudo-) massive \footnote{In a standard scalar theory, massive terms are those with fewer than two derivatives compared to the kinetic term, which behaves as $({\partial \phi})^2$. In membrane theory, the kinetic term behaves as $({\partial^2 r})^2$. Thus, by analogy, the terms proportional to $({\partial r})^2$ are referred to as having a ``pseudo-mass".}. These terms have squared pseudo-masses of order $(\lambda, \mu) \times \zeta^2$ and thus decouple at low-wavevectors (as detailed in section \ref{flatphaseeqs}). Moreover, Braghin and Hasselmann \cite{braghin10, hasselmann11} have demonstrated, using a wavevector-dependent approximation, that high-wavevector and high-order derivative terms do not contribute significantly. Their analysis found an exponent $\eta \sim 0.85$, which is almost indistinguishable from the value obtained within a derivative expansion at leading order ($\eta \sim 0.849$) \cite{kownacki09}. As a result, the combined field/field derivative expansion is highly effective for studying the flat phase of membranes.

\subsection{Monolayers}

Before turning to the bilayer case, we first recall how the NPRG works for a single membrane (monolayer), which has been studied in a variety of settings, including isotropic, anisotropic, quantum, and disordered cases \cite{kownacki09,braghin10,hasselmann11,essafi14,essafi11,coquand16a,coquand18}. In this case, the running Gibbs free energy is taken in the form
\begin{equation}
\begin{aligned}
 \hspace{-0.3cm} \Gamma_k[{\bf r}] = \int \mathrm{d}^2x \, \bigg\{ &
\frac{\kappa_k}{2} \big(\partial_\alpha^2{\bf r}\big)^2
+ \frac{\lambda_k}{2} u_{\alpha \alpha}^2
+ \mu_k\, u_{\alpha \beta}^2 \bigg\},
\label{actionp}
\end{aligned}
\end{equation}
with
\begin{equation}
u_{\alpha \beta} = \frac{1}{2}
\big(\partial_\alpha {\bf r}\cdot\partial_\beta{\bf r}
- \zeta_k^2 \delta_{\alpha\beta}\big).
\label{tensorel}
\end{equation}
In Eqs.~(\ref{actionp}) and (\ref{tensorel}), the couplings $\kappa_k$, $\lambda_k$, $\mu_k$ and $\zeta_k$ all depend on the running scale $k$.

It is convenient to redefine the field ${\bf r}$ and the couplings in such a way as to absorb $\kappa_k$ and make explicit the field renormalization factor $Z_k$. In terms of these rescaled quantities, one obtains
\begin{equation}
\begin{aligned}
\hspace{-0.3cm} \Gamma_k[{\bf r}] = \int \mathrm{d}^2x \, \bigg\{ &
\frac{Z_k}{2} \big(\partial_\alpha^2{\bf r}\big)^2
+ \frac{\lambda_k}{2} u_{\alpha \alpha}^2
+ \mu_k\, u_{\alpha \beta}^2 \bigg\}.
\label{actionfinale}
\end{aligned}
\end{equation}

\subsubsection{Derivation of the RG flow: general principle}

In order to derive the RG equations from the action Eq.~(\ref{actionfinale}), one must first define the running couplings unambiguously from $\Gamma_k$. Typically, for a generic coupling constant $g_k$ (including $\lambda_k$,  $\mu_k$ or $Z_k$), one proceeds as follows:
\begin{equation}
g_{k,ij} = \lim_{{\bf p}\rightarrow{\bf 0}}
\frac{1}{a!}\frac{1}{b!}
\frac{\partial^a}{\partial({\bf p}^2)^a}
\frac{\partial^b}{\partial(p_c^2)^b}
 \Gamma^{(2)}_{k,ij}[{\bf r};{\bf p},-{\bf p}]
\Big|_{{\bf r}_{0,k}},
\label{defflot}
\end{equation}
--  without summation over $a,b$ indices  -- for appropriately chosen values of the indices $a,b,c,i,j$. In Eq.~(\ref{defflot}), ${\bf r}_{0,k}$ denotes the running flat-phase configuration, see  Eq.~(\ref{flatmono}), which writes here: 
\begin{equation}
{\bf r}_{0,k} =  \zeta_k\, x_{\alpha}\, {\bf e}_{\alpha}\, .
\label{flatmono2}
\end{equation}

In order to obtain the corresponding RG flow, one acts on the right-hand side of (\ref{defflot}) with the operator $\partial_t$. This yields
\begin{widetext}
\begin{align}
\partial_t g_{k,ij} = \lim_{{\bf p}\rightarrow{\bf 0}}
\frac{1}{a!}\frac{1}{b!}
\frac{\partial^a}{\partial({\bf p}^2)^a}
\frac{\partial^b}{\partial(p_c^2)^b}
\Big\{
\partial_t \Gamma^{(2)}_{k,ij}[{\bf r};{\bf p},-{\bf p}]
\Big|_{{\bf r}_{0,k}}
+ \int_{\bf q}
\Gamma^{(3)}_{k,ijl}[{\bf r};{\bf p},-{\bf p},{\bf q}]\,
\partial_t r_l({\bf q})
\Big|_{{\bf r}_{0,k}}
\Big\}
\label{flowg}
\end{align}
\end{widetext}
--  without summation over $a,b$ indices  --  where  $r_l$ denotes the $l$-th component of  the position field ${\bf r}$. Eq.({\ref{flowg}}) couples the RG flow of $\Gamma^{(2)}_{k,ij}[{\bf r};{\bf p},-{\bf p}]$ to higher functional derivatives of $\Gamma_k[{\bf r}]$. The former is obtained by taking two functional derivatives of the Wetterich equation Eq.~(\ref{wetterich1}), which yields
\begin{widetext}
\begin{equation}
\begin{aligned}
\partial_t\Gamma^{(2)}_{k,ij}[{\bf r};{\bf p},-{\bf p}]
= \frac{1}{2}\,\widehat\partial_t \bigg\{
& \int_{\bf q} G_{k,ab}[{\bf r},{\bf q}]\,
\Gamma^{(4)}_{k,ijba}[{\bf r};{\bf p},-{\bf p},{\bf q},-{\bf q}] \\
&\quad - \int_{\bf q} G_{k,ab}[{\bf r},{\bf q}]\,
\Gamma^{(3)}_{k,ibc}[{\bf r};{\bf p},-{\bf q},{\bf q}-{\bf p}]\,
G_{k,cd}[{\bf r},{\bf q}-{\bf p}]\,
\Gamma^{(3)}_{k,jda}[{\bf r};-{\bf p},{\bf q},{\bf p}-{\bf q}]
\bigg\},
\label{flowgamma2}
\end{aligned}
\end{equation}
\end{widetext}
where $G_{k,ab}[{\bf r},{\bf q}]$  is the (modified) propagator : 
\begin{equation}
G_{k,ab}[{\bf r},{\bf q}]
= \big[\Gamma^{(2)}_{k,ab}[{\bf r};{\bf q},-{\bf q}]
  + {\cal R}_{k,ab}({\bf q})\big]^{-1}.
\end{equation}

The flow of the minimum ${\bf r}_{0,k}$ -- or, more precisely, of the extension factor $\zeta_k$ -- is defined in a slightly different way. We start from the condition that ${\bf r}_{0,k}$ minimizes $\Gamma_k$, namely
\begin{equation}
    \Gamma_{k,i}^{(1)}[{\bf r};{\bf p}]\Big|_{{\bf r}_{0,k}}
    = \frac{\delta\Gamma_k[{\bf r}]}{\delta r_i({\bf p})}\Bigg|_{{\bf r}_{0,k}} = 0,
\label{cond}
\end{equation}
with, in Fourier space,
\begin{equation}
{\bf r}_{0,k}({\bf p})
= - i\, \zeta_k\, {\bf e}_{\mu}\,
\frac{\partial}{\partial p_{\mu}} \delta({\bf p}) \, .
\label{defmin}
\end{equation}

Note, however, that because of  translational invariance any configuration of the form Eq.~(\ref{defmin}), with $\zeta_k$ replaced by an arbitrary constant $\theta_k$, would also satisfy Eq.~(\ref{cond}). To lift this ambiguity, it is more convenient to define $\zeta_k$ through a stationarity condition with respect to the tangent vector of the membrane, namely
\begin{equation}
\left.\frac{\delta\Gamma_k[\mathbf{r}]}{\delta \big(\partial_\alpha r_i(\mathbf{p})\big)}\right|_{\mathbf{r}_{0,k}} = 0.
\label{minn}
\end{equation}
Acting with $\partial_t$ on Eq.~(\ref{minn}) then yields
\begin{equation}
\left.\frac{\delta\, \partial_t \Gamma_k[\mathbf{r}]}{\delta(\partial_\alpha r_i)}\right|_{\mathbf{r}_{0,k}}
+ \int_{\mathbf{q}}
\left.\frac{\delta^2 \Gamma_k[\mathbf{r}]}{\delta(\partial_\alpha r_i)\,\delta(\partial_\beta r_j)}\,
\partial_t\big(\partial_\beta r_j\big)\right|_{\mathbf{r}_{0,k}}
= 0,
\label{flowzeta}
\end{equation}
where, using the explicit form Eq.~(\ref{defmin}) of ${\bf r}_{0,k}$, the second term is proportional to $\partial_t \zeta_k$, which can therefore be determined explicitly, see below.

As seen from Eq.~(\ref{flowgamma2}) -- which must be evaluated in the flat-phase configuration Eq.~(\ref{minn}) -- and Eq.~(\ref{flowzeta}), the propagator $G_{k,ab}[{\bf r},{\bf q}]$ in this configuration plays a central role in controlling the RG flow of all coupling constants. We now give its explicit expression.

\subsubsection{Propagators and spectrum of excitations}

The propagator at the minimum can be obtained by computing the second functional derivative of $\Gamma_k$ with respect to the fields $r_i({\bf q})$ and $r_j(-{\bf q})$ and then evaluating it in the configuration ${\bf r}_{0,k}$. One finds \cite{coquand16a}:
\begin{equation}
\Gamma^{(2)}_k[{\bf r};\mathbf{q},-\mathbf{q}]\Big|_{{\bf r}_{0,k}} \hspace{-0.3cm} =
\begin{pmatrix}
\Gamma^{(2)}_{k,p}[{\bf r};\mathbf{q},-\mathbf{q}] &  \mathbf{0}_{2\times 1} \\
\mathbf{0}_{1\times 2} &  \Gamma^{(2)}_{k,f}[{\bf r};\mathbf{q},-\mathbf{q}]
\end{pmatrix}\Bigg|_{{\bf r}_{0,k}}\hspace{-0.2cm},
\end{equation}
where the subscripts $p$ and $f$ refer to the phonon and flexural sectors, respectively. One has
\begin{equation}
\Gamma_{k,p}^{(2)}[{\bf r};\mathbf{q},-\mathbf{q}]\Big|_{{\bf r}_{0,k}} \hspace{-0.4cm} =
\begin{pmatrix}
Z_k \mathbf{q}^{4} + M_1(\mathbf{q}) & \zeta_k^{2}(\lambda_k+\mu_k)\, q_1 q_2 \\[2pt]
\zeta_k^{2}(\lambda_k+\mu_k)\, q_1 q_2 & Z_k \mathbf{q}^{4} + M_2(\mathbf{q})
\end{pmatrix},
\label{gamma2}
\end{equation}
with
\begin{equation}
M_i(\mathbf{q}) \equiv \zeta_k^2\big(\mu_k\, \mathbf{q}^2 + (\lambda_k + \mu_k)\, q_i^2\big),
\end{equation}
and
\begin{equation}
\Gamma_{k,f}^{(2)}[{\bf r};\mathbf{q},-\mathbf{q}]\Big|_{{\bf r}_{0,k}}
= Z_k\, \mathbf{q}^4 \, .
\end{equation}

The inverse, regularized propagator
\begin{equation}
\Gamma^{(2)}_k[{\bf r};\mathbf{q},-\mathbf{q}] + {\cal R}_k({\bf q})
\end{equation}
at ${\bf r}_{0,k}$ can be conveniently rewritten using projectors parallel and transverse to the wavevector ${\bf q}$ in the two-dimensional membrane:
\begin{equation}
\left\{
\begin{array}{ll}
\displaystyle
P^{\parallel}_{ij}(\boldsymbol{q})
= \delta_{\alpha i}\delta_{\beta j}\,\frac{q_{\alpha}q_{\beta}}{\boldsymbol{q}^2}, \\[0.35cm]
\displaystyle
P^{\perp}_{ij}(\boldsymbol{q})
= \delta_{\alpha i}\delta_{\alpha j}
- \delta_{\alpha i}\delta_{\beta j}\,\frac{q_{\alpha}q_{\beta}}{\boldsymbol{q}^2},
\end{array}
\right.
\label{projectors5}
\end{equation}
with
\begin{equation}
P^{\parallel}_{ij}(\boldsymbol{q}) + P^{\perp}_{ij}(\boldsymbol{q})
= \delta_{\alpha i}\delta_{\alpha j}\ ,
\end{equation}
and the flexural projector
\begin{equation}
P^{f}_{ij}(\boldsymbol{q}) = \delta_{ij} - \delta_{\alpha i}\delta_{\alpha j}.
\end{equation}
In terms of these projectors, one obtains
\begin{equation}
\begin{aligned}
\Gamma_{k,ij}^{(2)}[{\bf r};\mathbf{q},-\mathbf{q}]\Big|_{{\bf r}_{0,k}} + {\cal R}_k({\bf q})
&= G_{k,1}^{-1}({\bf q})\, P^{\perp}_{ij}(\boldsymbol{q})
\\
&\hspace{-2.2cm} + G_{k,2}^{-1}({\bf q})\, P^{\parallel}_{ij}(\boldsymbol{q})+ G_{0,k}^{-1}({\bf q})\, P^{f}_{ij}(\boldsymbol{q}) \, .
\label{gammaproj}
\end{aligned}
\end{equation}

The expression Eq.~(\ref{gammaproj}) is easily inversed according to the properties $(P^{\parallel})^2=(P^{\perp})^2=\indentit$ and $P^{\parallel}P^{\perp}=0$ and one gets: 
\begin{equation}
\begin{aligned}
\left[\Gamma_{k,ij}^{(2)}[{\bf r};\mathbf{q},-\mathbf{q}]\big |_{{\bf r}_{0,k}}+ {\cal R}_{k,ij}({\bf q})\right]^{-1}  &= G_{k,1}({\bf q}) P^{\perp}_{ij}(\boldsymbol{q}) 
\\
&\hspace{-3cm} + G_{k,2}({\bf q}) P^{\parallel}_{ij}(\boldsymbol{q}) +G_{k,0}({\bf q})\,P^{f}_{ij}(\boldsymbol{q})\ . 
\label{inversepropag}
\end{aligned}
\end{equation}

In Eqs.~(\ref{gammaproj}) and (\ref{inversepropag}), $G_{k,1}(\boldsymbol{q})$, $G_{k,2}(\boldsymbol{q})$ and $G_{k,0}(\boldsymbol{q})$ denote the propagators of the various excitations. More precisely, the spectrum consists of:
\begin{itemize}
    \item one transverse phonon mode with squared pseudo-mass 
    $m_{1,k}^2 = \zeta_k^2\, \mu_k$, whose propagator reads
    \begin{equation}
        G_{k,1}(\boldsymbol{q}) = \frac{1}{Z_k\,\boldsymbol{q}^4
        + m_{1,k}^2\,\boldsymbol{q}^2 + {\cal R}_k(\boldsymbol{q})}\,;
        \label{G1}
    \end{equation}
    \item one longitudinal phonon mode with squared pseudo-mass
    $m_{2,k}^2 = \zeta_k^2(\lambda_k + 2\,\mu_k)$, with propagator
    \begin{equation}
        G_{k,2}(\boldsymbol{q}) = \frac{1}{Z_k\,\boldsymbol{q}^4
        + m_{2,k}^2\,\boldsymbol{q}^2 + {\cal R}_k(\boldsymbol{q})}\,;
        \label{G2}
    \end{equation}
    \item one flexural (flexuron) mode with vanishing pseudo-mass, whose propagator is
    \begin{equation}
        G_{k,0}(\boldsymbol{q}) = \frac{1}{Z_k\,\boldsymbol{q}^4
        + {\cal R}_k(\boldsymbol{q})}\, .
        \label{G0}
    \end{equation}
\end{itemize}

\subsubsection{RG equations}

We now derive the RG equations for the various running couplings and for the field renormalization, which are defined as derivatives of $\Gamma_{k,ij}^{(2)}$ with respect to the external wavevector  ${\bf p}$. Explicitly, we set
\begin{equation}
\left\{
\begin{aligned}
    \mu_k &= \frac{1}{\zeta_k^2}
    \lim_{\boldsymbol{p} \to 0}
    \frac{\partial}{\partial\boldsymbol{p}^2}\ 
    \Gamma_{k,22}^{(2)}[{\bf r};\boldsymbol{p}, -\boldsymbol{p}]
    \Big|_{{\bf r}_{0,k}}\\[0.1cm]
    \lambda_k &= \frac{1}{\zeta_k^2}
    \lim_{\boldsymbol{p} \to 0}
    \frac{\partial}{\partial p_2^2}\,
    \Gamma_{k,22}^{(2)}[{\bf r};\boldsymbol{p},-\boldsymbol{p}]
    \Big|_{{\bf r}_{0,k}}
    - \mu_k\\[0.1cm]
    Z_k &= \lim_{{\bf p}\to 0}\,
    \frac{\partial^2}{\partial({\bf p}^2)^2}\,
    \Gamma^{(2)}_{k,33}[{\bf r};{\bf p},-{\bf p}]
    \Big|_{{\bf r}_{0,k}}
    \label{defz}
\end{aligned}
\right.
\end{equation}
where $p_2$ denotes the second component of $\boldsymbol{p}$.

Taking the RG-time derivative $\partial_t$ of these definitions and using the general flow identity Eq.~(\ref{flowg}) yields, for $\mu_k$,
\begin{equation}
\begin{aligned}
    \partial_t \mu_k
    &= \lim_{\boldsymbol{p} \to 0}
    \frac{\partial}{\partial \boldsymbol{p}^2}
    \bigg[
    \frac{1}{\zeta_k^2}\,
    \partial_t \Gamma_{k,22}^{(2)}[{\bf r};\boldsymbol{p},-\boldsymbol{p}]
    \\
    &\hspace{0.8cm} - \frac{2\, }{\zeta_k^3}\, \partial_t \zeta_k \, 
    \Gamma_{k,22}^{(2)}[{\bf r};\boldsymbol{p},-\boldsymbol{p}]
    \\
    &\hspace{0.8cm}
    + \frac{1}{\zeta_k^2}
    \int_{\bf q}
    \Gamma_{k,22j}^{(3)}[{\bf r};\boldsymbol{p},-\boldsymbol{p},\boldsymbol{q}]\,
    \partial_t r_j(\boldsymbol{q})
    \bigg]\Bigg|_{{\bf r}={\bf r}_{0,k}}\hspace{-0.8cm}
    \label{flowmu}
\end{aligned}
\end{equation}
where we have taken into account the $k$ -- or $t$ -- dependence of the extension parameter $\zeta_k$.

A similar expression holds for $\lambda_k$:
\begin{equation}
\begin{aligned}
    \partial_t \lambda_k
    &= \lim_{\boldsymbol{p} \to 0}
    \frac{\partial}{\partial p_2^2}
    \bigg[
    \frac{1}{\zeta_k^2}\,
    \partial_t \Gamma_{k,22}^{(2)}[{\bf r};\boldsymbol{p},-\boldsymbol{p}]
    \\
    &\hspace{1cm}
    - \frac{2\, }{\zeta_k^3}\, \partial_t \zeta_k \, 
    \Gamma_{k,22}^{(2)}[{\bf r};\boldsymbol{p},-\boldsymbol{p}]
    \\
    &\hspace{1cm}
    + \frac{1}{\zeta_k^2}
    \int_{\bf q}
    \Gamma_{k,22j}^{(3)}[{\bf r};\boldsymbol{p},-\boldsymbol{p},\boldsymbol{q}]\,
    \partial_t r_j(\boldsymbol{q})
    \bigg]\Bigg|_{{\bf r}={\bf r}_{0,k}}
    \\
    &\hspace{1.0cm}
    - \partial_t \mu_k\, .
    \label{flowlambda}
\end{aligned}
\end{equation}

Finally, the flow of the field renormalization $Z_k$ can be written as
\begin{equation}
\begin{aligned}
    \partial_t Z_k
    &= \lim_{\boldsymbol{p} \to 0}
    \frac{\partial^2}{\partial(\boldsymbol{p}^2)^2}
    \bigg[
    \partial_t \Gamma_{k,33}^{(2)}[{\bf r};\boldsymbol{p},-\boldsymbol{p}]
    \\
    &\hspace{1.3cm} + \int_{\bf q}
    \Gamma_{k,33j}^{(3)}[{\bf r};\boldsymbol{p},-\boldsymbol{p},\boldsymbol{q}]\,
    \partial_t r_j(\boldsymbol{q})
    \bigg]\Bigg|_{{\bf r}={\bf r}_{0,k}}\hspace{-0.8cm}\, .
    \label{flow1}
\end{aligned}
\end{equation}

Using the flow equation Eq.~(\ref{flowgamma2}), the flow of $\zeta_k$ obtained from Eq.~(\ref{flowzeta}), as well as the explicit expression of the three-point vertex
\begin{equation}
\begin{aligned}
    \Gamma_{k,ii j}^{(3)}[{\bf r};\boldsymbol{p},-\boldsymbol{p},\boldsymbol{q}]
    \Big|_{{\bf r}_{0,k}}
    &= - i\, \zeta_k \bigg[
      2\mu_k\, \theta(2-i)\, q_i\, \boldsymbol{p}^2\, \delta_{ij} \\
    &  \hspace{-0.6cm} + 2\, \theta(2-i)\, (\lambda_k + \mu_k)\, \delta_{ij}\,
      (\boldsymbol{p}\cdot\boldsymbol{q})\, p_i \\
    & \hspace{-0.6cm} + \theta(2-j)\,
      \big(2\,\mu_k\, p_j\, \boldsymbol{p}\cdot\boldsymbol{q}
           + \lambda_k\, \boldsymbol{p}^2\, q_j \big)
      \bigg] \, ,
\end{aligned}
\label{Gamma3}
\end{equation}
--  without summation over the indices  $i$ and $j$  -- one obtains the RG flow equations for the running couplings $\lambda_k$, $\mu_k$, $\zeta_k$ and for the field renormalization $Z_k$.

As usual, it is convenient to introduce dimensionless couplings constants $\bar g_k$. They are given by: 
\begin{equation}
    \zeta_k = k^{\eta_k}\, \bar{\zeta}_k, 
    \qquad
    \lambda_k = k^{2-2\eta_k}\, \bar{\lambda}_k,
    \qquad
    \mu_k = k^{2-2\eta_k}\, \bar{\mu}_k \, 
\label{dimless_couplings1}
\end{equation}
where  $\eta_k$ is the (scale-dependent) anomalous dimension such that:
\begin{equation}
Z_k = k^{-\eta_k}\ . 
\label{Z_scaling1}
\end{equation}

 One thus obtain \cite{kownacki09}: 
\begin{equation}
\begin{aligned}
    \partial_t \bar\l_k&=(-2+2\,\eta_k)\, \bar\l_k \\
    &\hspace{-0.4cm}+ \frac{A_2}{\bar\zeta_k^2(\bar\lambda_k +\bar\mu_k)} \bigg[\bar\lambda_k^2\, \bar\mu_k \, \bar\zeta_k^2\left(25 \bar L_{002}^6+10 \bar L_{020}^6+6 \bar L_{200}^6\right)  \\
   &\hspace{1cm} +\bar\lambda_k^3 \,\bar\zeta_k^2\left(7 \bar L_{002}^6+2 \bar L_{020}^6+ 2 \bar L_{200}^6\right)\\
   & \hspace{1cm}+ \bar\mu_k^3\, \bar\zeta_k^2 (9 \bar L_{002}^6 + 8\bar L_{020}^6+ \bar L_{200}^6) \\
   & \hspace{1cm}+ 4 \bar\mu_k^2 (\bar L_{001}^4-\bar L_{010}^4) + 2\bar\lambda_k\bar\mu_k(\bar L_{001}^4-\bar L_{010}^4)  \\
   & \hspace{1cm} + \bar\lambda_k\,\bar\mu_k^2\, \bar\zeta_k^2(27 \bar L_{002}^6 + 16 \bar L_{020}^6 +5 \bar L_{200}^6)\bigg]  \\
     \partial_t \bar\mu_k & = (-2+2\,\eta_k)\,\bar\mu_k\\
    &\hspace{0cm}+\frac{A_2}{\bar\zeta_k^2(\bar\l_k+\bar\mu_k)}\bigg[2\,\bar\mu_k\,(\bar\l_k+2\bar\mu_k)(\bar L_{010}^{4}-\bar L_{001}^{4})\\
    &\hspace{1cm}+ \,\bar\zeta_k^2\,(\bar\l_k+\bar\mu_k)(3\,\bar\mu_k+\bar\l_k)^2 \bar L_{002}^{6}\\
    &\hspace{1cm}+ \,\bar\zeta_k^2\,\bar\mu_k^2(\bar\l_k+\bar\mu_k)\,\bar L_{200}^{6}  \bigg] \\
    \partial_t\bar\zeta_k^2& =-\eta_k\,\bar\zeta_k^2 \\
    &\hspace{0.5cm}+\frac{2\,A_2}{\bar\mu_k+\,\bar\l_k}\bigg[\bar\l_k(2 \bar L_{001}^{4}+\bar L_{010}^{4}+L_{100}^{4}) \\
    &\hspace{0.5cm}+\bar\mu_k (3 \bar L_{001}^{4}+2 \bar L_{010}^{4}+\bar L_{100}^{4})\bigg]
\label{flowcouplings}
\end{aligned}
\end{equation}
while the expression of $\eta_k=-\partial_t \ln Z_k$ which is too long to be given here,  is provided in Appendix \ref{app: Running anomalous dimension}. In Eq.~(\ref{flowcouplings}) one has $A_2=1/8\pi$.

\medskip

In Eqs.~(\ref{flowcouplings}), we have introduced the so-called threshold functions
\cite{tetradis94,berges02,delamotte12,dupuis21}, which encode the contribution of fluctuations at scale $k$.
In the context of membranes they are defined as \cite{kownacki09}
\begin{equation}
    L_{abc}^{\,2+\alpha}
    = -\,\frac{1}{4A_2}\,\widehat{\partial}_t
    \int_{\boldsymbol{q}}
    \boldsymbol{q}^{\alpha}\,
    G_{k,0}(\boldsymbol{q})^{a}\,
    G_{k,1}(\boldsymbol{q})^{b}\,
    G_{k,2}(\boldsymbol{q})^{c},
\label{thresh}
\end{equation}
 where the propagators $G_{k,i}(\boldsymbol{q})$, $i=0\dots 2$ are given by Eqs.(\ref{G1}-\ref{G0}). 

The dimensionless counterparts of the threshold functions, $\bar L_{abc}^{\,2+\alpha}$, which are the ones entering
Eqs.~(\ref{flowcouplings}), are related to the $L_{abc}^{\,2+\alpha}$ by
\begin{equation}
    L_{abc}^{\,2+\alpha}
    = (Z_k k^4)^{-a-b-c}\, k^{2+\alpha}\,
      \bar L_{abc}^{\,2+\alpha}\,.
\end{equation}
--  without summation over the indices  $a$, $b$, $c$ and $\alpha$. 
Their explicit expressions in terms of dimensionless quantities are given in
App.~\ref{app:The monolayer case}.

\subsubsection{RG equations in the flat phase}
\label{flatphaseeqs}

The renormalization group equations relevant for the flat phase are obtained from the generic flow equations Eq.~(\ref{flowcouplings}) by taking the limit in which the phonon modes decouple from the theory. This corresponds to the regime where the running scale $k$ becomes small compared to the phonon pseudo-masses, denoted generically by $m_{p,k}$. These dimensionful pseudo-masses behave as
\begin{equation}
m_{p,k}^2 \propto  k^2\,\bar\zeta_k^2\,\bar g_k,
\end{equation}
where $\bar g_k$ stands for either $\bar\mu_k$ or $\bar\lambda_k + 2\,\bar\mu_k$, and $\bar\zeta_k$ is the dimensionless stretching factor. In practice, the phonon decoupling is thus realized on condition that
\begin{equation}
\bar\zeta_k^2\,\bar g_k \gg 1.
\end{equation}
Since the $\bar g_k$'s approach finite  values at the flat-phase fixed point, this condition is equivalent to taking  a very large dimensionless stretching factor,
\begin{equation}
\bar\zeta_k \gg 1\ .
\end{equation}
In this limit one recovers the flat-phase RG equations \cite{kownacki09}:
\begin{equation}
\left\{
\begin{aligned}
    &\partial_t \bar\lambda_k=\displaystyle 2(\,\eta_k-1)\,\bar\lambda_k+(2\,\bar\lambda_k^2+4\,\bar\lambda_k\,\bar\mu_k+\bar\mu_k^2)\,A_2\, \bar L_{200}^{6}\\
    &\partial_t \bar\mu_k=\displaystyle 2(\eta_k-1)\,\bar\mu_k+\bar\mu_k^2\, A_2\, \bar L_{200}^{6}\ . 
    \label{flatmonocouplings}
\end{aligned}
\right.
\end{equation}
 The only threshold that survives is a ``massless''  one $\bar L_{200}^6$. As for the running anomalous dimension $\eta_k$  in the flat phase it is given by \cite{kownacki09}:
\begin{equation}
\begin{aligned}
    \eta_k &= 2\, A_2 (\bar\lambda_k+2\,\bar\mu_k) \bar N_{200}^{4} \\
    &- \frac{2\, A_2}{(\bar\lambda_k+2\,\bar\mu_k)}(\bar\lambda_k^2+\bar\lambda_k\bar\mu_k+\bar\mu_k^2)\bar L_{100}^{2} 
    \label{flatmonoeta}
\end{aligned}
\end{equation}
with the massless  threshold functions given by \cite{kownacki09}:  
\begin{equation}
\left\{
\begin{aligned}
    &L_{a00}^{2+\alpha}=-\frac{1}{4\,A_2}\,\widehat{\partial}_t\int_q \boldsymbol{q}^\alpha \,P(\boldsymbol{q})^{-a} \\
    &N_{a00}^{2+\alpha}=-\frac{1}{4\,A_2}\,\widehat{\partial}_t\int_q \boldsymbol{q}^\alpha\,\frac{\partial P(\boldsymbol{q})}{\partial\boldsymbol{q}^2} \,P(\boldsymbol{q})^{-a}
\label{thresholds-complet}
\end{aligned}
\right.
\end{equation}
with  $P(\boldsymbol{q}) = Z_k\,\boldsymbol{q}^4 + {\cal R}_k(\boldsymbol{q})$. 
The  dimensionless threshold functions  $\bar N_{a00}^{2+\alpha}$  that  enter  in  Eq.~(\ref{flatmonoeta}) are related to the dimensionful threshold function by: 
{\begin{equation}
N_{abc}^{2+\alpha}=(Z_k k^4)^{-a-b-c} Z_k  k^{4+\alpha}\bar N_{abc}^{2+\alpha}
\end{equation}
--  without summation over the indices  $a$, $b$,  $c$ and $\alpha$. 
Their expressions  in terms of dimensionless quantities are  given in Appendix \ref{app:The monolayer case}.

Considering  the Litim  cut-off, see the Appendix \ref{app:litim}, the RG equations of the coupling constants and the running anomalous dimension read \cite{kownacki09}: 
\begin{equation}
\left\{
\begin{aligned}
    &\partial_t \bar\lambda_k=\displaystyle 2(\eta_k-1)\,\bar\lambda_k+\frac{(2\,\bar\lambda_k^2+4\,\bar\lambda_k\,\bar\mu_k+\bar\mu_k^2)\,(10-\eta_k)}{60 \pi}\\
    &\partial_t \bar\mu_k=\displaystyle 2(\eta_k-1)\,\bar\mu_k+\frac{\bar\mu_k^2 \,(10-\eta_k)}{60 \pi} \\
   &\eta_k=\frac{6\,\bar\mu_k\,(\bar\lambda_k+\bar\mu_k)}{\bar\mu_k\,(\bar\lambda_k+\bar\mu_k)+4\,\pi\,(\bar\lambda_k+2\,\bar\mu_k)}\, .
\label{monolayereq1}
\end{aligned}
\right.
\end{equation}

One can also introduce the Young modulus $Y_k=4\mu_k(\lambda_k+ \mu_k)/(\lambda_k+2\mu_k)$. In terms of this variable the flow reads : 
\begin{equation}
\left\{
\begin{aligned}
    &\partial_t \bar Y_k=\displaystyle 2(\eta_k-1)\bar Y_k +  \frac{(10-\eta_k)}{80 \pi}\bar Y_k^2 \\
   &\eta_k=\frac{6\,\bar Y_k}{16 \pi + \bar Y_k}\, .
\label{monolayereq2}
\end{aligned}
\right.
\end{equation}

The expressions Eqs.~(\ref{monolayereq1}) and (\ref{monolayereq2}) are remarkably simple. Nevertheless, they provide an excellent description of the flat phase of membrane monolayers. In particular, at the infrared fixed point characterizing the flat phase, one finds \cite{kownacki09}
\begin{equation}
\bar\mu_* \simeq 6.21,\qquad
\bar\lambda_* \simeq -3.10,\qquad Y_* \simeq 8.29,
\end{equation}
and, finally, an anomalous exponent \cite{kownacki09}
\begin{equation}
\eta \simeq 0.849,
\end{equation}
which is the value obtained both within the present approximation and in computations including wavevector-dependent elastic couplings \cite{braghin10,hasselmann11}.

\subsubsection{Cutoff dependence}

Note finally that, in this article, we use a specific cutoff function -- namely the Litim cutoff -- which yields extremely simple analytical expressions for the RG equations \eqref{monolayereq1} or  \eqref{monolayereq2}. In principle, one should vary this choice and even consider whole families of
cut-off functions, labeled by one or several parameters and optimized accordingly, see, e.g., Refs.~\cite{litim01,litim01b,canet03a}. Indeed, once  the effective average action is truncated, a residual
dependence on the regulator is indeed unavoidable, and optimizing the cut-off function then allows
one to minimize this dependence. However, the results obtained for the flat phase of membranes are
known to be very robust with respect to the choice of cut-off, see Ref.~\cite{kownacki09}.
Moreover, the extreme stability of the results upon enriching the ansatz -- either by including higher powers
of the order parameter~\cite{kownacki09} or by allowing the coupling constants to acquire a
wavevector dependence $g_k({\bf q})$~\cite{hasselmann11} -- provides additional support for this robustness.

\medskip 

\subsection{Bilayers}
Now, let us investigate the case of membrane bilayers. The running Gibbs free  energy, expressed in terms of fields $\boldsymbol{r}$ and $\boldsymbol{s}$ reads \cite{mauri21bis}:

\begin{equation}
\begin{aligned}
    \Gamma_k[&\boldsymbol{r},\boldsymbol{s}] = \int \mathrm{d}^2x \bigg\{ \frac{Z_k}{2}\left(2\, (\partial^2\boldsymbol{r})^2 + \frac{1}{2}\, (\partial^2\boldsymbol{s})^2\right)+ \lambda_k \, u_{\alpha\alpha}^2  \\
    & \hspace{-0.3cm} +  2\, \mu_k \, u_{\alpha\beta}^2 + \frac{\lambda_k}{4}\, (\partial_\alpha\boldsymbol{r}.\partial_\alpha\boldsymbol{s})^2 + \frac{\mu_k}{4}\, (\partial_\alpha\boldsymbol{s}.\partial_\beta\boldsymbol{r})^2\\
    & \hspace{-0.3cm}  +\frac{\lambda_k}{64}(\partial_\alpha\boldsymbol{s}.\partial_\alpha\boldsymbol{s})^2 + \frac{\mu_k}{32} (\partial_\alpha\boldsymbol{s}.\partial_\beta\boldsymbol{s})^2 + \frac{\lambda_k}{4}u_{\alpha \alpha}\,\partial_\beta\boldsymbol{s}.\partial_\beta\boldsymbol{s}\\
    &  \hspace{-0.3cm} + \frac{\mu_k}{2}u_{\alpha\beta}\,\partial_\alpha\boldsymbol{s}.\partial_\beta\boldsymbol{s} + \frac{\mu_k}{4} (\partial_\alpha \boldsymbol{r}.\partial_\beta \boldsymbol{s})(\partial_\alpha \boldsymbol{s}.\partial_\beta \boldsymbol{r}) \\
    & \hspace{-0.3cm}  +\frac{g_{1k}}{8\ell_k^4} (\boldsymbol{s}^2-\ell_k^2)^2 + \frac{g_{2k}}{2\ell_k^2}(\boldsymbol{s}.\partial_\alpha\boldsymbol{r})^2+\frac{g_{3k}}{4\ell_k^2} \big({\bf s}^2-\ell_k^2) u_{\alpha\alpha} \bigg\} 
\label{bilayeraction}
\end{aligned}
\end{equation} 
where  $u_{\alpha\beta}=1/2(\partial_\alpha {\bf r}.\partial_\beta{\bf r}-\partial_{\alpha}{\bf r}_0.\partial_{\beta}{\bf r}_0)$.

Starting from this expression, we will  perform a few simplifications. First, we will consider the case of an infinite coupling $g_{1k}$, so that the constraint $|{\bf S}|^2 = \ell_k^2$ is effectively enforced. It then follows automatically that the volume-area term $({\bf S}^2 - \ell_k^2)\,u_{\alpha\alpha}$, proportional to $g_{3k}$, does not contribute, so that one puts $g_{3k}\equiv 0$. Next, in the flat phase, we consider both a low-wavevector limit, ${\bf q}^2\ell^2\ll \zeta^2$ and a regime where $g_{2k}$ is very  large which simplifies the spectrum of excitations. Owing to the fact that these two limits do not commute, this procedure leads to a very specific form of the flexural propagator and, ultimately, to the mechanical rigidity crossover, see below.

Another approximation concerns the RG treatment of the bilayer action Eq.~(\ref{bilayeraction}).  
In the naive construction of $\Gamma_k[{\bf r},{\bf s}]$, no distinction is made between different
monomials: for instance, the same field renormalization factor $Z_k$ appears in front of both
$(\partial^2 {\bf r})^2$ and $(\partial^2 {\bf s})^2$, and the same coupling constants
$\lambda_k$ and $\mu_k$ are used for different invariants built from ${\bf r}$ and ${\bf s}$.
This is not the most general situation; in principle, one should assign distinct couplings,
indexed by the specific monomial to which they belong. For the sake of simplicity, however,
we shall restrict ourselves in this work to the simplified parametrization described above.

\subsubsection{Derivation of the RG flow: general }

The derivation of the RG flow proceeds exactly as in the monolayer case, except that the bilayer action Eq.~(\ref{bilayeraction}) now involves the additional couplings $g_{1,k}$ and $g_{2,k}$, which couple the two layers.

\subsubsection{Propagators}

We now give the expression of the propagator in the flat phase defined by
Eq.~(\ref{groundstatebi}) that writes in a RG context as : 
\begin{equation}
\left\{
\begin{array}{ll}
\displaystyle {\bf r}_{0,k}  = \zeta_k\, x_{\alpha}\, {\bf e}_{\alpha}, 
\\[0.35cm]
\displaystyle {\bf s}_{0,k} = \ell_k\, {\bf n} \ . 
\end{array}
\right.
\label{groundstatebi2}
\end{equation} 
One first needs the second functional derivative
$\Gamma_{k}^{(2)}[{\bf r},{\bf s};\mathbf{q},-\mathbf{q}]$ evaluated in this
configuration, hereafter denoted by the subscript $FP$:
\begin{widetext} 
\begin{equation}
\begin{aligned}
\Gamma^{(2)}_k[{\bf r},{\bf s};\mathbf{q},-\mathbf{q}]\Big|_{FP}
&=
\begin{pmatrix}
\Gamma_{k,rr}[{\bf r},{\bf s};\mathbf{q},-\mathbf{q}] & \mathbf{0}_{2\times 3} & \mathbf{0}_{2\times 1} \\
\mathbf{0}_{3\times 2} & \Gamma_{k,rs}[{\bf r},{\bf s};\mathbf{q},-\mathbf{q}] & \mathbf{0}_{3\times 1} \\
\mathbf{0}_{1\times 2} & \mathbf{0}_{1\times 3} & \Gamma_{k,ss}[{\bf r},{\bf s};\mathbf{q},-\mathbf{q}]
\end{pmatrix}\Bigg|_{FP} \, .
\end{aligned}
\end{equation}
\end{widetext} 
Here, $\Gamma_{k,rr}$ describes the pure center-of-mass sector, 
$\Gamma_{k,ss}$ the pure relative-displacement sector, and 
$\Gamma_{k,rs}$ (together with its transpose $\Gamma_{k,sr}$) encodes the
mixing between the ${\bf r}$ and ${\bf s}$ sectors.

One has: 
\begin{widetext} 
\begin{equation}
\Gamma_{k,rr}^{(2)}[{\bf r}, {\bf s};\mathbf{q},-\mathbf{q}]|_{FP} =
\begin{pmatrix}
2 Z_k{\bf q}^4 + 2M_1({\bf q}) & 2 \zeta_k^2 (\lambda_k + \mu_k) q_1 q_2 \\
\\
2 \zeta_k^2 (\lambda_k + \mu_k) q_1 q_2 & 2Z_k {\bf q}^4 + 2M_2({\bf q})
\end{pmatrix}
\end{equation}
\begin{equation}
\Gamma_{k,rs}^{(2)}[{\bf r}, {\bf s};\mathbf{q},-\mathbf{q}]|_{FP} =
\begin{pmatrix}
2 Z_k {\bf q}^4 + q_1^2 g_{2k} & -\frac{i g_{2k} \zeta_k q_1}{\ell_k} & \frac{i g_{2k} \zeta_k q_1}{\ell_k}  \\
\frac{i g_{2k} \zeta_k q_1}{\ell_k} & \frac{1}{2}Z_k {\bf q}^4 + \frac{1}{2}\zeta_k^2 \frac{2 g_{2k}}{\ell_k^2} + \frac{1}{2} M_1({\bf q}) & \frac{1}{2} \zeta_k^2 (\lambda_k + \mu_k) q_1 q_2  \\
\frac{i g_{2k} \zeta_k q_2}{\ell_k} & \frac{1}{2} \zeta_k^2 (\lambda_k + \mu_k) q_1 q_2 & \frac{1}{2}Z_k {\bf q}^4 +  \frac{1}{2}\zeta_k^2 \frac{2 g_{2k}}{\ell_k^2} + \frac{1}{2} M_2({\bf q})
\end{pmatrix}
\end{equation}
\end{widetext}
with $M_i({\bf q})=\zeta_k^2\big(\mu_k {\bf q}^2 + (\lambda_k + \mu_k) q_i^2\big)$ 
and 
\begin{equation}
 \Gamma_{k,ss}^{(2)}[{\bf r}, {\bf s};\mathbf{q},-\mathbf{q}]\big|_{FP} = {Z_k\over 2}  {\bf q}^4+ {g_{1k}\over \ell_k^2}\ . 
 \end{equation}

The eigenvalues of $\left[\Gamma_{k,ij}^{(2)}|_{FP}+{\cal R}_k(\mathbf{q})\right]^{-1}$ can be easily extracted. 

\vspace{0,2cm}
$\bullet$ In  the $rr$-sector one gets: 
\begin{equation}
\begin{aligned}
    &\left\{
        \begin{array}{ll}
            \displaystyle G_{k,1rr}(\boldsymbol{q}) = \frac{1}{2 \big(Z_k\, \boldsymbol{q}^4 + m_{1,k}^2 \boldsymbol{q}^2 +{\cal R}_k(\boldsymbol{q})\big)}\\
            ~\\
            \displaystyle G_{k,2rr}(\boldsymbol{q}) = \frac{1}{2 \big(Z_k\, \boldsymbol{q}^4 + m_{2,k}^2 \boldsymbol{q}^2  +{\cal R}_k(\boldsymbol{q})\big)}\ . 
        \end{array}
    \right.
    \end{aligned}
\end{equation}
The propagators $G_{k,1rr}(\boldsymbol{q})$ and $G_{k,2rr}(\boldsymbol{q})$ are associated with the phonon modes of the mean membrane and are characterized by squared pseudo-masses $m_{1,k}^2 = \zeta_k^2\,\mu_k$ and $m_{2,k}^2 = \zeta_k^2\,(\lambda_k + 2\,\mu_k)$, respectively, which are identical to those of a membrane monolayer. Up to overall factors of $1/2$, these propagators coincide with the monolayer expressions given in Eqs.~(\ref{G1}) and (\ref{G2}).

\vspace{0,2cm} 
$\bullet$  In the $ss$-sector one has: 
\begin{equation}
        G_{k,ss}(\boldsymbol{q}) = \frac{2}{Z_k\,\boldsymbol{q}^4 + 2m_{ss,k}^{2}+{\cal R}_k(\boldsymbol{q})}\, .
    \end{equation}
The propagator $G_{k,ss}(\boldsymbol{q})$ is associated with the flexural mode of the relative membrane. It is characterized by a genuine squared mass
\begin{equation}
m_{ss,k}^{2} = \frac{g_{1,k}}{\ell_k^2},
\end{equation}
which reflects the breaking of translational invariance along the direction orthogonal to the membranes. In the following, we shall focus on the regime where the coupling $g_{1,k}$ is much larger than the other couplings. We therefore take the limit $g_{1,k} \to \infty$, in which $G_{k,ss}(\boldsymbol{q})$ vanishes and the interlayer distance remains fixed.

\vspace{0,2cm}

\noindent\textbullet\;In the $rs$ sector, one obtains a first propagator
\begin{equation}
  G_{k,1rs}(\boldsymbol{q})
  = \frac{2}{Z_k\, \boldsymbol{q}^4
    + m_{1,k}^2 \boldsymbol{q}^2
    + m_{rs,k}^{2}
    + {\cal R}_k(\boldsymbol{q})}\,,
\end{equation}
with
\begin{equation}
  m_{rs,k}^{2} = \frac{2\,\zeta_k^2\, g_{2,k}}{\ell_k^2}\, . 
\end{equation}
Because of the presence of the coupling $g_{2,k}$, the explicit expressions of the two remaining propagators in this sector are rather lengthy and are therefore relegated to App.~\ref{app: Propagators}.

Since we are interested in long-distance physics, we may consider the low-wavevector limit, ${\bf q}^2\,\ell_k^2 \ll \zeta_k^2$  at leading order in the parameter $u=q^2 l_k^2/\zeta_k^2$. In this regime, a second propagator takes the form
\begin{equation}
  G_{k,2rs}(\boldsymbol{q})
  = \frac{2}{Z_k\,\boldsymbol{q}^4
    + m_{2,k}^2 \boldsymbol{q}^2
    + m_{rs,k}^{2}
    + {\cal R}_k(\boldsymbol{q})}\, .
\end{equation}

The propagators $G_{k,1rs}$ and $G_{k,2rs}$ are associated with phonon modes of the relative membrane and exhibit two distinct kinds of masses: the squared pseudo-masses $m_{1,k}^2$ and $m_{2,k}^2$ as well as  the genuine squared masses $m_{rs,k}^{2}$  arising from the coupling between the membranes. Again the propagators $G_{k,1rs}$ and $G_{k,2rs}$ deduce from those of monolayers by a factor 2.

 The third, and most important, propagator in the $rs$ sector, in the low-wavevector limit, reads 
\begin{equation}
  G_{k,3rs}(\boldsymbol{q})
  = \frac{1}{2\big((Z_k + c_k)\,\boldsymbol{q}^4 + {\cal R}_k(\boldsymbol{q})\big)} \, ,
  \label{Grs_def}
\end{equation}
with
\begin{equation}
  c_k = \ell_k^2\,\frac{\lambda_k + 2\,\mu_k}{4} \, .
  \label{ck_def}
\end{equation}

The propagator $G_{k,3rs}(\boldsymbol{q})$ is associated with the flexural mode of the mean membrane and carries neither a mass nor a pseudo-mass. As can be seen from Eq.~(\ref{Grs_def}), its explicit dependence on $g_{2,k}$ has disappeared (see App.~\ref{app: Propagators}). However, as the low-wavevector expansion, $\boldsymbol{q}^2 \ell_k^2 \ll\zeta_k^2$, is \emph{not uniform} in $g_{2k}$, the limits $\boldsymbol{q}\to 0$ and $g_{2,k}\to 0$ do not commute so that  the signature of a finite interlayer coupling $g_{2,k}$ remains encoded in the specific form of $G_{k,3rs}$; see App. \ref{app: G3rs} for details. 

Indeed, Eqs.~(\ref{Grs_def}) and (\ref{ck_def}) show that the bending rigidity is shifted with respect to the monolayer case, compare Eq.~(\ref{G0}), according to
\begin{equation}
    \kappa \;\longrightarrow\; \kappa_{\mathrm{eff}} = \kappa + c_k \, .
    \label{modif}
\end{equation}
This is precisely the modification obtained by de~Andr\'es {\it et al.} \cite{andres12}  within a harmonic treatment. It constitutes the main effect of adding the interlayer coupling $g_{2,k}$ to the monolayer membrane theory and is responsible for the crossover of the running coupling constants along the RG flow.

\subsubsection{Physical discussion: origin of the combination $\ell^2(\lambda+2\mu)/2$}

We now discuss the origin of the combination $\ell^2(\lambda+2\mu)/2$ appearing in the effective bending rigidity $\kappa_{\mathrm{eff}}$. It reflects the additional in-plane elastic cost generated by bending when the two sheets are constrained to stay at fixed separation $\ell$ and to bend without lateral slip. The large-$g_1$ constraint enforces $|\boldsymbol s|\simeq \ell$, while the $g_2$ term penalizes the tangential component $\boldsymbol s_\parallel$; thus, for large $g_2$, $\boldsymbol s\simeq \ell\,\boldsymbol n$ and the two physical layers are parallel surfaces,

\begin{equation}
\boldsymbol r_{1,2}=\boldsymbol r \pm \frac{\ell}{2}\boldsymbol n.
\end{equation}
where the indices 1 and 2 refer to bilayer 1 and 2. 

Using the parallel-surface metric expansion (App.~\ref{geometry})
\begin{equation}
g_{\alpha\beta}(d)=g_{\alpha\beta}-2d\,K_{\alpha\beta}+{\cal O}(d^2),
\end{equation}
where $K_{\alpha\beta}$ is the second fundamental form. In our case one has $d=+{\ell/ 2}$ and  $d=-{\ell/ 2}$  and the layer metrics at $d=\pm \ell/2$ read
\begin{equation}
g^{(1,2)}_{\alpha\beta}=g_{\alpha\beta}\mp \ell\,K_{\alpha\beta}+{\cal O}(\ell^2),
\end{equation}
hence the corresponding strains
\begin{equation}
u^{(1,2)}_{\alpha\beta}
=\frac12\big(g^{(1,2)}_{\alpha\beta}-\delta_{\alpha\beta}\big)
= u_{\alpha\beta}(0)\mp \frac{\ell}{2}K_{\alpha\beta}+{\cal O}(\ell^2),
\label{eq:strain_layers_keepu0}
\end{equation}
where $u_{\alpha\beta}(0)\equiv \tfrac12(g_{\alpha\beta}-\delta_{\alpha\beta})$ is the mean-surface strain.

The in-plane elastic energy density of one sheet is
\begin{equation}
f_{\rm el}=\lambda\,(u_{\gamma\gamma})^2+2\mu\,u_{\alpha\beta}u_{\alpha\beta}.
\end{equation}
Substituting \eqref{eq:strain_layers_keepu0} and summing the two layers, all terms linear in $K_{\alpha\beta}$ cancel by symmetry, and one obtains 
\begin{equation}
\begin{array}{ll}
f_{\rm el}^{\rm (bilayer)}
&= \displaystyle \sum_{i=1,2}\Big[\lambda\,(u^{(i)}_{\gamma\gamma})^2+2\mu\,u^{(i)}_{\alpha\beta}u^{(i)}_{\alpha\beta}\Big] \nonumber\\
\\
&= 2\Big[\lambda\,u_{\gamma\gamma}(0)^2+2\mu\,u_{\alpha\beta}(0)u_{\alpha\beta}(0)\Big]\\
\\
&+\displaystyle  \frac{\ell^2}{2}\Big[\lambda\,(K_{\gamma\gamma})^2+2\mu\,K_{\alpha\beta}K_{\alpha\beta}\Big]
+{\cal O}(\ell^4).
\label{eq:bilayer_energy_decomp}
\end{array}
\end{equation}
The first bracket is simply the in-plane elastic energy of the mean surface. The second bracket is the curvature-induced contribution: it is quadratic in $K_{\alpha\beta}$ and therefore renormalizes the bending sector of the effective theory.

For a uniaxial bending mode (Monge gauge, $h=h(x)$), the dominant component is $K_{xx}\simeq \kappa$, with $K_{yy}\simeq K_{xy}\simeq 0$, so that
\begin{equation}
(K_{\gamma\gamma})^2=K_{xx}^2=\kappa^2,
\qquad
K_{\alpha\beta}K_{\alpha\beta}=K_{xx}^2=\kappa^2,
\end{equation}
and the curvature part of \eqref{eq:bilayer_energy_decomp} reduces to
\begin{equation}
f_{\rm el,curv}^{\rm (bilayer)}
=\frac{\ell^2}{2}(\lambda+2\mu)\,\kappa^2
\;\sim\;
\frac{\ell^2}{2}(\lambda+2\mu)\,(\partial_x^2 h)^2.
\end{equation}
This identifies the induced elastic bending rigidity
\begin{equation}
\kappa_{\rm el}=\frac{\ell^2}{2}(\lambda+2\mu)
\end{equation}
which contributes additively to $\kappa_{\mathrm{eff}}$.
    
\subsubsection{RG equations in the flat phase }

We now derive the RG equation, considering straight away the flat phase limit. 

The definitions of the coupling constants does not change -- up to factors 1/2 -- with  respect the monolayer case and are given by:
\begin{equation}
    \left\{
\begin{aligned}
    &\mu_k=\frac{1}{2\,\zeta_k^2}\lim_{\boldsymbol{p} \to 0}\frac{\partial}{\partial\boldsymbol{p}^2}\left( \Gamma_{k,2\, 2}^{(2)}[\boldsymbol{r},\boldsymbol{s};\boldsymbol{p},-\boldsymbol{p}]\right)\Big|_{FP}
    \\
     & \lambda_k=\frac{1}{2\,\zeta_k^2}\lim_{\boldsymbol{p} \to 0}\frac{\partial}{\partial p_2^2}\left(\Gamma_{k,2\, 2}^{(2)}[\boldsymbol{r},\boldsymbol{s};\boldsymbol{p},-\boldsymbol{p}]\right)\Big|_{FP}-\mu_k
     \\
     &  Z_k=\frac{1}{2}\lim_{\boldsymbol{p} \to 0}\frac{\partial}{\partial\boldsymbol{p}^4}\left( \Gamma_{k,3\,3}^{(2)}[\boldsymbol{r},\boldsymbol{s};\boldsymbol{p},-\boldsymbol{p}]\right)\Big|_{FP}
    \label{definitions}
\end{aligned}
\right.
\end{equation}
while $\zeta_k$ is defined in the same way as in Eq.(\ref{minn}).

We proceed in the same way as in the monolayer case and introduce dimensionless couplings. We define
\begin{equation}
\begin{aligned}
\lambda_k &= k^{2-2\eta_k}\,\bar{\lambda}_k, 
\hspace{1.1cm}\mu_k = k^{2-2\eta_k}\,\bar{\mu}_k,
\\
\zeta_k & = k^{\eta_k}\,\bar{\zeta}_k \hspace{1.85cm} \ell_k = k^{(-2+\eta_k)/2}\,\bar{\ell}_k
\label{dimless_bilayer_couplings2}
\end{aligned}
\end{equation}
and   the running anomalous dimension as
\begin{equation}
Z_k = k^{-\eta_k}.
\label{Z_scaling2}
\end{equation}

We are only interested in the flat phase of the system. We can therefore take, prior to the calculation, the limit $\bar\zeta_k \gg 1$. The RG equations that control the infrared behavior of membrane bilayers are then given by:
\begin{equation}
\hspace{-0.3cm}\left\{
\begin{aligned}
    &\partial_t\bar\lambda_k=\displaystyle 2(\eta_k-1)\,\bar\lambda_k + (2\,\bar\lambda_k^2+4\,\bar\lambda_k\,\bar\mu_k+\bar\mu_k^2)\,A_2\,{}_{B}{}\bar{L}_{200}^{6}
    \\
    &\partial_t\bar\mu_k=\displaystyle 2(\eta_k-1)\,\bar\mu_k+\frac{1}{2}\,\bar\mu_k^2\,A_2\,{}_{B}{}\bar{L}_{200}^{6}
    \\
    &\eta_k   = (\bar\l_k+2\,\bar\mu_k)\,A_2\,_B\bar{N}_{200}^{4}\\
    & \hspace{3cm}-\frac{(\bar\l_k^2+\bar\l_k\,\bar\mu_k+\bar\mu_k^2)\,A_2}{\bar\l_k+2\,\bar\mu_k}{}_B\bar{L}_{100}^{2} \ . 
\end{aligned}
\right.
\label{floflat}
\end{equation}

Once again, these equations are remarkably simple. In fact, they differ from those of monolayer membranes, Eqs.~(\ref{flatmonocouplings}) and (\ref{flatmonoeta}), only by the rescalings
\begin{equation}
\lambda_k \;\to\; \frac{\lambda_k}{2}, 
\qquad
\mu_k \;\to\; \frac{\mu_k}{2},
\end{equation}
and by the replacement of the threshold functions
\begin{equation}
\bar T_{a00}^{2+\alpha} \;\to\; {}_B\bar T_{a00}^{2+\alpha},
\quad T = L, N.
\end{equation}
The new, bilayer-specific threshold functions are defined as, see App.\ref{thresholdbilayers} 
\begin{equation}
\left\{
\begin{aligned}
{}_B L_{a00}^{\,2+\alpha}
&= -\,\frac{1}{4 A_2}\,\widehat{\partial_t}
   \int_{\boldsymbol{q}} \boldsymbol{q}^{\alpha}\,
   P_{c_k}(\boldsymbol{q})^{-a},
\\[0.2cm]
{}_B N_{a00}^{\,2+\alpha}
&= -\,\frac{1}{4 A_2}\,\widehat{\partial_t}
   \int_{\boldsymbol{q}} \boldsymbol{q}^{\alpha}\,
   \frac{\partial P_{c_k}(\boldsymbol{q})}{\partial \boldsymbol{q}^2}\,
   P_{c_k}(\boldsymbol{q})^{-a},
\end{aligned}
\right.
\end{equation}
where the only difference between the monolayer and bilayer cases lies in the definition of
$P_{c_k}(\boldsymbol{q})$, which now reads
\begin{equation}
    P_{c_k}(\boldsymbol{q})
    = (Z_k + c_k)\,\boldsymbol{q}^4 + {\cal R}_k(\boldsymbol{q}) \, .
\label{P}
\end{equation}
\bigskip

Using the Litim cut-off (see App.~\ref{app:litimbilayer}), the RG flow equations for the coupling constants in the flat phase take the form:
\begin{equation}
\left\{
\begin{aligned}
    &\partial_t \bar\lambda_k=  2(\eta_k - 1)\,\bar\lambda_k
- H_k\,\bigl(2\,\bar\lambda_k^{\,2}
             + 4\,\bar\lambda_k\,\bar\mu_k
             + \bar\mu_k^{\,2}\bigr), \\
    &\partial_t \bar\mu_k
= 2(\eta_k - 1)\,\bar\mu_k
- H_k\,\bar\mu_k^{\,2} 
\label{flowbil}
\end{aligned}
\right.
\end{equation}
with: 
\begin{equation}
\begin{aligned}
H_k &\equiv 
\frac{1}{128\pi\,\bar c_k^{5/2}(1+\bar c_k)^2}
\Big[
\sqrt{\bar c_k}\,\Big((1+\bar c_k)(3+\bar c_k)\,\eta_k  \\
& -4(\bar c_k-1)\,\bar c_k\Big)
+(1+\bar c_k)^2\Big(\bar c_k(\eta_k-4)-3\,\eta_k\Big)\, \\
&\times\text{ArcTan}\Big[\sqrt{\bar c_k}\Big] \Big]\ . 
\label{H}
\end{aligned}
\end{equation}

For the Young modulus, one has: 
\begin{equation}
\partial_t \bar Y_k
= 2(\eta_k - 1)\, \bar Y_k
- \frac{3}{4}\,H_k\,\bar Y_k^{2},
\label{FlotY}
\end{equation}
while $\eta_k$, expressed in terms of $\bar Y_k$, is given by: 
\begin{widetext}
\begin{equation}
\eta_k
=
\frac{
12\,\bar Y_k\,\bar c_k\,
\Big(
\sqrt{\bar c_k}
+
(1+\bar c_k)\,\text{ArcTan}[\sqrt{\bar c_k}]
\Big)
}{
(1+\bar c_k)\Big[
3\,\bar Y_k\,(\bar c_k-1)\,\text{ArcTan}[\sqrt{\bar c_k}]
+
\sqrt{\bar c_k}\,\big(3\, \bar Y_k+128\pi\,\bar c_k\big)
\Big]
}\ . 
\label{etabil}
\end{equation}
\end{widetext}
In Eqs.~({\ref{H}) and (\ref{etabil})  one has
\begin{equation}
\bar c_k = \bar \ell_k^{\,2}\,\frac{\bar \lambda_k + 2\,\bar \mu_k}{4},
\end{equation}
which induces a nontrivial dependence of the flow on $\bar \ell_k$, $\bar \lambda_k$ and $\bar \mu_k$.
We finally give the flow equation for the parameter $\ell_k$, which characterizes the rest distance
between the two membranes. Since the physical separation $\ell$ is fixed, the flow of $\bar \ell_k$
is purely dimensional and is simply given by
\begin{equation}
    \partial_t \bar \ell_k = \frac{2-\eta_k}{2}\, \bar  \ell_k 
    \label{flow_l}
\end{equation}
and that of $\bar c_k$ by:
\begin{equation}
\partial_t \bar c_k
=
\eta_k\,\bar c_k
- H_k\,\bar c_k\,
\frac{2\bar \lambda_k^{\,2} + 4\bar \lambda_k\bar \mu_k + 3\bar \mu_k^{\,2}}
     {\bar \lambda_k+2\bar \mu_k}
\,.
\end{equation}

It is easy to check that in the limit $\bar \ell_k\to 0$, thus $\bar c_k\to 0$,  the equations Eqs.~(\ref{FlotY}) and $(\ref{etabil})$ give those of monolayers Eq.~(\ref{monolayereq2}).

\section{Physical results}

We now turn to a discussion of our results. We begin in Sec. IV A with a qualitative analysis of the
short- and long-distance behavior of the effective bending rigidity, from which we obtain an estimate
of the mechanical crossover scale $k_c$. In Sec. IV B, we specify the initial conditions of the RG flow for
realistic (graphene) parameter values. We then discuss, in Sec. IV C, the Ginzburg scale $k_G$ associated
with the harmonic-anharmonic crossover. Finally, in Sec. IV D, we study the flow of the
running anomalous dimension and  Young modulus.

\subsection{Mechanical crossover scale}

Here we analyze the short- and long-distance behavior of the effective bending rigidity and provide an estimate of the mechanical crossover length scale.

As recalled above, the effective bending rigidity of the mean surface contains two contributions:
\begin{equation}
\kappa_{\text{eff}}(k) \;\simeq\;
\underbrace{2\,\kappa_k}_{\text{microscopic bending}}
\;+\;
\underbrace{\frac{\ell^2}{2}\big(\lambda_k+2\mu_k\big)}_{\text{elastic bilayer contribution}}.
\label{eq:kappa_eff_split}
\end{equation}

 It is worth stressing at this point that our NPRG approach naturally provides
a scale-dependent effective bending rigidity $\kappa_{\text{eff}}(k)$, while
most experimental determinations of $\kappa$ extract an effective value
averaged over a finite wave-vector window set by the sample geometry and the
probing technique (nanoindentation, snap-through, modal analysis, or
scattering). In this sense, measurements typically probe
$\kappa_{\text{eff}}(k\simeq k_{\text{geom}})$, with $k_{\text{geom}}\sim 1/L$
the inverse size of the probed membrane or the typical wave vector of the
relevant mode, rather than the strict $k\to 0$ limit. More precisely, in a
finite sample of lateral size $L$, the infrared sector is cut off by the
smallest accessible wave vector $q_{\min}\sim \alpha/L$, where the numerical
factor $\alpha$ depends on the geometry and boundary conditions (e.g.
$\alpha=2\pi$ for periodic boundaries, while for a circular drum $\alpha$ is
set by the relevant Bessel-mode wave number). Therefore the identification
$k\sim 1/L$ is meant at the scaling level and the proportionality constant is
not universal; it can be fixed for a given experimental (or numerical) setup
either by matching to the dominant eigenmode(s) of the corresponding
geometry, or empirically by calibrating $\alpha$ from measurements/simulations
where the probed wave-vector window is known. The RG description in terms of
$\kappa_{\text{eff}}(k)$ is therefore particularly well suited to interpret
these experiments, since it directly encodes how the apparent bending
rigidity changes when the probed length scale is varied.

For bilayers, the crossover between the ``thick-plate'' regime and the
two-monolayer regime predicted by our RG flow should manifest itself as a
strong dependence of the effective bending rigidity on the probed length
scale, in qualitative agreement with the thickness- and size-dependent
stiffness reported in recent experiments and simulations on few-layer
graphene.

\subsubsection{Short- and long-distance behavior of the effective bending rigidity}

In the RG framework, it is convenient to work with dimensionless
couplings $\bar\lambda_k,\bar\mu_k,\bar\kappa_k$ defined as :
\begin{equation}
\lambda_k = k^{2-2\eta_k}\,\bar\lambda_k,\hspace{0.3cm}
\mu_k     = k^{2-2\eta_k}\,\bar\mu_k,\hspace{0.3cm}
\kappa_k  = k^{-\eta_k}\,\bar\kappa_k,
\label{eq:scaling_dimful}
\end{equation}
with $\{\bar\lambda_k,\bar\mu_k,\bar\kappa_k,\eta_k\}\to
\{\bar\lambda_\ast,\bar\mu_\ast,\bar\kappa_\ast,\eta\}$ as $k\to 0$. The physical interlayer distance,
$\ell$ is  kept fixed, so that we do not make $\ell$ dimensionless.
Using the scaling forms \eqref{eq:scaling_dimful}, the  two pieces of $\kappa_{\text{eff}}(k)$ in Eq.(\ref{eq:kappa_eff_split})  read
\begin{equation}
\left\{
\begin{aligned}
\kappa_{\text{mi}}(k)
&\equiv 2\,\kappa_k
 \;=\; 2\,\bar\kappa_k\,k^{-\eta_k},
\\
\kappa_{\text{el}}(k)
&\equiv \frac{\ell^2}{2}\big(\lambda_k+2\mu_k\big)
 \;=\; \frac{\ell^2}{2}\big(\bar\lambda_k+2\bar\mu_k\big)\,k^{2-2\eta_k}.
\end{aligned}
 \right.
 \end{equation}

\paragraph{Short-distance regime.}

At high RG scales $k$, the
geometric elastic contribution associated with the finite interlayer
distance $\ell$ is already present, while the microscopic bending
rigidity has not yet undergone its full anomalous growth. 
If the bare combination
$\ell^2(\bar\lambda_k+2\bar\mu_k)/2$ is sufficiently large at the
microscopic scale, the UV effective bending rigidity is dominated by
the elastic contribution,
\begin{equation}
\kappa_{\text{eff}}(k) \;\simeq\; \kappa_{\text{el}}(k)
  = \frac{\ell^2}{2}\big(\bar\lambda_k+2\bar\mu_k\big)\,k^{2-2\eta_k}.
\end{equation}

\paragraph{Long-distance regime.}

At long distance  the dimensionless coupling constants $\bar\lambda_k,\bar\mu_k$ and $\bar\kappa_k$ approach  the flat-phase fixed point, so the $k$-dependence is fully contained in the powers of $k$:
\begin{equation}
\kappa_{\text{mi}}(k) \propto k^{-\eta},
\qquad
\kappa_{\text{el}}(k) \propto \ell^2\,k^{2-2\eta}.
\end{equation}
In $D=2$ one has $0<\eta<1$, so
\begin{equation}
-\eta < 0 ,\qquad  2-2\eta > 0.
\end{equation}
Therefore, as $k\to 0$,
\begin{equation}
\kappa_{\text{mi}}(k)\;\longrightarrow\; \infty,
\qquad
\kappa_{\text{el}}(k)\;\longrightarrow\; 0,
\end{equation}
and the effective bending rigidity is dominated by the renormalized
microscopic bending term,
\begin{equation}
\kappa_{\text{eff}}(k)\;\simeq\; 2\,\kappa_k
\;\propto\; k^{-\eta}\ . 
\end{equation}

\subsubsection{Mechanical crossover scale : first analysis}
\label{crossover}

The crossover scale $k_c$ between these two regimes is determined
by the condition
\begin{equation}
\frac{\ell^2}{2}\big(\lambda_{k_c}+2\mu_{k_c}\big)
\;\simeq\; 2\,\kappa_{k_c},
\label{kcexact}
\end{equation}
which expresses the competition, along the RG flow, between the elastic
bending stiffness generated by the in-plane couplings
$\bar\lambda_k,\bar\mu_k$ at fixed physical separation $\ell$ and
the anomalously growing microscopic bending rigidity $\kappa_k$.

Note that, in terms of purely dimensionless coupling, this is equivalent to: 
\begin{equation}
\frac{\ell_k^2}{2}\big(\bar\lambda_{k_c}+2\bar\mu_{k_c}\big)
\;\simeq\; 2\,\bar\kappa_{k_c} 
\label{kcexact2}
\end{equation}
which  gives : 
\begin{equation}
\bar c_{k_c}\simeq 1\ . 
\label{conditionck}
\end{equation}

If $k_c$ lies in the scaling regime of the flat phase, the
dimensionless couplings are already close to their fixed-point values:
\begin{equation}
\bar\lambda_{k_c}\simeq \bar\lambda_*,\ \
\bar\mu_{k_c}\simeq \bar\mu_*,\ \
\bar\kappa_{k_c}\simeq \bar\kappa_*,\ \
\eta_{k_c}\simeq \eta\ . 
\end{equation}
Under this approximation, Eq.~(\ref{kcexact2}) gives, by restoring the dimension of $\ell$, the simple
estimate
\begin{equation}
\ell^2\big(\bar\lambda_*+2\bar\mu_*\big)\,
k_c^{2-\eta}
\simeq 4\,\bar\kappa_*,
\end{equation}
so that
\begin{equation}
k_c
\sim
\left[
\frac{4\,\bar\kappa_*}
     {\ell^2\big(\bar\lambda_*+2\bar\mu_*\big)}
\right]^{\!\frac{1}{2-\eta}}.
\label{eq:kstar_estimate}
\end{equation}
Numerically, with the bare values of graphene  \cite{mauri21bis}  $\kappa_*=\kappa_g=1,$ and $\ell=\ell_g=3.25\, \mathrm{\mathring{A}}$ and the fixed point values \cite{kownacki09}: $\bar\mu_*\simeq 6.21, \bar\lambda_*\simeq -3.10$ and   $\eta=0.85$  one gets $k_c\simeq 0.06 \mathrm{\mathring{A}}^{-1}$ and the RG-time -- taking $a=a_g=2.46\,   \mathrm{\mathring{A}}$ --   $t_c=-\ln (k_c a) =1.88$. 

Alternatively one could use the bare coupling  constants
\begin{equation}
k_c
\sim
a^{-1}\left[
\frac{4\,\kappa}
     {\ell^2\big(\lambda+2\mu\big)}
\right]^{\!\frac{1}{2-\eta}}.
\label{eq:kstar_estimatebis}
\end{equation}
If one considers  those of the graphene  \cite{mauri21bis}: 
\begin{equation}
\begin{aligned}
&\ \lambda_g=3.8\,\text{eV} \mathrm{\mathring{A}}^{-2}~~~~\mu_g=9.3\,\text{eV} \mathrm{\mathring{A}}^{-2} 
\label{graphene_values}
\end{aligned}
\end{equation}
one gets: $k_c\simeq 0.012 \mathrm{\mathring{A}}^{-1}$ and the RG-time $t_c=-\ln k_c a=3.54$. 

 Note that, when the crossover scale $k_c$ is estimated from the fixed-point
values of the dimensionless couplings, Eq.(\ref{eq:kstar_estimate}) the microscopic lattice spacing $a$
does not appear explicitly: the condition defining $k_c$ compares two
contributions to $\kappa_{\mathrm{eff}}(k)$ at the same running scale $k$
and is therefore formulated entirely in terms of renormalized, dimensionless
quantities. The dependence on the ultraviolet cutoff $\Lambda\sim a^{-1}$
enters only through the choice of origin of the RG “time” $t=\ln(k/\Lambda)$
and thus drops out of the crossover condition itself, in contrast with bare
estimates,  Eq.(\ref{eq:kstar_estimatebis}),  written directly in terms of microscopic parameters.

\medskip 

An improved estimate of $k_c$ is provided below. 

\subsubsection{Mechanical crossover scale :  comparison with prior  work}

It is interesting to consider the mechanical cross-over scales obtained in presence of an explicit $g_2$-dependence as considered by Mauri {\it et al.} \cite{mauri21bis}. One has \cite{mauri21bis}:
\begin{equation}
\left\{
\begin{aligned}
k_{1c} &= \sqrt{\frac{g_2}{2\kappa}}\,, \\[0.2cm]
k_{2c} &= \frac{1}{\ell}\,\sqrt{\frac{2 g_2}{\lambda + 2\mu}} \, .
\end{aligned}
\right.
\label{qc}
\end{equation}
For the representative value $g_2 = 0.11$ used in Ref.~\cite{mauri21bis}, these expressions yield the orders of magnitude
\begin{equation}
k_{1c} \simeq 0.23\,\mathrm{\mathring{A}}^{-1},
\qquad
k_{2c} \simeq 0.03\,\mathrm{\mathring{A}}^{-1},
\end{equation}
and the corresponding RG ``times''
\begin{equation}
t_{1c} = -\ln(k_{1c} a) \simeq 0.55,
\qquad
t_{2c} = -\ln(k_{2c} a) \simeq 2.60.
\label{deftc}
\end{equation}

However, within the present approach these scales are only indicative, since (i) the coupling
$g_2$ drops out at leading order, and (ii) we work with renormalized rather than bare quantities. Nevertheless it is interesting to note that the bare quantity $\bar c_{k=a^{-1}}$ is nothing but the squared
ratio $(k_{1c}/k_{2c})^2$. Thus:

\begin{itemize}
  \item When $\bar c_{k=a^{-1}}$ is large, i.e. $k_{1c} \gg k_{2c}$, the effective rigidity
  is dominated by the elastic contribution,
 \begin{equation}
    \kappa_{\mathrm{eff}} \simeq \frac{\ell^2}{2}\,(\lambda + 2\mu)\,.
  \end{equation}

  \item When $\bar c_{k=a^{-1}}$ is of order unity, i.e. $k_{1c} \sim k_{2c}$, the bending
  rigidity $\kappa$ and the elastic couplings $\lambda$ and $\mu$ contribute on an
  equal footing, and
  \begin{equation}
    \kappa_{\mathrm{eff}} \simeq 2\kappa + \frac{\ell^2}{2}\,(\lambda + 2\mu)\,.
  \end{equation}

  \item When $\bar c_{k=a^{-1}}$ is small, i.e. $k_{1c} \ll k_{2c}$, in-plane elasticity becomes
  negligible for bending and the two monolayers bend together; one then recovers a regime
  dominated by the bending rigidity of two independent monolayers,
 \begin{equation}
    \kappa_{\mathrm{eff}} \simeq 2\kappa\,.
 \end{equation}
\end{itemize}

We therefore have a natural correspondence between the parameter 
$c_k$ and the problem's  natural crossover scales.

\subsubsection{Mechanical crossover scale : RG analysis}

We finally provide an estimate of the mechanical crossover scale $k_c$ based on an RG analysis.  We define the crossover RG-time $t_c$ as the scale at which $\bar c_{k_c} \simeq 1$, see Eq.(\ref{conditionck}).  For a temperature $T = 500$~K, one finds $k_c \simeq 0.0028\,\mathrm{\mathring{A}}^{-1}$ and $t_c \simeq 4.96$, see Fig.~\ref{ckt}, which is larger than both $t_{1c}$ and $t_{2c}$.

As seen in Fig.~\ref{ckt}, the typical crossover RG-time $t_c$ lies well above the two scales
$t_{1c}$ and $t_{2c}$, see Eq.(\ref{deftc}), between which the rigidity crossover would naively be expected to occur.
The origin of this discrepancy is the following. When $g_2$ is kept finite, the mechanical
crossover observed in Ref.~\cite{mauri21bis} is driven by interlayer shear, characterized by $g_2$,
and naturally takes place between $t_{1c}$ and $t_{2c}$. In the present case, we define the crossover scale by
\begin{equation}
\frac{\ell^2}{2}\bigl(\lambda_{k_c} + 2\mu_{k_c}\bigr)\simeq   2\,\kappa_{k_c},
\end{equation}
which marks the scale at which the microscopic (running) bending contribution $2\kappa_k$ and the
(running) elastic thick-plate contribution $\ell^2(\lambda_k + 2\mu_k)/2$ to the effective bending
rigidity become equal. In the flat phase, $\kappa_k$ increases under the RG flow whereas
$\lambda_k$ and $\mu_k$ decrease, so that this balance is reached only deep in the infrared.
As a consequence, the crossover RG-time $t_c = \ln(k_c/k_0)$ is much larger than the bare
mechanical crossover times $t_{1c}$ and $t_{2c}$ defined at the harmonic level in terms of the bare coupling constants. It therefore reflects a fully renormalized internal bending/elastic balance rather than a bare sliding crossover.

\begin{figure}[h!]
\centering
\includegraphics[width=0.5\textwidth, trim=10mm 88mm 1mm 45mm, clip]{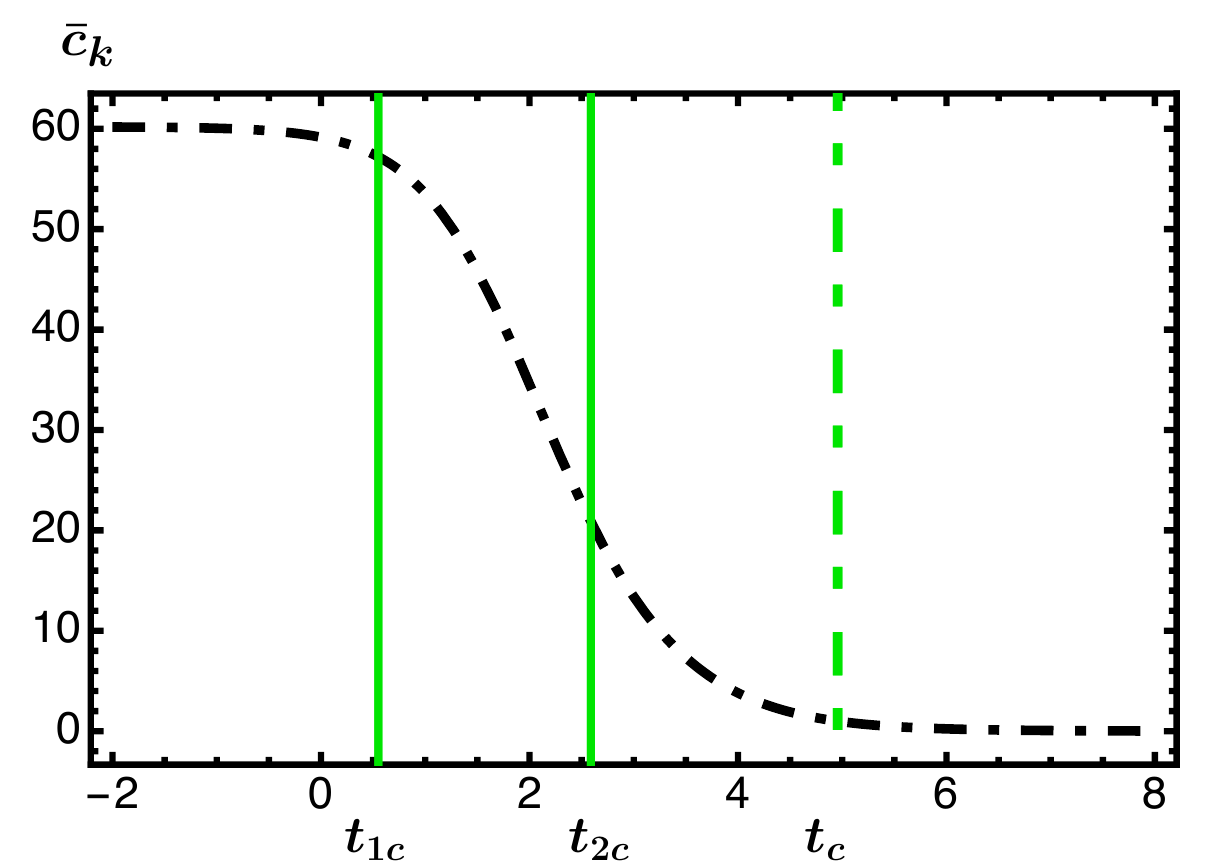}
\caption{Behavior of $\bar c_k$ -- dashed-dot curve -- as a function of $t=-\ln ka$ for $T=500$ K. One has displayed the typical RG-time $t_c$ (dashed-dot vertical line) associated with the bending rigidity  crossover and the RG-times  $t_{1c}$ and $t_{2c}$ (full vertical lines). }
\label{ckt}
\end{figure} 

\medskip

The different wavevector scales given above should be compared with the wave-vector
windows (or, equivalently, length scales) effectively probed in experiments
and simulations, rather than with absolute values of the bending rigidity.
In particular, the crossover between the elasticity-dominated and
curvature-dominated regimes encoded in $\kappa_{\mathrm{eff}}(k)$ provides a
natural framework to rationalize the strong thickness dependence observed in
bilayer and few-layer graphene \cite{lindahl12,Han20,Jiang23} as well
as the size- and temperature-dependent effective rigidities inferred from
modal analysis and helium-atom scattering \cite{Sajadi18,Eder21}.

\subsection{Initial conditions RG flow: graphene parameters}

We give here the  initial conditions of the flow. We restore the physical units of the coupling constants and consider the microscopic scale  $k=a^{-1}$: 
\begin{equation}
\begin{aligned}
    &~\bar{\lambda}_{k=a^{-1}}=k^{-2+2\,\eta_k}Z_k^2\,\beta^{-1}\kappa^{-2}\lambda_g\vert_{k=a^{-1}}=\frac{k_B T\,a^2}{\kappa^2} \lambda_g\\
    &~\bar{\mu}_{k=a^{-1}}=k^{-2+2\,\eta_k}Z_k^2\,\beta^{-1}\kappa^{-2}\mu_g\vert_{k=a^{-1}}=\frac{k_B T\,a^2}{\kappa^2} \mu_g\\
    &~\bar{\ell}_{k=a^{-1}}=k^{(4-D+\eta_k)/2}\sqrt{\frac{\beta\,\kappa}{Z_k}}\,\ell_g\big\vert_{k=a^{-1}}=\sqrt{\frac{\kappa}{k_B T}}\,{\ell_g\over a} 
\end{aligned}
\end{equation}
 Let us remark that  the present dimensionless formulation, the RG flow equations are
temperature independent, see e.g. Eq.(\ref{floflat}) while  the temperature $T$ enters  through the microscopic initial conditions at $k=\Lambda$, since restoring physical units
amounts to setting the initial dimensionless couplings proportional to
$\beta=1/k_B T$.

If we again use the physical coupling constants of graphene, we obtain
\begin{equation}
\bar c_{k=a^{-1}}
= \frac{\ell_g^2}{4\kappa}\,\big(\lambda_g + 2\,\mu_g\big)
\simeq 59.15,
\end{equation}
 see Fig.\ref{ckt},  which is much larger than unity. We are therefore precisely in the conditions required to observe a mechanical crossover.

\subsection{Ginzburg scale}

One finally  define the Ginzburg scale which sets the harmonic-to-anharmonic crossover.  For a monolayer, is given by \cite{zakharchenko10,los17}
\begin{equation}
k_{G, \text{mono}} = \sqrt{\frac{3 T Y}{16 \pi \kappa^2}}\,.
\end{equation}

For bilayers,  Mauri  {\it et al.} \cite{mauri21bis} have defined: 
\begin{equation}
\left\{
\begin{aligned}
&k_{1G}=\sqrt{{3 T\over {16\pi}} {(2Y)\over (2 \kappa)^2}}\\
\\
&k_{2G}=\sqrt{{3 T\over {16\pi}} {(2Y)\over (\kappa_0)^2}}
\label{qG12}
\end{aligned}
\right.
\end{equation}
where $Y=4\mu(\lambda+ \mu)/(\lambda+2\mu)$ is the Young modulus and   $\kappa_0=2 \kappa+\ell^2\,\displaystyle \frac{\lambda+2\,\mu}{2}$. 

We introduce the corresponding RG-times
\begin{equation}
t_{1G} = -\ln(k_{1G} a), \qquad
t_{2G} = -\ln(k_{2G} a)\ .
\end{equation}
These scales are associated with the thermal crossover between the harmonic regime,
for $t \ll t_{1G}, t_{2G}$, and the anharmonic (renormalized) regime, for
$t \gg t_{1G}, t_{2G}$.

Numerically, for $T = 500\,\mathrm{K}$, one finds
\begin{equation}
k_{1G} \simeq 0.17\,\mathrm{\mathring{A}}^{-1},
\qquad
k_{2G} \simeq 0.0028\,\mathrm{\mathring{A}}^{-1},
\end{equation}
corresponding to
\begin{equation}
t_{1G} \simeq 0.89,
\qquad
t_{2G} \simeq 4.98.
\end{equation}

A harmonic-to-anharmonic RG-time scale that is more directly relevant to the present RG study  is given, for instance, by the scale at which the anomalous dimension, obtained from the RG equation Eqs.~(\ref{flowbil}),  (\ref{FlotY}) and (\ref{etabil}), becomes of order its fixed-point value $\eta$. More precisely,
we impose
\begin{equation}
\eta_{k_G} = \omega \,\eta,
\label{defetag}
\end{equation}
with $\omega$ a number of order unity (e.g. $\omega=1/2$), so that for $k\gg k_G$ the flow
remains close to the harmonic (Gaussian) regime with $\eta_k\simeq 0$, while for
$k\ll k_G$ it is governed by the nontrivial flat-phase fixed point with
$\eta_k\simeq \eta$.

At a temperature $T=500$K one finds  $k_G = 0.09\mathrm{\mathring{A}}^{-1}$, as defined in Eq.({\ref{defetag}}),  and $t_G=1.50$,  that lies between $t_{1G}$ and $t_{2G}$ as defined in Eq.({\ref{qG12}}), see Fig.{\ref{etat500}. This turns out to be true at all temperatures. From now on, we will consider  $t_G$ as the  harmonic-to-anharmonic crossover RG-time scale.

\begin{figure}[h!]
\centering 
\includegraphics[width=0.5\textwidth, trim=10mm 88mm 1mm 45mm, clip]{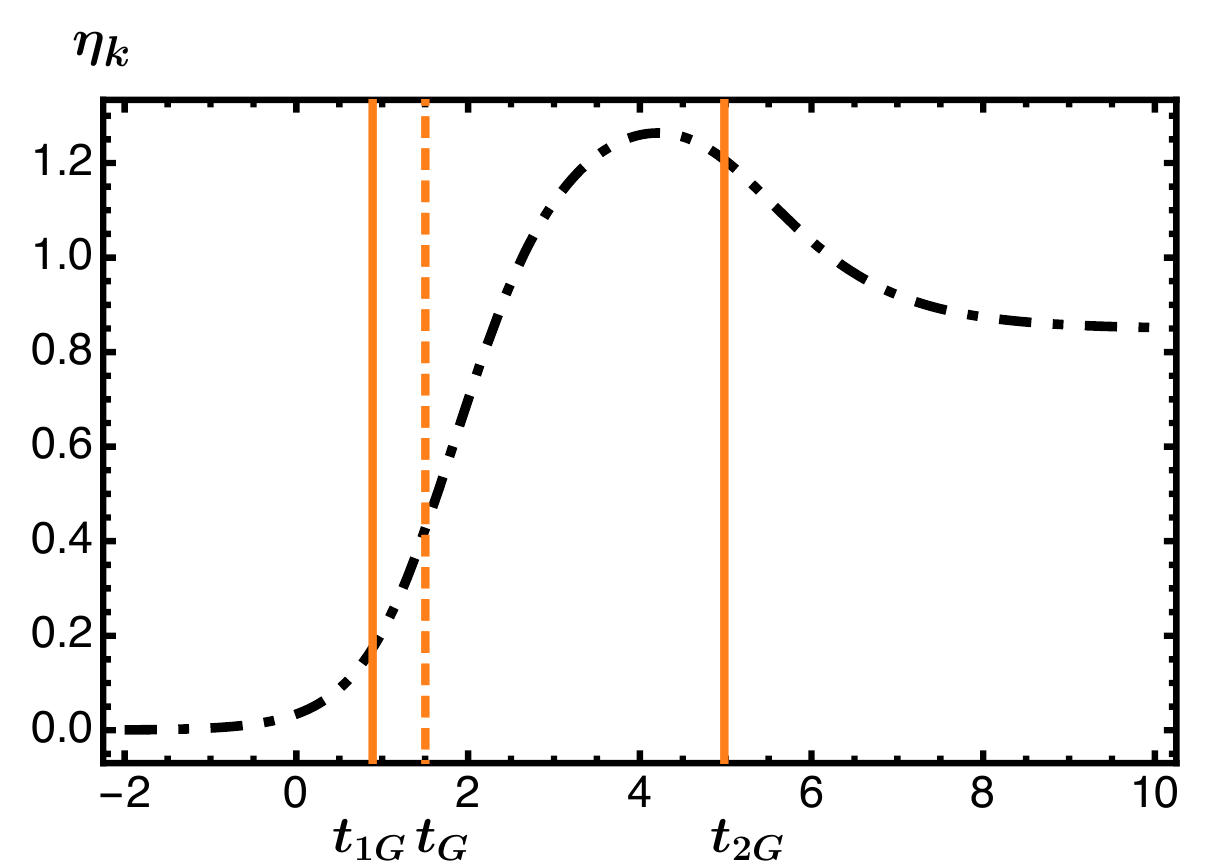}
\caption{Behavior of $\eta_k$ -- dashed curve --  as a function of $t=-\ln ka$ for $T=500$ K;  One gives also the typical Ginzburg RG-time scale $t_G$ (dashed vertical line) and the bare Ginzburg  RG-time scales   $t_{1G}$ and $t_{2G}$ (full vertical lines). }
\label{etat500}
\end{figure} 

\subsection{RG analysis}

We now present our results, which concern the RG evolution of both  the running anomalous dimension $\eta_k$, given by Eq.~(\ref{etabil}) and the running Young modulus $\bar Y_k$, given by Eq.~(\ref{FlotY}), for various temperatures.
We have considered $T = 10\,\text{K}$, $T = 100\,\text{K}$ and $T = 1000\,\text{K}$. In each case,
we indicate the Ginzburg scale $t_G$ by a vertical dotted line and the mechanical crossover scale
$t_c$ by a vertical dot-dashed line.

As in Ref.~\cite{mauri21bis}, it is instructive to compare the flows for bilayers with those for
monolayers, in order to disentangle the rigidity crossover specific to bilayers from the
harmonic-to-anharmonic crossover that occurs in both systems.

\medskip 

As a preliminary remark,  we find that the harmonic-anharmonic crossover scale $t_G$ and the
mechanical crossover scale $t_c$ shift with temperature in a strongly
correlated way, so that their separation along the RG flow remains essentially
constant. This is naturally understood within the RG picture: both $t_G$ and
$t_c$ are defined by fixed conditions on functions of the same running
dimensionless couplings (e.g. $\eta_k$ and the ratio
$\ell^2(\lambda_k+2\mu_k)/(4\kappa_k)$), along a unique renormalized trajectory
in coupling space. Changing the microscopic parameters (such as $T$) mainly
shifts the entry point on this trajectory, which translates all characteristic
scales in the RG-time $t$ by approximately the same amount. As a
result, the difference $\Delta t = t_G-t_c$, and hence the relative
position of the two crossovers, is nearly temperature independent.

\subsubsection{Behavior of the anomalous dimension $\eta_k^B$}

In Fig.\ref{Eta10t}-\ref{Eta1000t} we have represented the anomalous dimensions  of a bilayer $\eta_k^B$ and that of a monolayer $\eta_k^m$ at the temperatures $T=10\text{K}$, $T=100\text{K}$ and $T=1000\text{K}$. First, one observes that at large times, the RG flow  of  the  anomalous dimension   of a bilayer converges towards that of a monolayer that, itself, reaches  the value taken at the  IR fixed point of the flat phase:  $\eta\simeq 0.85$. Second, one sees that  harmonic-to-anharmonic crossover takes place long before the mechanical one, and that, at all temperatures. Third,  one observes that the two crossovers occur at earlier and earlier RG-times when the temperature increases, as expected.

\begin{figure}[h!]
\centering
\includegraphics[width=0.5\textwidth, trim=20mm 88mm 1mm 45mm, clip]{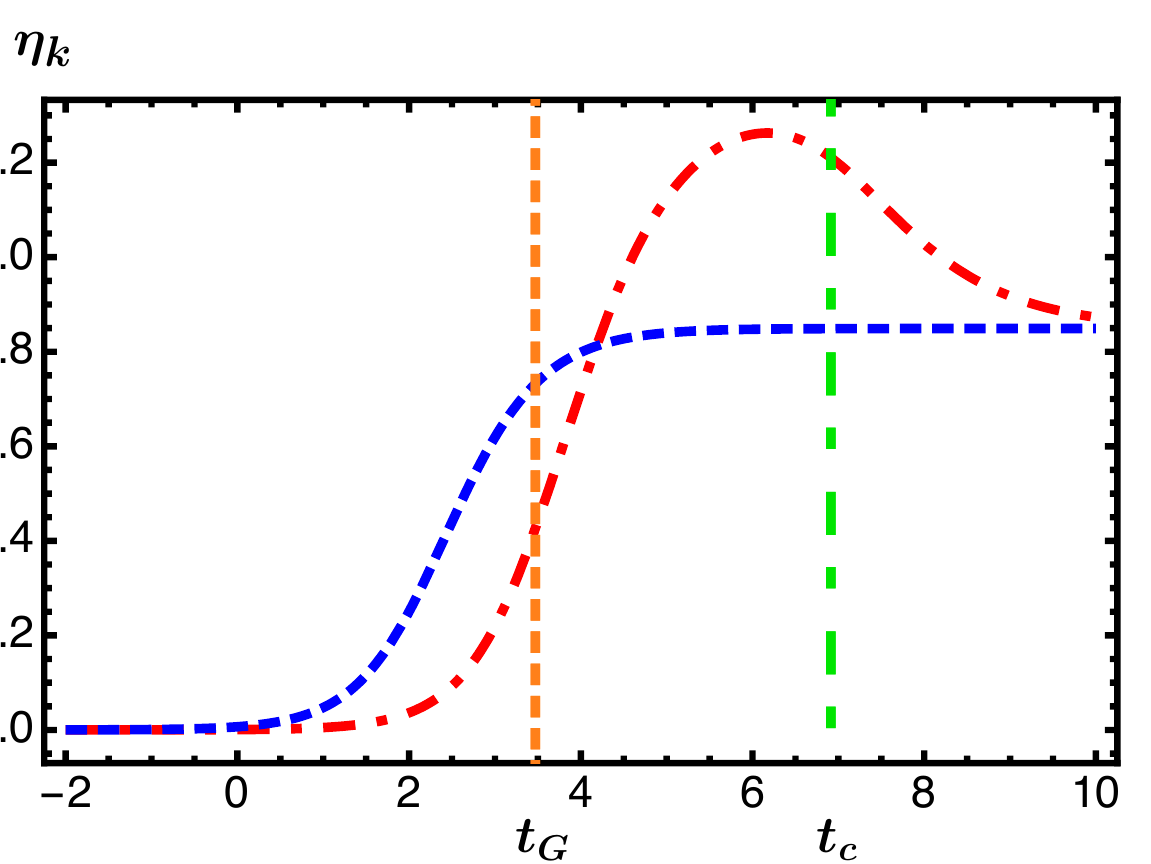}
\caption{The flow of the anomalous dimension at $T=10\text{K}$. Dashed-dot curve :  $\eta_k^B$ of a bilayer; dashed curve :  $\eta_k^m$ of a  monolayer; dotted vertical lines  :  the Ginzburg RG-time $t_{G}$; dashed-dot  vertical line : the  mechanical crossover RG-time  $t_c$.}
\label{Eta10t}
\end{figure} 

\begin{figure}[h!]
\centering
\includegraphics[width=0.5\textwidth, trim=20mm 88mm 1mm 45mm, clip]{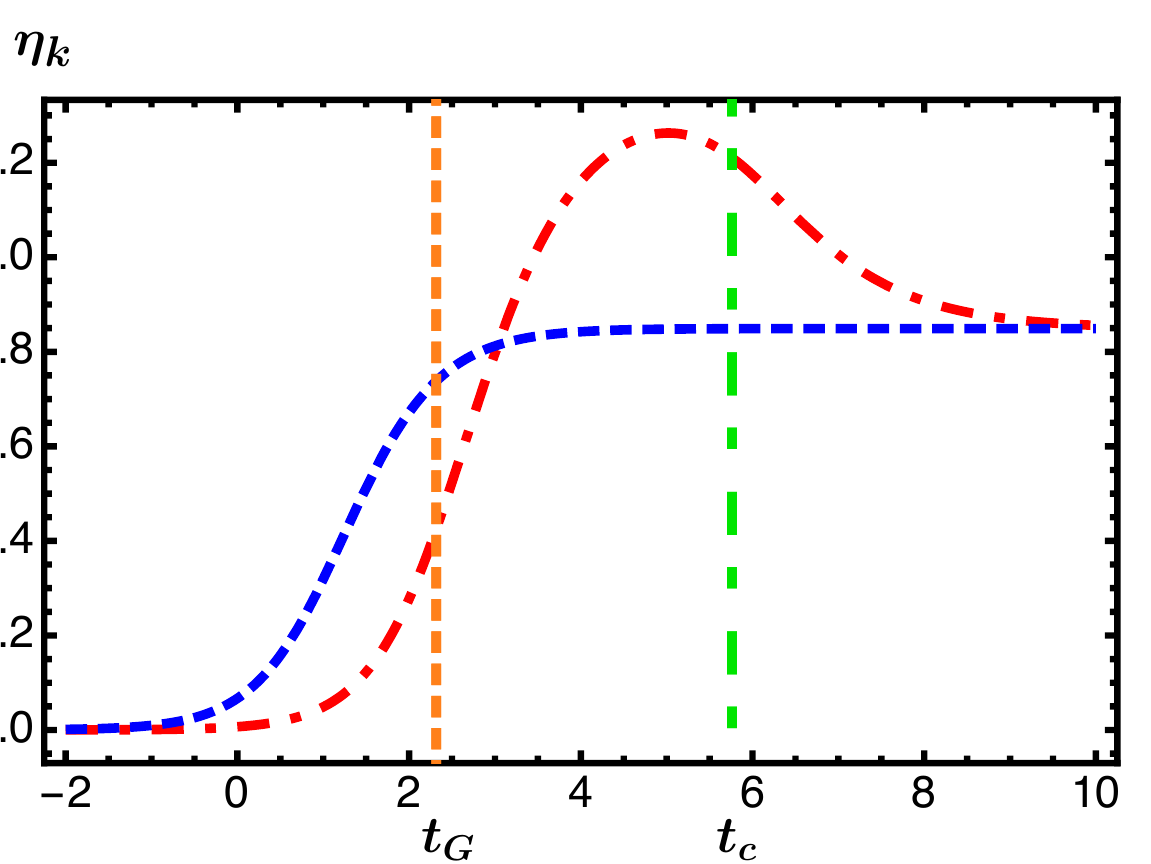}
\caption{The flow of the anomalous dimension at $T=100\text{K}$.  Dashed-dot curve :  $\eta_k^B$ of a bilayer; dashed curve :  $\eta_k^m$ of a  monolayer; dotted vertical lines  :  the Ginzburg RG-time $t_{G}$; dashed-dot  vertical line : the  mechanical crossover RG-time $t_c$.}
\label{Eta100t}
\end{figure} 

\begin{figure}[h!]
\centering
\includegraphics[width=0.5\textwidth, trim=20mm 88mm 1mm 45mm, clip]{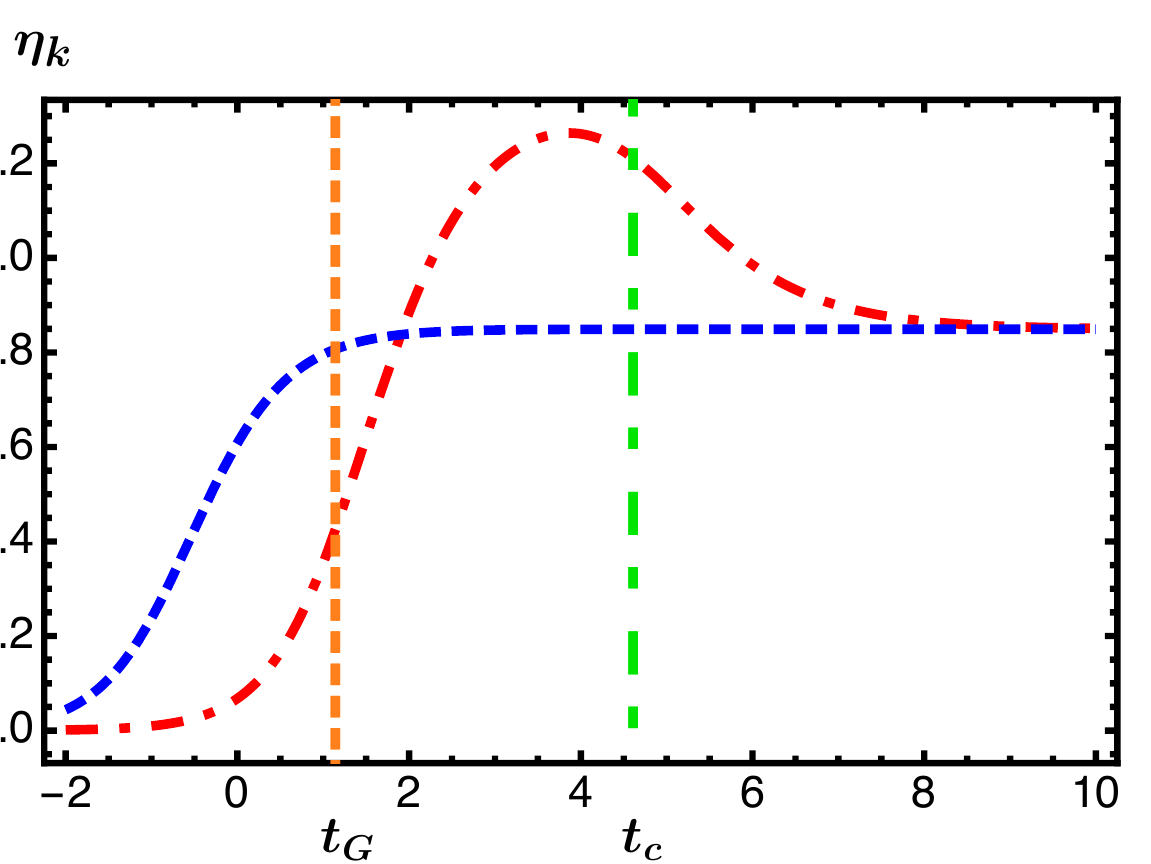}
\caption{The flow of the anomalous dimension at $T=1000\text{K}$.  Dashed-dot curve :  $\eta_k^B$ of a bilayer; dashed curve :  $\eta_k^m$ of a  monolayer; dotted vertical lines  :  the Ginzburg RG-time $t_{G}$; dashed-dot  vertical line : the  mechanical crossover RG-time $t_c$.}
\label{Eta1000t}
\end{figure} 

Then, comparing the behavior of  $\eta_k^B$ and  $\eta_k^m$ one also observes  that the former  stays longer in the  harmonic  regime ($\eta_k=0$)  than the latter one. This relies on  the very structure of $\eta_B^k$ which is such that, at the beginning of the flow, {\it i.e.} at  large $\bar c_k$,  one has
\begin{equation}
\eta_k^B \underset{\bar c_k \gg 1}=
\frac{3}{64}\,\frac{\bar Y_k}{\sqrt{\bar c_k}}
+ {\cal{O}}\!\left(\frac{1}{\bar c_k}\right)
\end{equation}
and is thus  small. 
This effect is obviously all the more pronounced at low temperature as fluctuations are small.
Then one  observes that $\eta_k^B$  varies in a relatively moderate way with $\bar c_k$ with only a slight peak -- of order 50$\%$ with respect to the monolayer case -- in the region bounded by the harmonic-to-anharmonic crossover and the mechanical crossover. 
 Expanding $\eta_k^B$ at  small $\bar c_k$  one gets: 
\begin{equation}
\eta_k^B \underset{\bar c_k \ll 1}=
\frac{6 \bar Y_k}{\bar Y_k+32 \pi }+{\cal{O}}\left(\bar c_k\right)\ . 
\end{equation}
 The dependence on  $\bar c_k$ is therefore smooth and only appears at linear order in $\bar c_k$ near $\bar c_k=0$:   when going from $\bar c_k$ to a moderately nonzero $\bar c_k$, $\eta_k^B$   changes only moderately. 

\subsubsection{Behavior of the Young modulus  $\bar Y_k^B$}

In Figs.~\ref{Y10t}-\ref{Y1000t}, we show the running Young modulus
$\bar Y_k^{B}$ of a bilayer together with the quantity $2\,\bar Y_k^{m}$ corresponding to two
uncoupled monolayers, for temperatures $T = 10\,\text{K}$, $T = 100\,\text{K}$ and
$T = 1000\,\text{K}$. Once again, we observe that at large RG-times the flow of the bilayer Young
modulus $\bar Y_k^{B}$ converges towards $2\,\bar Y_k^{m}$, which itself approaches the value at
the infrared fixed point of the flat phase, characterized by
$\bar\mu_*^{m} \simeq 6.21$, $\bar\lambda_*^{m} \simeq -3.10$, and hence
$2\,\bar Y_*^{m} \simeq 16.57$.

\begin{figure}[h!]
\centering
\includegraphics[width=0.5\textwidth, trim=20mm 88mm 1mm 45mm, clip]{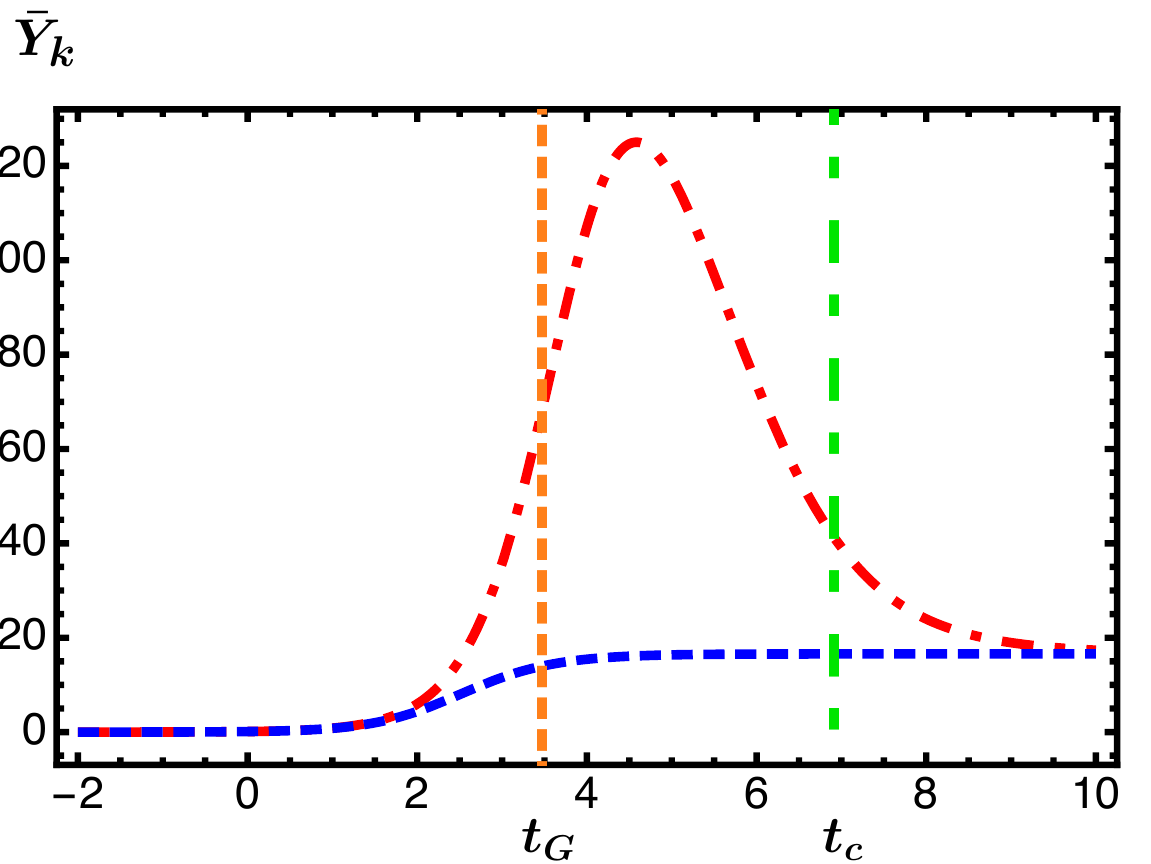}
\caption{The flow of the Young modulus at $T=10\text{K}$.  Dashed-dot curve :  $\bar Y_k^B$ of a bilayer; dashed curve :  $\bar Y_k^m$ of a  monolayer; dotted vertical lines  :  the Ginzburg RG-time $t_{G}$; dashed-dot  vertical line : the  mechanical crossover RG-time $t_c$.}
\label{Y10t}
\end{figure} 
\begin{figure}[h!]
\centering
\includegraphics[width=0.5\textwidth, trim=20mm 88mm 1mm 45mm, clip]{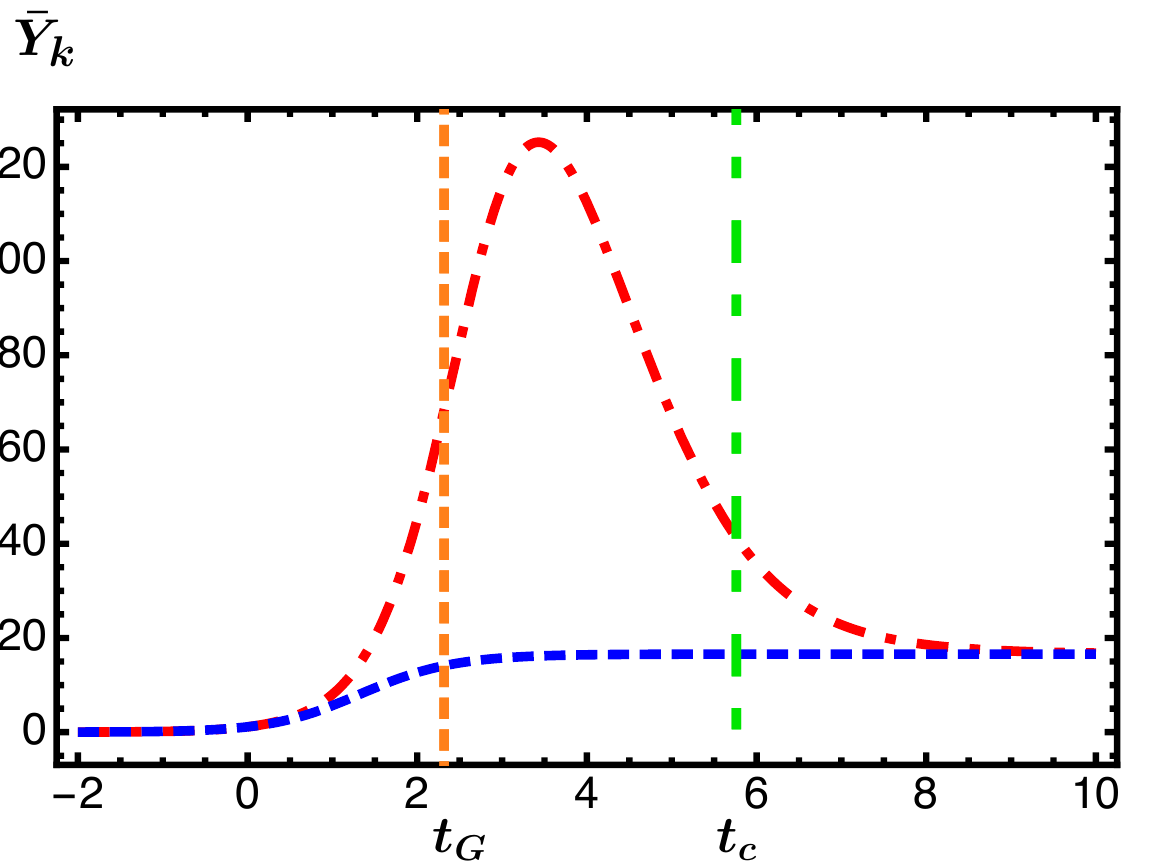}
\caption{The flow of the Young modulus at $T=100\text{K}$.  Dashed-dot curve :  $\bar Y_k^B$ of a bilayer; dashed curve :  $\bar Y_k^m$ of a  monolayer; dotted vertical lines  :  the Ginzburg RG-time $t_{G}$; dashed-dot  vertical line : the  mechanical crossover RG-time $t_c$.}
\label{Y100t}
\end{figure} 
\begin{figure}[h!]
\centering
\includegraphics[width=0.5\textwidth, trim=20mm 88mm 1mm 45mm, clip]{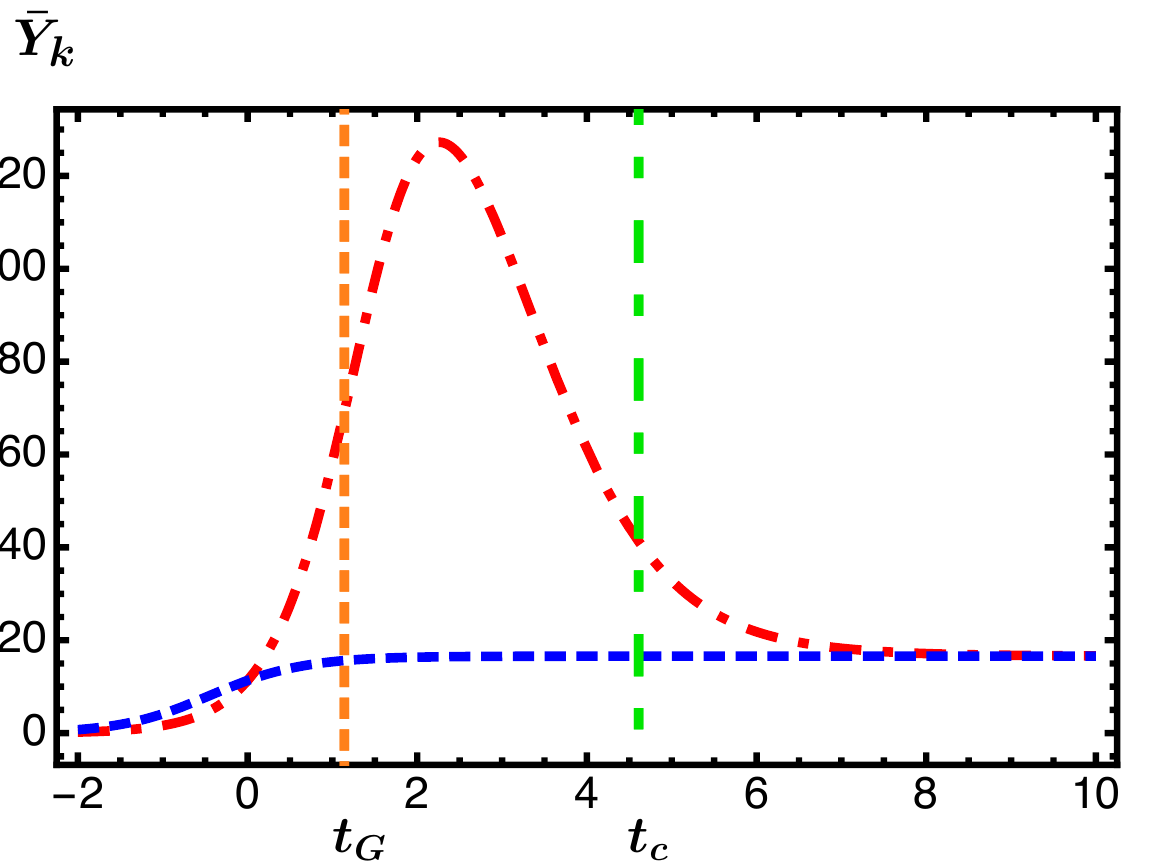}
\caption{The flow of the Young modulus at $T=1000\text{K}$. Dashed-dot curve :  $\bar Y_k^B$ of a bilayer; dashed curve :  $\bar Y_k^m$ of a  monolayer; dotted vertical lines  :  the Ginzburg RG-time $t_{G}$; dashed-dot  vertical line : the  mechanical crossover RG-time  $t_c$.}
\label{Y1000t}
\end{figure} 

As for the anomalous dimension, the harmonic-to-anharmonic crossover takes place well before the
mechanical crossover, and both crossover scales are shifted to earlier RG-times as the temperature
 $T$ is increased.

In this case, we see that $\bar Y_k^{B}$ exhibits a behavior that differs drastically from that of a
monolayer: it displays a very pronounced peak of order $650\%$ with respect to the monolayer
value  in the window between the harmonic-to-anharmonic crossover and the mechanical crossover.
This phenomenon is again straightforward to analyze. To do so, let us consider the RG flow equation
for $\bar Y_k^{B}$
\begin{equation}
\partial_t \bar Y_k
= 2(\eta_k - 1)\,\bar Y_k
- \frac{3}{4}\,H_k\,\bar Y_k^{2},
\label{flotY2}
\end{equation}
and examine how the interplay between the running couplings $\lambda_k$, $\mu_k$ and the scale
dependence of $\kappa_{\mathrm{eff},k}$ amplifies the transient growth of the bilayer Young modulus
before it eventually relaxes towards its fixed-point value.

In Eq.~(\ref{flotY2}) the  dependence with respect to $\bar c_k$ appears both in $\eta_k$ and $H_k$. For large $\bar c_k$, $\eta_k$ is, as seen above, only slightly modified. However the behavior of $H_k$  is given by 
\begin{equation}
H_k\underset{\bar c_k \gg 1}= -\frac{1}{64\, c_k^{3/2}} +{\cal{O}}\Big(\frac{1}{\bar c_k^2}\Big)\ . 
\end{equation}
Thus, for an initial value of $\bar c_k$ relatively large, $H_k$ is very small so that the nonlinear term in the flow of $\bar Y_k$,  Eq.~(\ref{flotY2}), is strongly suppressed at the beginning of the flow  which is quasi-linear: 
\begin{equation}
\partial_t \bar Y_k \simeq  2(\eta_k - 1)\,\bar Y_k \ . 
\end{equation}
On the other hand for $\bar c_k\ll 1$, $H_k$  behaves as:
\begin{equation}
H_k\underset{\bar c_k \ll 1}= -\frac{\bar Y_k+80\pi}{30 \pi(\bar Y_k+32\pi)}+ {\cal{O}}\left({\bar c_k}\right)
\end{equation}
and is thus of order unity, the nonlinear term in Eq.~(\ref{flotY2}) becomes important as soon as $\bar Y_k$ becomes large.  From these two behaviors one  can thus expect a rapid change (a jump) of $\bar Y_k$ in a transient phase.

\section{Conclusion}

In this work, we have studied elastic bilayers  within a nonperturbative renormalization group framework, starting from the simplest possible action written in terms of the mean field ${\bf R}$  and the relative field ${\bf S}$.  We have shown that this description reproduces in a controlled way the crossover of the effective bending rigidity previously identified within a self-consistent approach: at short distances the rigidity is dominated by the combination $\ell^2(\lambda+2\mu)/2$, while at large distances it tends to 2$\kappa$. The key difference is that, within the RG formalism, the elastic nonlinearities neglected in the self-consistent treatment are now explicitly included, providing a more coherent  picture of the crossover regime. More fundamentally, the formalism used here, unlike the self-consistent method, in no way requires nonlinearities to be neglected at the structural level. On the contrary, it is possible to consider an action as general as possible and to treat all the associated fluctuations. We also recall the rotationally invariant form of the action used, which stands in clear contrast to those employed in the perturbative framework.

To return to the explicit results, from a formal point of view, the flow equations obtained for the bilayer have exactly the same structure as those of a single polymerized membrane but with a propagator for the flexuron field modified by an additional term.  This shows that, in our formalism,  going from a monolayer to a bilayer does not require any new conceptual machinery: it simply amounts to applying the same flow equations to an enlarged set of fields and combinations of  fields. 

These remarks naturally  open several directions for future work:

(i) As a first step, one can relax the assumption of a common field renormalization for both fields {\bf R} and {\bf S} and, more generally, of identical couplings for different kinds of monomials  allowed by symmetry and entering in the running Gibbs free energy Eq.(\ref{bilayeraction}). In particular, it would make it possible to follow separately the renormalization of the symmetric and antisymmetric sectors and to test the robustness of the crossover scenario when the two modes acquire different anomalous dimensions. In the same spirit, the action can be enriched by terms involving higher powers of the fields {\bf R} and {\bf S}. Although simple power-counting and RG relevance arguments suggest that such contributions should remain subdominant for large-scale elasticity, including them explicitly would provide an a posteriori check of this expectation and a test of the stability of the critical exponents and crossover lengths.

(ii)  Another natural extension is to treat wavevector-dependent couplings, that is, to work with a genuinely ${\bf q}$-dependent ansatz for the effective Lam\'e coefficients and bending rigidities of the {\bf R} and {\bf S} modes. This would give a more refined description of the crossover region between the regimes dominated by projected elasticity and by the renormalized bending rigidity. 

(iii)  One can then include explicitly the dependence on $g_2$  and, more generally,  on various interlayer couplings like  $g_1$ and $g_3$ in the flow equations. This would allow a more quantitative characterization of how interlayer couplings  feeds into the effective infrared bending rigidity of the bilayer.

(iv) The present formalism can also be generalized to asymmetric bilayers, with different elastic and curvature properties for each sheet. Allowing for different bare rigidities and Lam\'e coefficients on the two layers would bring the model closer to realistic systems where the two membranes are not mechanically identical. Finally, one can introduce spontaneous curvatures (possibly distinct for each layer or for specific modes) and investigate their impact on the flow of couplings and on the phase structure, in particular on the locking of the relative mode and on the average geometry of the bilayer.

From a phenomenological perspective, it is natural to view the
short-scale regime $\kappa_{\mathrm{eff}}\simeq \ell^2(\lambda+2\mu)/2$ and
the long-scale regime $\kappa_{\mathrm{eff}}\simeq 2\kappa$ as the
continuum counterparts of the distinct behaviors reported in recent
experiments and simulations on few-layer graphene. The large apparent
bending rigidity measured for bilayers in buckling and snap-through
experiments \cite{lindahl12} the pronounced softening and change of
scaling under strong bending in trilayers and thicker stacks because of interlayer slip \cite{Han20} and the observed size- and
temperature-dependent effective stiffness in suspended membranes all point
to a complex crossover structure \cite{Sajadi18,Eder21,Jiang23}. In addition,
atomistic Monte Carlo and molecular-dynamics simulations of bilayer and
few-layer graphene also reveal a thickness-dependent flexural response and
a crossover from correlated to uncorrelated layer fluctuations, further
highlighting the role of interlayer coupling in the long-wavelength
mechanical behavior \cite{zakharchenko10,herrero23}.  Within our NPRG
framework, this complexity is encoded in the flow of $\kappa_{\mathrm{eff}}(k)$
and the associated crossover scale $k_c$, which can be directly compared to
the wave-vector windows effectively probed in these measurements and
numerical studies.

The present work provides a first coherent nonperturbative RG framework for elastic bilayers,
simple enough to remain tractable, yet already rich enough to support a broad range of systematic
extensions able to tackle a wide variety of experimental and numerical situations.

\acknowledgements

L.D. and D.M.  greatly thank A. Mauri for illuminating discussions. 

\bigskip 

\appendix
\setcounter{section}{0}
\renewcommand{\thesection}{\Alph{section}}

\section{Running anomalous dimension}
\label{app: Running anomalous dimension}

We  give here the expression of the running anomalous dimensions in the monolayer case in $D$ dimensions. One has: 
\begin{widetext} 
\begin{equation}
\begin{aligned}
        \eta_k=&\frac{A_D}{\bar\zeta_k
        ^4 \bar\mu_k  (\bar\lambda_k +2 \bar\mu_k )} \left(-2 \bar\zeta_k^4 \bar\mu_k \left(\bar\lambda_k ^2+\bar\lambda_k  \bar\mu_k +\bar\mu_k ^2\right) \, \bar L^D_{100}+\bar\zeta_k ^8 \bar\lambda_k\, \bar\mu_k^4 \bar L^{D+4}_{030}+2\, \bar\zeta_k^8 \bar\mu_k^5 \bar L^{D+4}_{030} 
        +\bar\zeta_k^6 \bar\mu_k^3 (\bar\lambda_k +2 \bar\mu_k )\, \bar L^{D+2}_{020} \right.\\ 
        &\hspace{-0.5cm}+3\, \bar\zeta_k^8 \bar\lambda_k^4 \bar\mu_k\, \bar L^{D+4}_{003}+24\, \bar\zeta_k^8 \bar\lambda_k^3 \bar\mu_k^2 \bar L^{D+4}_{003}+72\, \bar\zeta_k^8 \bar\lambda_k^2 \bar\mu_k^3 \bar L^{D+4}_{003}+96\, \bar\zeta_k^8 \bar\lambda_k\, \bar\mu_k^4 \bar L^{D+4}_{003}+48\, \bar\zeta_k^8 \bar\mu_k^5 \bar L^{D+4}_{003}-\bar\zeta_k^6 \bar\lambda_k^3 \bar\mu_k\, \bar L^{D+2}_{002} \\
        &\hspace{-0.5cm} -8 \bar\zeta_k^6 \bar\lambda_k^2 \bar\mu_k^2 \bar L^{D+2}_{002}-20\, \bar\zeta_k^6 \bar\lambda_k\,  \bar\mu_k^3 \bar L^{D+2}_{002}-16\,\bar\zeta_k^6 \bar\mu_k^4 \bar L^{D+2}_{002}+2\, \bar\zeta_k^4 \bar\mu_k \left(\bar\lambda_k^2+\bar\lambda_k\,\bar\mu_k +\bar\mu_k^2\right) \bar L^{D}_{001}+\bar\zeta_k^2 \bar\lambda_k\, \bar\mu_k \, \bar M^{D+2}_{020}\\
        &\hspace{-0.5cm} +2\, \bar\zeta_k^2 \bar\mu_k^2 \bar M^{D+2}_{020}+2\, \bar\lambda_k\, \bar M^{D}_{010}+4\, \bar\mu_k\, \bar M^{D}_{010}+\bar\zeta_k^6 \bar\lambda_k\, \bar\mu_k^3 \bar N^{D+4}_{030}+2\, \bar\zeta_k^6 \bar\mu_k^4 \bar N^{D+4}_{030}+3\, \bar\zeta_k^4 \bar\lambda_k\,\bar\mu_k^2 \bar N^{D+2}_{020}+6\, \bar\zeta_k^4 \bar\mu_k^3 \bar N^{D+2}_{020}\\
        &\hspace{-0.5cm}  +5\, \bar\zeta_k^2 \bar\lambda_k\, \bar\mu_k\, \bar N^{D}_{010}+10\, \bar\zeta_k^2 \bar\mu_k^2 \bar N^{D}_{010}+3\,\bar\zeta_k^2 \bar\lambda_k\, \bar\mu_k\, \bar M^{D+2}_{002}+6\, \bar\zeta_k^2 \bar\mu_k^2 \bar M^{D+2}_{002}+6\,\bar\mu_k\, \bar M^{D}_{001}+3\, \bar\zeta_k^6 \bar\mu_k\, (\bar\lambda_k +2\, \bar\mu_k)^3 \bar N^{D+4}_{003}\\
        &\hspace{-0.5cm} +5\, \bar\zeta_k^4 \bar\lambda_k^2 \bar\mu_k\, \bar N^{D+2}_{002} +18\, \bar\zeta_k^4 \bar\lambda_k\, \bar\mu_k^2 \bar N^{D+2}_{002}+16\,\bar\zeta_k^4 \bar\mu_k^3 \bar N^{D+2}_{002}+7\,\bar\zeta_k^2 \bar\lambda_k\,\bar\mu_k\, \bar N^{D}_{001}+10\, \bar\zeta_k^2 \bar\mu_k^2 \bar N^{D}_{001} +4\, \bar\zeta_k^2 \bar\lambda_k\,\bar\mu_k\, \bar M^{D+2}_{200}\\
        &\hspace{-0.5cm}  +8\, \bar\zeta_k^2 \bar\mu_k^2 \bar M^{D+2}_{200}-2\, \bar\lambda_k\, \bar M^{D}_{100}-10\, \bar\mu_k\, \bar M^{D}_{100}+2\, \bar\zeta_k^4 \bar\lambda_k^2 \bar\mu_k\,  \bar N^{D+2}_{200}  +8\,\bar\zeta_k^4 \bar\lambda_k\,  \bar\mu_k^2 \bar N^{D+2}_{200} +8\, \bar\zeta_k^4 \bar\mu_k^3 \bar N^{D+2}_{200}-12\, \bar\zeta_k^2 \bar\lambda_k\,  \bar\mu_k\,  \bar N^{D}_{100} \\
        &\left. -20\, \bar\zeta_k^2 \bar\mu_k^2 \bar N^{D}_{100}\right)
\label{eta_thres_mon}
\end{aligned}
\end{equation}
\end{widetext}
where the threshold functions $\bar L_{abc}^{D+\alpha}$, $\bar N_{abc}^{D+\alpha}$ and $\bar M_{abc}^{D+\alpha}$ are given in the following appendix. 

\section{Threshold functions}
\label{app: Threshold functions}

Here we provide  the expressions of the dimensionful and dimensionless threshold functions in $D$ dimensions, the latter ones being used in practical computations to identify fixed points.

\newpage

\subsection{The monolayer case}
\label{app:The monolayer case}

\subsubsection{Definitions}

The thresholds functions involving three propagators thus with three pseudo-masses $m_0$, $m_1$ and $m_2$ are defined as:
\begin{widetext}
\begin{equation}
\left\{
\begin{aligned}
    &L_{abc}^{D+\alpha}=-\frac{1}{4\,A_D}\, \widehat{\partial_t}\int_q \boldsymbol{q}^\a \, \left(P(\boldsymbol{q})+\boldsymbol{q}^2\,m_0^2\right)^{-a}\,\left(P(\boldsymbol{q})+\boldsymbol{q}^2\,m_1^2\right)^{-b}\,\left(P(\boldsymbol{q})+\boldsymbol{q}^2\,m_2^2\right)^{-c}\\
    &N_{abc}^{D+\a}=-\frac{1}{4\,A_D}\,\widehat{\partial_t}\int_q \, \boldsymbol{q}^\a\,\frac{\partial P(\boldsymbol{q})}{\partial\boldsymbol{q}^2} \, \left(P(\boldsymbol{q})+\boldsymbol{q}^2\,m_0^2\right)^{-a}\,\left(P(\boldsymbol{q})+\boldsymbol{q}^2\,m_1^2\right)^{-b}\,\left(P(\boldsymbol{q})+\boldsymbol{q}^2\,m_2^2\right)^{-c}\\
    &M_{abc}^{D+\a}=-\frac{1}{4\,A_D}\, \widehat{\partial_t}\int_q \, \boldsymbol{q}^\a\,\left(\frac{\partial P(\boldsymbol{q})}{\partial\boldsymbol{q}^2}\right)^2 \, \left(P(\boldsymbol{q})+\boldsymbol{q}^2\,m_0^2\right)^{-a}\,\left(P(\boldsymbol{q})+\boldsymbol{q}^2\,m_1^2\right)^{-b}\,\left(P(\boldsymbol{q})+\boldsymbol{q}^2\,m_2^2\right)^{-c}
\end{aligned}
\right.
\end{equation}
\end{widetext}
with $A_D=1/2^{D+1}\pi^{D/2}\Gamma(D/2)$, $P(\boldsymbol{q})=Z_k\,\boldsymbol{q}^4+{\cal R}_k(\boldsymbol{q})$, $\widehat{\partial_t}=\partial_t {\cal R}_k(\boldsymbol{q}) \partial_{{\cal R}_k}$. 

Below we give the threshold functions $L_{a00}^{D+\a}$, $N_{a00}^{D+\a}$ and $M_{a00}^{D+\a}$  which  are relevant to study the flat phase which is governed by massless flexurons modes. We thus put $m_0=0$. One has:
\begin{equation}
\left\{
\begin{aligned}
 &L_{a00}^{D+\a}=-\frac{1}{4\,A_D}\, \widehat{\partial_t}\int_q \boldsymbol{q}^\a \, P(\boldsymbol{q})^{-a} \\
    \\
& N_{a00}^{D+\a}=-\frac{1}{4\,A_D}\, \widehat{\partial_t}\int_q \, \boldsymbol{q}^\a\,\frac{\partial P(\boldsymbol{q})}{\partial \boldsymbol{q}^2} \, P(\boldsymbol{q})^{-a}\\
\\
&M_{a00}^{D+\a}=-\frac{1}{4\,A_D}\, \widehat{\partial_t}\int_q\, \boldsymbol{q}^\a\, \left(\frac{\partial P(\boldsymbol{q})}{\partial \boldsymbol{q}^2}\right)^2 \, P(\boldsymbol{q})^{-a} 
\end{aligned}
\right.
\end{equation}
with $\displaystyle \int_q=\int {\text{d}^D q\over (2\pi)^D}\ := \int \dbar^D q$ 
and $\displaystyle \dbar^Dq=4\,A_D\,q^{D-1}\text{d}q$. 

\medskip

We now set $x=\boldsymbol{q}^2$ so that  $\text{d}^Dq=2\,A_D\,x^{(D-2)/2}\text{d}x$ and  apply $\widehat{\partial}_t= \partial_t {\cal R}_k(\boldsymbol{q}) \partial_{{\cal R}_k}$. One gets: 
\begin{equation}
\left\{
\begin{aligned}
& L_{a00}^{D+\a}=\frac{1}{2}\int \text{d}x\,x^{{D+\a\over 2}-1} \frac{a\,\partial_t {\cal R}_k(x)} {(Z_k\,x^2+{\cal R}_k(x))^{a+1}}\\
&\hspace{-0.1cm}N_{a00}^{D+\a}=-\frac{1}{2}\int \text{d}x\,x^{{D+\a\over 2}-1} \left[\frac{\partial_t \partial_x P(x)}{(Z_k\,x^2+{\cal R}_k(x))^{a}}\right. \\
&\hspace{3.5cm} \left.-\,\frac{a\, \partial_x P(x)\, \partial_t {\cal R}_k(x)}{(Z_k\,x^2+{\cal R}_k(x))^{a+1}} \right] \\
\\
& M_{a00}^{D+\a}=-\frac{1}{2}\int \text{d}x\,x^{{D+\a\over 2}-1} \left[\frac{2\,\partial_x P(x)\, \partial_t \partial_x P(x)}{(Z_k\,x^2+{\cal R}_k(x))^{a}}\right.\\
&\hspace{3.5cm}  \left. -\frac{a\,(\partial_xP(x))^2\,\partial_t {\cal R}_k(x)}{(Z_k\,x^2+{\cal R}_k(x))^{a+1}} \right]\, .
\label{threholdx}
\end{aligned}
\right.
\end{equation}

\subsubsection{Dimensionless threshold functions}

We now express the thresholds functions  in terms dimensionless quantities. One writes: 
\begin{equation}
    {\cal R}_k({\bf q})[y]=Z_k\, k^4\, y^2\,r_k(y)~~~~~\text{with}~~~~~ y=\frac{x}{k^2}\, ,
\label{rk}
\end{equation}
where $r_k(y)$ is the dimensionless cut-off function and $y$ is a dimensionless variable. With this expression, one  can evaluate: 
\begin{equation}
    P(\boldsymbol{q})[y]=Z_k\,\boldsymbol{q}^4+{\cal R}_k(\boldsymbol{q})=Z_k\, k^4\ y^2\big(1+r_k(y)\big)
\end{equation}
and can compute all the quantities needed for the evaluations of the threshold functions Eq.~(\ref{threholdx}). One has: 

\begin{equation}
    \partial_t {\cal R}_k(x)[y] =-Z_k\, k^4\big[\eta_k\,y^2 r_k(y)+2 y^3\partial_y r_k(y)\big]\\
\end{equation}

 \begin{equation}
     \partial_x P(x)[y] =Z_k\, k^2 \big[2 y (1+r_k(y))+y^2 \partial_y r_k(y)\big]\\
\end{equation}
and 
\begin{equation}
\begin{aligned}
& \widehat\partial_t\partial_y P(x)[y] =\partial_y \widehat\partial_t P(x)[y]= \partial_y\partial_t {\cal R}_k(x)[y]\\ 
&  \hspace{1.3cm}=-Z_k k^2\big[2\eta_k y r_k(y) + y^2(\eta_k+6)\; \partial_yr_k(y)  \\
& \hspace{1.7cm}+2y^3\partial_y^2r_k(y)\big]\. 
\label{deriv-rk}
\end{aligned}
\end{equation}

This leads to the dimensionless threshold functions:

\begin{widetext}
\begin{equation}
\left\{
\begin{aligned}
    &\bar{L}_{a00}^{D+\a}=-\frac{a}{2}\int \text{d}y\,y^{{D+\alpha-2\over 2}-2a}\left[ \frac{\eta_k\,r_k(y)+2\,y\,\partial_y r_k(y)}{(1+r_k(y))^{a+1}} \right]
\\
\\
    &\bar N_{a00}^{D+\a}=\frac{1}{2}\int \text{d}y\,y^{{D+\alpha\over 2}-2a}\left[\frac{2\, \eta_k\,r_k(y)+y(\eta_k+6)\,\partial_y r_k(y)+2\,y^2\,\partial^2_y r_k(y)}{(1+r_k(y))^{a}}\right.\\
    &\hspace{1cm}\left.-a\big(2(1+\,r_k(y))+y\,\partial_y r_k(y)\big){\eta_k\,r_k(y)+2\,y\,\partial_y r_k(y)\over (1+r_k(y))^{a+1}}\right]\
\\
\\
   & \bar M_{a00}^{D+\a}=\frac{1}{2}\int \text{d}y\,y^{{D+\alpha+2\over 2}-2a}\big(2(1+\,r_k(y))+y\,\partial_y r_k(y)\big)\left[\frac{2\big(2\, \eta_k\,r_k(y)+y(\eta_k+6)\, \partial_y r_k(y)+2\,y^2\,\partial^2_y r_k(y)\big)}{(1+r_k(y))^a}\right.\\
    &\hspace{1cm}-\left.  a\big(2(1+\,r_k(y))+y\partial_y r_k(y)\big)\frac{\eta_k\,r_k(y)+2\,y\,\partial_y r_k(y)}{(1+r_k(y))^{a+1}}\right] 
\end{aligned}
\right.
\end{equation}
\end{widetext}
while  the dimensionful and dimensionless threshold functions are related by: 
\begin{equation}
\left\{
\begin{aligned}
&L_{abc}^{2+\alpha}=(Z_k k^4)^{-a-b-c} k^{D+\alpha}\bar L_{abc}^{2+\alpha}\\
&N_{abc}^{2+\alpha}=(Z_k k^4)^{-a-b-c} Z_k\;  k^{D+2+\alpha}\bar N_{abc}^{2+\alpha}\\
&M_{abc}^{2+\alpha}=(Z_k k^4)^{-a-b-c} Z_k^2\; k^{\alpha}\bar M_{abc}^{D+4+\alpha}
\end{aligned}
\right.
\end{equation}
--  without summation over $a$, $b$, $c$ and $\alpha$ indices. 

\subsubsection{Litim (or $\Theta$) cut-off}

\label{app:litim}

We have chosen to use the Litim cut-off as cut-off function ${\cal R}_k(\boldsymbol{q})$:
\begin{equation}
    {\cal R}_k(\boldsymbol{q})[y]=Z_k(k^4-|\boldsymbol{q}|^4)\Theta(k^2-|\boldsymbol{q}|^2)=Z_k k^4 r_k(y)
    \label{Litimdimless}
\end{equation}
with: $r_k(y)={{1-y^{-2}}\over y^2}\Theta(1-y^2)$. 

\medskip 

We compute the derivatives: 
\begin{equation}
    \left\{
\begin{aligned}
    &\partial_y r_k(y)=-2\,y^{-3}\Theta(1-y^2)\\
    \\
    &\partial^2_y r_k(y)=6\,y^{-4}\Theta(1-y^2)+4\,y^{-3}\delta(1-y^2)\ . 
    \end{aligned}
\right.
\end{equation}
We finally obtain the following form for the threshold functions:

\begin{equation}
\left\{
\begin{aligned}
    &\bar L_{a00}^{D+\a}={4 a\over D+\alpha} -\eta_k{4a\over (D+\a)(D+4+\a)} \\
    &\bar N_{a00}^{D+\a}=2-\eta_k{2\over D+2+\a}\\
    &\bar M_{a0 0}^{D+\a}=4\, .
\end{aligned}
\right.
\end{equation}
where one has taken $\Theta(0)=1/2$. 

\medskip

\subsection{The bilayer case}

\label{thresholdbilayers} 

\subsubsection{Definitions}

In the case of bilayer membranes in the flat phase the relevant threshold functions are:

\begin{equation}
\left\{
\begin{aligned}
    &{}_BL_{a00}^{D+\a}=-\frac{1}{4\,A_D}\, \widehat{\partial_t}\int_q \boldsymbol{q}^\a \,  P_{c_k}(\boldsymbol{q})^{-a}\\
    & {}_BN_{a00}^{D+\a}=-\frac{1}{4\,A_D}\, \widehat{\partial_t}\int_q \, \boldsymbol{q}^\a\,\frac{\partial P_{c_k}(\boldsymbol{q})}{\partial \boldsymbol{q}^2} \, P_{c_k}(\boldsymbol{q})^{-a} 
    \end{aligned} 
    \right.
\end{equation}
\vspace{0.5cm}

with $P_{c_k}({\bf q})=\displaystyle\left(Z_k + \ell_k^2 {{\l_k^2+2\mu_k^2}\over 4}\right){\bf q}^4+{\cal R}_k({\bf q})$.

\subsubsection{Dimensionless threshold functions}

Proceeding in the same way as in the monolayer case, one  obtains for the dimensionless threshold functions:

\begin{equation}
{}_{B}{}\bar{L}_{a00}^{D+\a}=-\frac{a}{2}\int \text{d}y\,y^{{{D+\a-2}\over 2}-2a}\left[ \frac{\eta_k\,r_k(y)+2\,y\, \partial_y r_k(y)}{(1+\bar c_k+r_k(y))^{a+1}}\right]\\
\end{equation}
\begin{equation}
\begin{aligned}
&{}_B\bar{N}_{a00}^{D+\a}=\frac{1}{2}\int \text{d}y\, y^{{{D+\a}\over 2}-2a} \\
 &\left[\frac{2\,\eta_k\,r_k(y)+y(\eta_k+6)\partial_yr_k(y)+2\,y^2\partial^2_yr_k(y)}{(1+\bar c_k+r_k(y))^{a}}\right.\\
&\left. \hspace{-0.2cm}  -  a\big(2(1+r_k(y)+\bar c_k)\,+y\,\partial_yr_k(y)\big)\frac{\eta_k\,r_k(y)+2\,y\,\partial_yr_k(y)}{(1+\bar c_k+r_k(y))^{a+1}}\right]
\end{aligned}
\end{equation}
with $\displaystyle \bar c_k=\bar \ell_k^2{\bar\l_k+2\bar\mu_k\over 4}$.

\subsubsection{Litim cut-off}

\label{app:litimbilayer}

Using then the Litim cut-off one gets: 
\begin{equation}
\left\{
\begin{aligned}
{}_{B}\bar{L}_{a00}^{D+\a} &=a(4-\eta_k)\ {{}_2F_1\left(1+a,\frac{D+\alpha}{4}, \frac{D+\alpha+4}{4},-\bar c_k\right)\over D+\a}\\
& + a \eta_k \frac{\, _2F_1\left(1+a, {{D+\alpha +4}\over 4};\frac{D+\alpha+8}{4},-\bar c_k\right)}{D+\alpha+4}
\\
{}_{B}\bar{N}_{a00}^{D+\a} &=  {2\over (1+\bar c_k)^a} \\
& \hspace{-0.4cm} - 2\eta_k{{}_2F_1\left(a,{D+\alpha+2\over4},{D+\alpha+6\over 4}, -\bar c_k\right)\over D+\alpha+2} \\
&\hspace{-0.4cm} -2a\bar c_k(\eta_k-4) {{}_2F_1\left(1+a,{D+\alpha+2\over4},{{D+\alpha+6}\over4},-\bar c_k\right)\over D+\alpha+2}\\
& \hspace{-0.4cm} +2a\bar c_k \ \eta_k {{}_2F_1\left(1+a,{D+\alpha+6\over4},{D+\alpha+10\over 4},-\bar c_k\right)\over D+\alpha+6} 
&\displaystyle 
\end{aligned}
\right.
\end{equation}

\section{Propagators of bilayer membranes}
\label{app: Propagators}

We  give here the expressions of the full propagators:

\begin{widetext}
\begin{equation}
\begin{aligned}
\begin{cases}
G_{k,1rr}(\boldsymbol{q}) = \dfrac{1}{2 (Z_k\, \boldsymbol{q}^4+ \boldsymbol{q}^2 \zeta_k^2 \mu_k+{\cal R}_k(\boldsymbol{q}))},\\
G_{k,2rr}(\boldsymbol{q}) = \dfrac{1}{2 (Z_k\, \boldsymbol{q}^4 + \boldsymbol{q}^2 \zeta_k^2 (\lambda+2\,\mu_k) +{\cal R}_k(\boldsymbol{q}))}\\
G_{k,1rs}(\boldsymbol{q}) = \dfrac{2}{Z_k\, \boldsymbol{q}^4 + \zeta_k^2\,(2\,g_{2k}/{\ell_k^2+\boldsymbol{q}^2\,\mu_k)})+ {\cal R}_k(\boldsymbol{q})},\\
G_{k,2rs}(\boldsymbol{q}) = \dfrac{4\,\ell_k^2}{2\,g_{2k}\,(\l_k^2\boldsymbol{q}^2+\zeta_k^2) +\ell_k^2\big(5\,Z_k\,\boldsymbol{q}^4+\boldsymbol{q}^2\,\zeta_k^2\,(\lambda_k+2\mu_k)+5\,{\cal R}_k(\boldsymbol{q}) \big)+\sqrt{\Delta(\boldsymbol{q})}},\\
G_{k,3rs}(\boldsymbol{q}) = \dfrac{4\,\ell_k^2}{2\,g_{2k}\,(\ell_k^2\boldsymbol{q}^2+\zeta_k^2) +\ell_k^2\big(5\,Z_k\,\boldsymbol{q}^4+\boldsymbol{q}^2\,\zeta_k^2\,(\lambda_k+2\mu_k)+5\,{\cal R}_k(\boldsymbol{q}) \big)-\sqrt{\Delta(\boldsymbol{q})}}\\
G_{k,ss}(\boldsymbol{q}) = \dfrac{2}{Z_k\,\boldsymbol{q}^4 + 2\,g_{1k}/\ell_k^2+{\cal R}_k(\boldsymbol{q})} 
\end{cases}
\end{aligned}
\end{equation}
with 
\begin{equation}
\begin{aligned}
    \Delta(\boldsymbol{q})&=4\,g_{2k}^2\,(\ell_k^2\boldsymbol{q}^2+\zeta_k^2)^2+4\,g_{2k}\,\ell_k^2\boldsymbol{q}^2\,(\ell_k^2\boldsymbol{q}^2-\zeta_k^2)\Big(3\,Z_k\,\boldsymbol{q}^2-\zeta_k^2\,(\l_k+2\mu_k)\Big)+\ell_k^4\boldsymbol{q}^4\Big(3\,Z_k\,\boldsymbol{q}^2 -\zeta_k^2\,(\l_k+2\mu_k)\Big)^2\\
    &\ +3\,\ell_k^2\,{\cal R}_k(\boldsymbol{q})\,\Big(4\,g_{2k}\,(\ell_k^2\boldsymbol{q}-\zeta_k^2)+2\,\ell_k^2\boldsymbol{q}^2\,(3\,Z_k\,\boldsymbol{q}^2-\zeta_k^2\,(\l_k+2\mu_k))+3\,\ell_k^2\,{\cal R}_k(\boldsymbol{q}) \Big)\, .
\end{aligned}
\end{equation}
\end{widetext}

\section{Low-wavevector expansion of the propagator $G_{k,3rs}({\bf q})$}
\label{app: G3rs}

One explains here the non-commutativity of the limits  $\boldsymbol{q}^2 \ell_k^2\ll \zeta_k^2$ and $g_{2k}\to 0$.  One considers  the propagator $G_{k,3rs}$ in the form
\begin{widetext}
\begin{equation}
G_{k,3rs}(\boldsymbol{q}) \;=\;
\frac{4\,\ell_k^2}{
2\,g_{2k}\,(\ell_k^2\boldsymbol{q}^2+\zeta_k^2)
+\ell_k^2\!\left(5\,Z_k\,\boldsymbol{q}^4+\boldsymbol{q}^2\,\zeta_k^2\,(\lambda_k+2\mu_k)+5\,{\cal R}_k(\boldsymbol{q}) \right)
-\sqrt{\Delta(\boldsymbol{q})}}\;,
\label{eq:Gkminus}
\end{equation}
\end{widetext} 
with $\Delta(\boldsymbol{q})$ the corresponding discriminant.
The non-commutativity arises because the low-wavevector expansion, $\boldsymbol{q}^2 \ell_k^2\ll \zeta_k^2$ is \emph{not uniform} in $g_{2k}$,  because of a cancellation between the explicit $\boldsymbol{q}^2\zeta_k^2(\lambda_k+2\mu_k)$ term and the leading small-$\boldsymbol{q}$ contribution of $\sqrt{\Delta}$ when $g_{2k}\to 0$.

\medskip
a. \noindent{\it Limit $g_{2k}\to  0$ first.}
\par\medskip

At $g_{2k}=0$, the square root  -- taking ${\cal R}_k({\bf q})=0$ --    admits the form: 
\begin{equation}
\sqrt{\Delta(\boldsymbol{q})}
=
\ell_k^2\,\boldsymbol{q}^2\,\zeta_k^2(\lambda_k+2\mu_k)
-3Z_k\,\ell_k^2\,\boldsymbol{q}^4\ . 
\label{eq:sqrtDelta_g20}
\end{equation}
Inserting this expression into Eq.\eqref{eq:Gkminus} the ${\cal O}(\boldsymbol{q}^2)$ terms cancel in the denominator of $G_{k,3rs}(\boldsymbol{q})$ \ which is  given by: 
\begin{equation}
\begin{array}{ll}
\mathrm{Den}(\boldsymbol{q}) \; & = 8Z_k\,\ell_k^2\,\boldsymbol{q}^4,
\end{array}
\end{equation}
and therefore
\begin{equation}
G_{k,3rs}(\boldsymbol{q})
\;\xrightarrow[g_{2k}\to 0]{}\;
\frac{4\ell_k^2}{8Z_k\,\ell_k^2\,\boldsymbol{q}^4}
=
\frac{1}{2Z_k\,\boldsymbol{q}^4}
\label{eq:G_g20}
\end{equation}
as expected.

\medskip
b. \noindent{\it Low-wavevector limit first, with fixed $g_{2k}>0$.}
\par\medskip

For fixed $g_{2k}>0$ and $\boldsymbol{q}^2 \ell_k^2\ll \zeta_k^2$, the discriminant is dominated by its $g_{2k}^2$ contribution and
\begin{equation}
\begin{array}{ll}
\sqrt{\Delta(\boldsymbol{q})}
& = 2g_{2k}\,(\ell_k^2\boldsymbol{q}^2+\zeta_k^2)
+\ell_k^2\,\boldsymbol{q}^2\,\zeta_k^2(\lambda_k+2\mu_k)\\
\\
& -\ell_k^2\,\boldsymbol{q}^4\bigl(3 Z_k+ 2\ell_k^2(\lambda_k+2\mu_k)\bigr)
+{\cal O}(\boldsymbol{q}^6),
\label{eq:sqrtDelta_lowq_g2pos}
\end{array}
\end{equation}
so that, in the denominator of~\eqref{eq:Gkminus}, both the ${\cal O}(g_{2k})$ terms and the
${\cal O}(\boldsymbol{q}^2)$ term cancel, leaving a leading ${\cal O}(\boldsymbol{q}^4)$ contribution. One obtains
\begin{equation}
\mathrm{Den}(\boldsymbol{q})
\;\sim\;
2\,\ell_k^2\,\boldsymbol{q}^4\bigl(4Z_k+\ell_k^2(\lambda_k+2\mu_k)\bigr)
\end{equation}
hence
\begin{equation}
\begin{array}{ll}
G_{k,3rs}(\boldsymbol{q})
& \;\sim\; \displaystyle \frac{4\ell_k^2}{2\,\ell_k^2\,\boldsymbol{q}^4\bigl(4Z_k+\ell_k^2(\lambda_k+2\mu_k)\bigr)}\\
\\
 &  =\displaystyle \frac{1}{2\bigg(Z_k+{\ell_k^2(\lambda_k+2\mu_k)\over 4}\bigg)\boldsymbol{q}^4}
\label{eq:G_lowE}
\end{array}
\end{equation}
that corresponds to Eq.(82).

\vspace{0.5cm} 

\section{Geometry of the bilayer: parallel surfaces}
\label{geometry}

We give here the expression for the induced metric on ($\Sigma_d$), the hypersurface obtained by translating a hypersurface ($\Sigma$) by a (signed) distance $d$ along its unit normal in $\mathbb{R}^3$.

We describe the mean surface of the bilayer as an embedding
\begin{equation}
\boldsymbol{r}(x^1,x^2) \in \mathbb{R}^3,
\end{equation}
with local coordinates $x^\alpha$ ($\alpha=1,2$).

\bigskip
a. \noindent{\it Tangent basis, metric, normal and curvature.}
\par\medskip

The tangent vectors are
\begin{equation}
\boldsymbol{e}_\alpha = \partial_\alpha \boldsymbol{r},
\end{equation}
and the induced (first fundamental) metric is
\begin{equation}
g_{\alpha\beta} = \boldsymbol{e}_\alpha \cdot \boldsymbol{e}_\beta.
\end{equation}

We choose a unit normal field $\boldsymbol{n}$ such that
\begin{equation}
\boldsymbol{n}\cdot\boldsymbol{e}_\alpha = 0,\qquad \boldsymbol{n}\cdot\boldsymbol{n}=1.
\end{equation}

The second fundamental form (curvature tensor) is defined as
\begin{equation}
K_{\alpha\beta} = \boldsymbol{n}\cdot\partial_\alpha\partial_\beta\boldsymbol{r}.
\end{equation}

It is convenient to introduce the mixed tensor (shape operator)
\begin{equation}
K_\alpha^{\ \gamma} = g^{\gamma\delta} K_{\alpha\delta},
\end{equation}
with $g^{\alpha\beta}$ the inverse of $g_{\alpha\beta}$. The Weingarten relation expresses the derivative of the normal in the tangent basis:
\begin{equation}
\partial_\alpha \boldsymbol{n}
= -\,K_\alpha^{\ \gamma}\,\boldsymbol{e}_\gamma.
\end{equation}

\medskip
b. \noindent{\it Parallel surfaces at signed distance $d$.}
\par\medskip

We now consider the surface obtained by shifting $\boldsymbol{r}$ a signed distance $d$ along the normal:
\begin{equation}
\boldsymbol{r}_d(x) = \boldsymbol{r}(x) + d\,\boldsymbol{n}(x).
\end{equation}

Its tangent vectors are
\begin{equation}
\partial_\alpha \boldsymbol{r}_d
= \partial_\alpha \boldsymbol{r} + d\,\partial_\alpha \boldsymbol{n}
= \boldsymbol{e}_\alpha + d\,\partial_\alpha \boldsymbol{n}.
\end{equation}

Using the Weingarten relation,
\begin{equation}
\partial_\alpha \boldsymbol{n}
= -\,K_\alpha^{\ \gamma}\,\boldsymbol{e}_\gamma,
\end{equation}
we find
\begin{equation}
\partial_\alpha \boldsymbol{r}_d
= \boldsymbol{e}_\alpha - d\,K_\alpha^{\ \gamma}\,\boldsymbol{e}_\gamma.
\end{equation}

The induced metric on the shifted surface is
\begin{equation}
g_{\alpha\beta}(d)
= \partial_\alpha \boldsymbol{r}_d \cdot \partial_\beta \boldsymbol{r}_d.
\end{equation}

Inserting the expression for $\partial_\alpha \boldsymbol{r}_d$,
\begin{equation}
\begin{aligned}
g_{\alpha\beta}(d)
&= (\boldsymbol{e}_\alpha - d\,K_\alpha^{\ \gamma}\boldsymbol{e}_\gamma)
   \cdot
   (\boldsymbol{e}_\beta - d\,K_\beta^{\ \delta}\boldsymbol{e}_\delta) \nonumber\\
& = \boldsymbol{e}_\alpha\cdot\boldsymbol{e}_\beta
   - d\,K_\alpha^{\ \gamma}(\boldsymbol{e}_\gamma\cdot\boldsymbol{e}_\beta)
   - d\,K_\beta^{\ \delta}(\boldsymbol{e}_\alpha\cdot\boldsymbol{e}_\delta) \\
& \hspace{0.3cm}  + d^2 K_\alpha^{\ \gamma}K_\beta^{\ \delta} (\boldsymbol{e}_\gamma\cdot\boldsymbol{e}_\delta). 
\end{aligned}
\end{equation}

Using $\boldsymbol{e}_\gamma\cdot\boldsymbol{e}_\beta = g_{\gamma\beta}$, this becomes
\begin{equation}
\begin{aligned}
g_{\alpha\beta}(d)
&= g_{\alpha\beta}
   - d\,K_\alpha^{\ \gamma} g_{\gamma\beta}
   - d\,K_\beta^{\ \delta} g_{\alpha\delta}
   + d^2 K_\alpha^{\ \gamma} K_\beta^{\ \delta} g_{\gamma\delta}. 
   \nonumber
\end{aligned}
\end{equation}

The linear terms simplify using
\begin{equation}
K_{\alpha\beta} = \boldsymbol{n}\cdot\partial_\alpha\partial_\beta\boldsymbol{r}
= g_{\beta\gamma} K_\alpha^{\ \gamma},
\end{equation}
and the symmetry $K_{\alpha\beta} = K_{\beta\alpha}$, which implies
\begin{equation}
K_\alpha^{\ \gamma} g_{\gamma\beta} = K_{\alpha\beta},
\qquad
K_\beta^{\ \delta} g_{\alpha\delta} = K_{\beta\alpha} = K_{\alpha\beta}.
\end{equation}

Hence the linear part in $d$ is $-2 d\,K_{\alpha\beta}$. The quadratic part can be written as
\begin{equation}
(K^2)_{\alpha\beta} = K_\alpha^{\ \gamma} b_{\gamma\beta}.
\end{equation}

Altogether we obtain the standard formula for the metric of a parallel surface:
\begin{equation}
g_{\alpha\beta}(d)
= g_{\alpha\beta} - 2d\,K_{\alpha\beta} + d^2 (K^2)_{\alpha\beta}.
\label{eq:metric_parallel}
\end{equation}


%

\end{document}